\documentclass[prx,twocolumn,floatfix,superscriptaddress, longbibliography]{revtex4-1}
\usepackage[pdftex,plainpages=false,colorlinks=true,linkcolor=blue, citecolor=blue, urlcolor=blue]{hyperref}
\usepackage{amsfonts}
\usepackage{amsmath}
\usepackage{amssymb}
\usepackage{graphicx}
\usepackage{natbib}
\usepackage{color}
\usepackage{placeins}
\usepackage{physics}

\begin{document}
 
\title{Entanglement and classical correlations at the doping-driven Mott transition in the two-dimensional Hubbard model}
\author{C. Walsh}
\affiliation{Department of Physics, Royal Holloway, University of London, Egham, Surrey, UK, TW20 0EX}
\author{P. S\'emon}
\affiliation{Computational Science Initiative, Brookhaven National Laboratory, Upton, NY 11973-5000, USA}
\author{D. Poulin}
\email[Deceased]{}
\affiliation{D\'epartement de physique \& Institut quantique, Universit\'e de Sherbrooke, Sherbrooke, Qu\'ebec, Canada J1K 2R1}
\affiliation{Canadian Institute for Advanced Research, Toronto, Ontario, Canada, M5G 1Z8}
\author{G. Sordi}
\email[corresponding author: ]{giovanni.sordi@rhul.ac.uk}
\affiliation{Department of Physics, Royal Holloway, University of London, Egham, Surrey, UK, TW20 0EX}
\author{A.-M. S. Tremblay}
\affiliation{D\'epartement de physique \& Institut quantique, Universit\'e de Sherbrooke, Sherbrooke, Qu\'ebec, Canada J1K 2R1}

\date{\today}

\begin{abstract}
Tools of quantum information theory offer a new perspective to characterize phases and phase transitions in interacting many-body quantum systems. The Hubbard model is the archetypal model of such systems and can explain rich phenomena of quantum matter with minimal assumptions. 
Recent measurements of entanglement-related properties of this model using ultracold atoms in optical lattices hint that entanglement could provide the key to understanding open questions of the doped Hubbard model, including the remarkable properties of the pseudogap phase. These experimental findings call for a theoretical framework and new predictions. 
Here we approach the doped Hubbard model in two dimensions from the perspective of quantum information theory. We study the local entropy and the total mutual information across the doping-driven Mott transition within plaquette cellular dynamical mean-field theory. 
We find that upon varying doping these two entanglement-related properties detect the Mott insulating phase, the strongly correlated pseudogap phase, and the metallic phase. Imprinted in the entanglement-related properties we also find the pseudogap to correlated metal first-order transition, its finite temperature critical endpoint, and its supercritical crossovers. 
Through this footprint we reveal an unexpected interplay of quantum and classical correlations. Our work shows that sharp variation in the entanglement-related properties and not broken symmetry phases characterizes the onset of the pseudogap phase at finite temperature. 
\end{abstract}

\maketitle

\section{Introduction}

Quantum information theory~\cite{CoverInformation, watrous2018} provides new concepts, based on the nature of the entanglement, for characterising phases of matter and phase transitions in correlated many-body systems~\cite{amicoRMP2008, EisertArea2010, Laflorencie:PhysRep2016, zhengBOOK}. 
Entanglement properties can even describe new kinds of orders beyond the Landau theory, such as quantum-topological phases~\cite{Wen:RMP2017}. 
But entanglement properties are in general elusive for measurements. However, advances with ultracold atoms have removed this barrier: recent experiments have demonstrated the ability to directly probe entanglement properties in bosonic and fermionic quantum many-body systems~\cite{greinerNat2015, Kaufman:Science2016, Cocchi:PRX2017, Lukin:Science2019, Brydges:Science2019}. 

In particular, the fermionic Hubbard model in two dimensions has a prominent role in the study of correlated many-body systems with ultracold atoms~\cite{jzHM, Esslinger:2010, Jordens:2008, Schneider:2008, Hofrichter:PRX2016, Cocchi:PRL2016, Cheuk:2016, Parsons:2016, Boll:2016, DrewesPRL2016, drewesPRL2017, Nichols:2018, Brown:Science2019}. On the theory side, this is because this model captures the essence of the correlation problem on a lattice. Experimentally, this is because its behavior can be linked to the phenomenology of high-temperature cuprate superconductors~\cite{Anderson:1987}. 

Recent experimental work of Cocchi et al.~\cite{Cocchi:PRX2017} with ultracold atoms has probed key measures of quantum correlations in the two-dimensional fermionic Hubbard model, paving the way for probing the role of the entanglement in the description of the complex phases of the model. 
On the theory side, in our work of Ref.~\cite{Caitlin:PRL2019} using the cellular extension of dynamical mean-field theory (CDMFT), we showed that two indicators of entanglement and classical correlations --the local entropy and the mutual information-- detect the first-order nature of the Mott metal-insulator transition, the universality class of the Mott endpoint, and the crossover emanating from it in the supercritical region. The local entropy is a measure of entanglement between a single site and its environment~\cite{amicoRMP2008}. The mutual information measures the total (quantum and classical) correlations between a single site and its environment~\cite{groisman2005, watrous2018}. 

However, the theoretical framework of Ref.~\cite{Caitlin:PRL2019} has so far been restricted to half filling. Hence, the exploration of the richer region of finite doping using entanglement-related properties  remains an open frontier, one that bears relevance for the unconventional superconductivity problem in cuprates~\cite{Anderson:1987, LeeRMP:2006, tremblayR}. Indeed, upon hole-doping the Mott insulating state, electrons start to delocalise but correlations due to Mott physics persist up to large doping levels~\cite{sht}. This gives rise to complex electronic behavior, epitomised by the pseudogap state. 

Here we generalise the study of Ref.~\cite{Caitlin:PRL2019}, using the same methodology, to the  region of finite doping. By tuning the level of doping, we use local entropy and mutual information to characterise the Mott insulator, the strongly correlated pseudogap phase, and the metallic state. By tuning the temperature, we address the interplay between quantum and classical correlations, which is an open research challenge of growing interest~\cite{groisman2005, wolfPRL2008, amicoRMP2008, EisertArea2010, Laflorencie:PhysRep2016} especially close to critical points~\cite{Amico_Patan_2007, Melko:2010, Singh:2011, Kallin:2011,  Wilms_Troyer_Verstraete_2011, Wilms:2012, Iaconis:2013, Gabbrielli:2019, Frerot:2019, Wald:JSM2020}. Specifically, in CDMFT the pseudogap to metal transition induced by doping is first-order, terminates in a second-order endpoint at finite doping and finite temperature, and is followed by crossover lines in the supercritical region~\cite{sht,sht2,ssht}. We found that all these features are imprinted in the entanglement-related properties of a single site, without the need to look for area law properties or topological terms~\cite{EisertArea2010}. Hence signatures at the level of entanglement persist from the low-temperature first-order quantum limit, up to the finite-temperature endpoint and beyond into the supercritical region.  
Our study corroborates the viewpoint that quantum and classical correlations resulting from Mott physics  -- and not a low-temperature symmetry-breaking order -- are at the origin of the opening of the pseudogap at finite temperature~\cite{ssht,sshtRHO,Alexis:2019} that is observed, for example, in NMR experiments~\cite{Alloul:1989}. 

Even though our calculations are performed on the square lattice to minimize Monte Carlo sign problems, the phenomena would be best observed on frustrated lattices where they would not be masked by antiferromagnetic fluctuations~\cite{reymbaut2020crossovers}.

We start in Sec.~\ref{sec:Model_and_method} by discussing the model and the quantum information tools we are using to characterize phases and phase transitions in the model. Then in Sec.~\ref{sec:PhaseDiagram} we briefly review the thermodynamic description of the normal-state phase diagram of the two-dimensional Hubbard model. This sets the stage for the information theory description of the doping-driven transition using entanglement entropy in Sec.~\ref{sec:s1}, thermodynamic entropy in Sec.~\ref{sec:s}, and mutual information in \ref{sec:I1}. We conclude with an outlook on the implications for the pseudogap problem in cuprates in Sec.~\ref{sec:Conclusions}. Additional figures for the analysis of local entropy, thermodynamic entropy, and total mutual information can be found in the appendices.

\section{Model and method}
\label{sec:Model_and_method}

\subsection{Solving the two-dimensional Hubbard model}

We study the single band Hubbard model in two dimensions on a square lattice. The Hamiltonian describing the Hubbard model in two dimensions is
\begin{align}
H & =-\sum_{\langle ij\rangle \sigma}t_{ij}c_{i\sigma}^\dagger c_{j\sigma}
+U\sum_{i} n_{i\uparrow } n_{i\downarrow }
-\mu\sum_{i\sigma} n_{i\sigma}, 
\label{eq:HM}
\end{align}
where $t_{ij}$ is the nearest-neighbour hopping amplitude, $c_{i\sigma}^\dagger$ and $c_{i\sigma}$ are the creation and annihilation operators for an electron at site $i$ with spin $\sigma$, $U$ is the Coulomb repulsion felt by an electron on site $i$, the number operator is $n_{i\sigma}=c_{i\sigma}^\dagger c_{i\sigma}$, and the chemical potential is $\mu$.

We study this model using the cellular extension~\cite{maier, kotliarRMP, tremblayR} of dynamical mean-field theory~\cite{rmp}. In this approach, the self-energy of the lattice Green's function is obtained from that of a cluster immersed in a self-consistent bath of non-interacting electrons. The properties of the bath are determined by requiring that the lattice Green's function projected on the cluster equal the cluster Green's function. We solve the plaquette (i.e. $2\times 2$ cluster) in a bath problem as a quantum impurity problem. We use the continuous-time quantum Monte Carlo method~\cite{millisRMP,patrickSkipList} based on the hybridization expansion of the impurity action. We set $t=1$ as our energy unit, and we take $k_B=1$ and $\hbar=1$. We also set the lattice spacing equal to $1$ as our distance unit. 

We follow the numerical protocol described in Refs.~\cite{Caitlin:PRL2019, CaitlinSb}. In particular, we extract the occupation $n$ and the double occupation $D$ from the empty band up to half filling, in the temperature range $1/100 \lesssim T \lesssim 1$, and in a broad $U$ range. 
All the results in this paper were obtained with a small chemical potential step (down to $\Delta\mu =0.0025$) so that derivatives can be easily calculated with the simplest finite difference. 

Let us briefly comment on the main strengths and limitations of the CDMFT method which are relevant for the results of this work. Like other cluster methods, such as dynamical cluster approach (DCA), CDMFT is a nonperturbative method that exactly treats local and nonlocal (spatial) correlations within the cluster. The inclusion of spatial correlations is a strength of the study on low dimensional systems such as the two-dimensional case studied here. It allows one to go beyond the single-site DMFT description which takes into account local correlations only. 

This strength comes with the limitation that cluster size sets the length scale of the spatial correlations that are treated exactly. To overcome this limitation, one can in principle increase the cluster size, although in practice this is a challenging computational task. In other words, CDMFT - along with other cluster methods - can be viewed as a controlled approximation where the control parameter is the cluster size. In this work we do not explicitly assess the convergence of the CDMFT results with the cluster size. Instead, by considering CDMFT with a $2\times 2$ cluster only, we constrain the correlations to be short ranged. 
Nonlocal correlations coming from long-wavelength spin fluctuations could hide certain short-range phenomena, including the Mott transition considered in this work. This is true on the square lattice~\cite{Shafer2015}. However, calculations on a small cluster of the square lattice are a good proxy for the Mott transition on the triangular lattice where frustration is strong enough to avoid the influence of long-wavelength antiferromagnetic fluctuations. We perform our calculations on the square lattice because on frustrated lattices, such as the triangular lattice, the continuous-time Quantum-Monte Carlo method that we use (CT-HYB) develops a severe sign problem~\cite{millisRMP}, related to Fermi statistics, that prevents us from reaching very low temperatures.

Note also that within CDMFT, when the correlation length remains finite, then the local observables - like the occupation $n$ and the double occupation $D$ calculated in this work - converge exponentially fast with increasing the cluster size~\cite{BiroliExp2005}. Useful benchmarks of CDMFT results with other numerical techniques can be found for the half-filled model in Refs.~\cite{Schafer:arXiv2020, LorenzoAF} and away from half filling in Ref.~\cite{LeBlancPRX}. 

Based on these considerations, we expect our results to be quantitatively correct at high temperatures, or at large $U$ (where electrons are more localized), or at large $n$ (where the effects of correlations are smaller), and only qualitatively correct in the other regions of the phase diagram (figure~\ref{figSMexp} discussed below shows a comparison with cold atom experiments).

\subsection{Extracting entanglement-related properties}
\label{sec:EntProp}

We focus on two entanglement-related properties, the local entropy and the mutual information between a single site and the rest of the lattice, averaged over all sites. 
Following our Refs.~\cite{Caitlin:PRL2019, CaitlinSb}, these two entanglement-related properties can be derived from the entanglement entropy, as explained below. 

By partitioning the lattice sites in subsystem $A$ and its complement $B=\overline{A}$, the  entanglement entropy associated to subsystem $A$ in a pure state is $s_A=-\Tr_A [\rho_A \ln \rho_A]$, where $\rho_A$ is the reduced density matrix obtained by tracing the density matrix of the total system $AB=A \cup B$ over $B$, $\rho_A = \Tr_B[\rho_{AB}]$. In information theory language, $s_A$ measures the uncertainty, or lack of information, in the state of subsystem $A$. At zero temperature $T=0$, the entanglement between $A$ and $B$ is the source of that uncertainty, and $s_A$ is a quantitative measure of the entanglement between $A$ and $B$, hence the name `entanglement entropy'. Instead, at finite temperature $T\neq 0$, $s_A$ contains thermal contributions and hence it is no longer a quantitative measure of the quantum entanglement alone~\cite{Cardy:2017, Vedral:T2004, Anders:2007}. 

At finite temperature, it is therefore useful to also consider the concept of mutual information between subsystems $A$ and $B$, which measures the total (quantum and classical) correlations shared between $A$ and $B$. The mutual information is defined as $I(A:B) = s_A +s_B -s_{AB}$, where $s_X$ is the entanglement entropy~\cite{CoverInformation, watrous2018}. Subadditivity of the entropy implies that $I(A:B)$ is non-negative. It equals $0$ if and only if the subsystems $A$ and $B$ are uncorrelated, $\rho_{AB}= \rho_A \otimes \rho_B$. Hence a non-zero value of $I(A:B)$ means that the sum of the entropy of the subsystems is larger than the entropy of the total system. In information theory language, a non-zero value of $I(A:B)$ means that the total system $AB$ contains more information than the sum of its subsystems $A$ and $B$ -- that is, $A$ and $B$ are correlated. 

Entanglement entropy and mutual information can be used to identify and characterize phases in many-body systems, and their phase transitions~\cite{amicoRMP2008, EisertArea2010}. 
For many-body systems on a lattice, a first step consists of calculating the entanglement entropy and mutual information between a single site (i.e. subsystem $A$ is just a site) and the rest of the lattice. One goal is to identify and characterize phase transitions while varying the tuning parameters of the phase transitions. A second step consists of analyzing how entanglement entropy and mutual information scale when the subsystem $A$ grows in size. One of the goals in this case is to characterize phases of matter as different structures in the correlations, and phase transitions as rearrangements of correlation patterns. 

Here we confine our study to the first step, i.e. we analyze the behavior of the single-site entanglement entropy and the mutual information between a single site and the rest of the lattice upon doping the Mott insulator within the two-dimensional Hubbard model. 
This task has the advantage that it can be implemented readily in numerical simulations and in experiments with ultracold atoms in optical lattices. 

\begin{figure}[ht!]
\centering{
\includegraphics[width=0.999\linewidth]{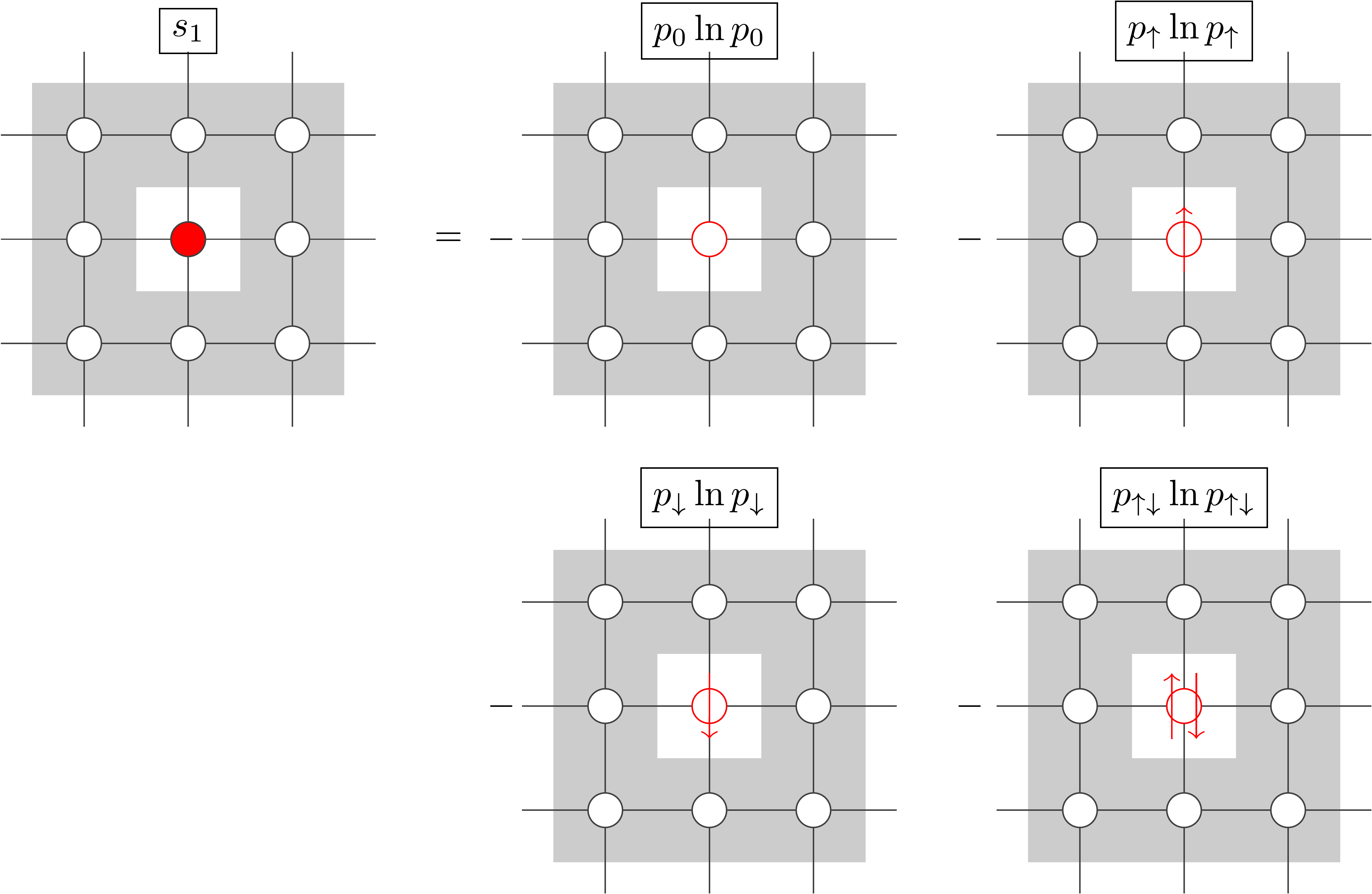}
}
\caption{Sketch visually showing how the local entropy $s_1$ is calculated. Shaded grey region indicates the degrees of freedom that are traced over to obtain $\rho_A=\Tr_B (\rho_{AB})$. The red dot indicates the site where the density matrix is used to compute $s_A$, i.e. $\rho_A = \rho_1$. 
}
\label{fig1a}
\end{figure}

First, let us discuss how to find the single-site entanglement entropy, which we shall call from now on `local entropy' $s_1$ (since we study finite temperatures, and $A$ is just a single site $A=1$). Particle and spin conservation require the reduced density matrix to be diagonal~\cite{zanardi2002}, $\rho={\rm diag}(p_0, p_{\uparrow}, p_\downarrow, p_{\uparrow\downarrow})$, where the $p_i$ are the probabilities of finding the site doubly occupied, occupied with a spin up or down particle, or empty. We have $p_{\uparrow \downarrow}=\langle n_{i\uparrow} n_{i\downarrow} \rangle=D$, $p_{ \uparrow} = p_{\downarrow} = \langle n_{i\uparrow} -n_{i\uparrow} n_{i\downarrow} \rangle$ and $p_{0}=1-2p_\uparrow -p_{\uparrow\downarrow}$.  Hence $s_1$ is 
\begin{align}
s_1 & = -\Tr_A [\rho_A \ln \rho_A] \nonumber \\
& = -\sum_i p_i \ln(p_i).
\label{eq:s1}
\end{align}
Fig.~\ref{fig1a} illustrates graphically the construction of $s_1$ using Eq.~\ref{eq:s1}. 

Hence knowledge of the occupancy $n$ and double occupancy $D$ suffice to determine $s_1$. Both $n$ and $D$ can be accurately calculated in numerical simulations~\cite{zanardi2002, amicoRMP2008, Gu:2004, larssonPRL:2005, larssonPRA2006, Anfossi_Giorda_Montorsi_Traversa_2005, Anfossi:2007, byczukPRL2012, lanataEnt, Udagawa_Motome:2015} and measured with ultracold atoms, thanks to advances in single site microscopy~\cite{Bakr:2009, Sherson:2010, Cheuk:PRL2015, Haller:NatPhys2015, Parsons:PRL2015, Omran:PRL2015, Edge:PRA2015, GrossScience2017, Hartke:2020}. 

\begin{figure}[ht!]
\centering{
\includegraphics[width=0.999\linewidth]{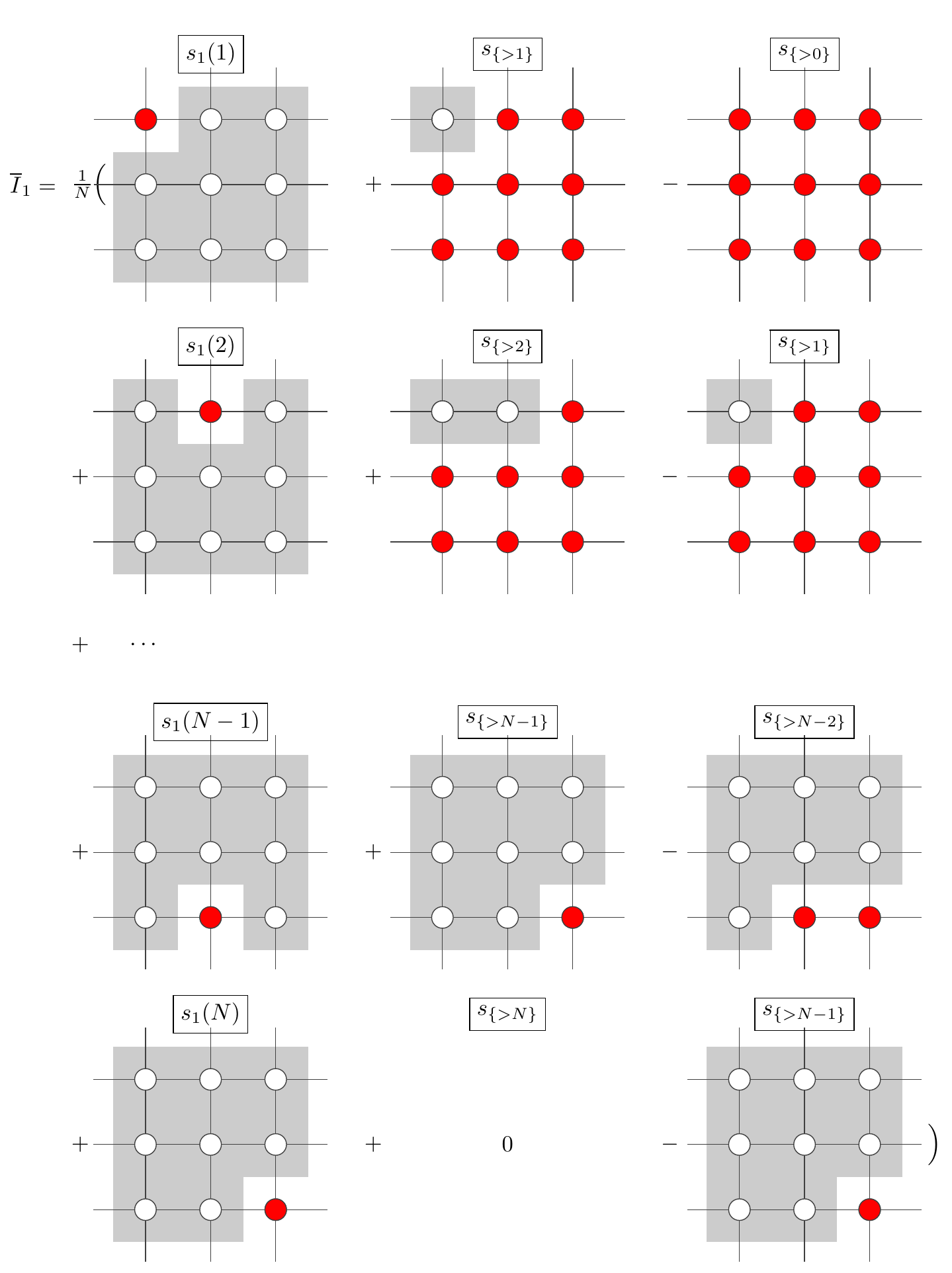}
}
\caption{Sketch visually showing how the mutual information between a single site and the rest of the lattice, averaged over all sites (that we call total mutual information $\overline{I}_1$), is calculated. Shaded grey regions indicate the degrees of freedom traced over to obtain $\rho_A=\Tr_B (\rho_{AB})$. Red dots indicate the ensemble of sites where the density matrix $\rho_A$ is used to compute $s_A$. Each term in the sum of Eq.~\ref{eq:TMI} is represented by a square lattice, starting with (left to right) $s_1(1)$, $s_{\left\{ >1 \right\}}$, $s_{\left\{ >0 \right\}}$, and so on - finishing with $s_1(N)$, $s_{\left\{ >N \right\}}=0$,  and $s_{\left\{ >N-1 \right\}}$. We assume that the lattice is infinite. 
}
\label{fig1b}
\end{figure}

Next we turn to the calculation of the mutual information between a single site and the rest of the lattice. For the site $i=1$, it is defined as $I(1: \left\{ >1 \right\}) = s_1+s_{\left\{ >1 \right\}} -s_{\left\{ >0 \right\}}$, where $\left\{ >k \right\}$ is the set of sites with indices greater than $k$ (the set $\left\{ >0 \right\}$ means the whole lattice). 
We previously showed in our work of Refs.~\cite{Caitlin:PRL2019, CaitlinSb} that further simplifications and comparison with experiments become possible if we consider the mutual information between a single site and the rest of the lattice, {\it averaged over all sites}. We called this quantity total mutual information. It is defined as 
\begin{align}
\overline{I}_1= \frac{1}{N}\sum_{i=1}^N I(i: \left\{ >i \right\}) = \frac{1}{N} \sum_{i=1}^N \left( s_1(i) +s_{\left\{ >i \right\}} -s_{\left\{ >i-1 \right\}} \right).
\label{eq:TMI}
\end{align}

Fig.~\ref{fig1b} illustrates graphically the terms in the sum of Eq.~\ref{eq:TMI}. 
The first line of Fig.~\ref{fig1b} is the standard definition of mutual information between the site $i=1$ and the rest of the lattice. The second line is the mutual information between the site $i = 2$ and the rest of the lattice, but where the site $i=1$ has been traced over to avoid double counting the correlations between sites 1 and 2. This correlation has already been taken into account in $I(1: \left\{ >1 \right\})$. 
Note that the last term of the second line cancels the second term in the first line. 
This pattern repeats and the only terms in Fig.~\ref{fig1b} that survive are $\overline{I}_1 = \left(  \sum_{i=1}^N s_1(i) \right)/N - s$, where $s=s_{\left\{ >0 \right\}}/N$ is the thermodynamic entropy per site. 
In translationally invariant systems all $s_1(i)$ are equal, so $\overline{I}_1$ simplifies further to the difference of local entropy and thermodynamic entropy, $\overline{I}_1=s_1-s$. 
Since $s$ can be obtained using the Gibbs-Duhem relation (see Sec.~\ref{sec:s}) from the knowledge of the occupancy $n(\mu)$, then $\overline{I}_1$ also can be readily computed in numerical simulations. 

Since mutual information is always positive, the total mutual information $\overline{I}_1=s_1-s$ is a sum of positive terms. We have then an information-theoretic meaning for this quantity that was first proposed in an experimental context~\cite{Cocchi:PRX2017}.

\begin{figure}[h!]
\centering{
\includegraphics[width=0.999\linewidth]{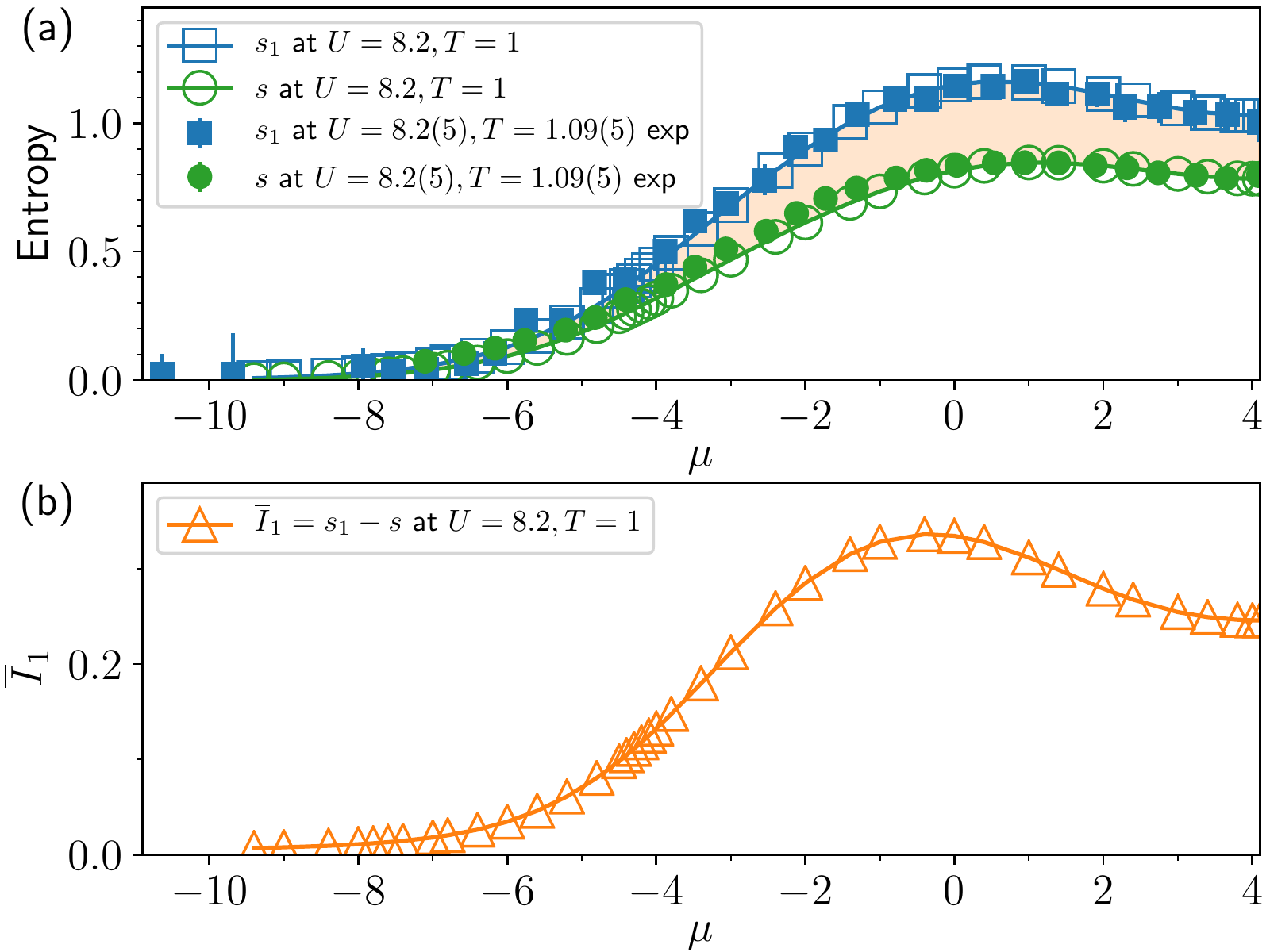}
}
\caption{(a) Local entropy $s_1$ (blue squares) and thermodynamic entropy $s$ (green circles) as a function of the chemical potential $\mu$ for $U=8.2$ and $T=1$. Open symbols are our results from CDMFT calculations. Filled symbols are experimental results of Ref.~\cite{Cocchi:PRX2017} using ultracold atoms, which show strong agreement with our data. The shaded area indicates the total mutual information $\overline{I}_1 = s_1 - s$, also shown in panel (b). 
(b) Total mutual information $\overline{I}_1$ as a function of $\mu$ for $U=8.2$ and $T=1$, defined as the mutual information between a single site and the rest of the lattice, averaged over all sites. 
}
\label{figSMexp}
\end{figure}
To illustrate our method, in Fig.~\ref{figSMexp}a we compare our CDMFT results for the entropies $s_1$ and $s$ with the experimental data of Cocchi et al.~\cite{Cocchi:PRX2017} using ultracold atoms. The overall agreement is excellent over the whole range of chemical potential $\mu$. The shaded area indicates the total mutual information $\overline{I}_1=s_1-s$, and is shown in Fig.~\ref{figSMexp}b.

\section{Phase diagram} 
\label{sec:PhaseDiagram}

\begin{figure*}[ht!]
\centering{
\includegraphics[width=0.999\linewidth]{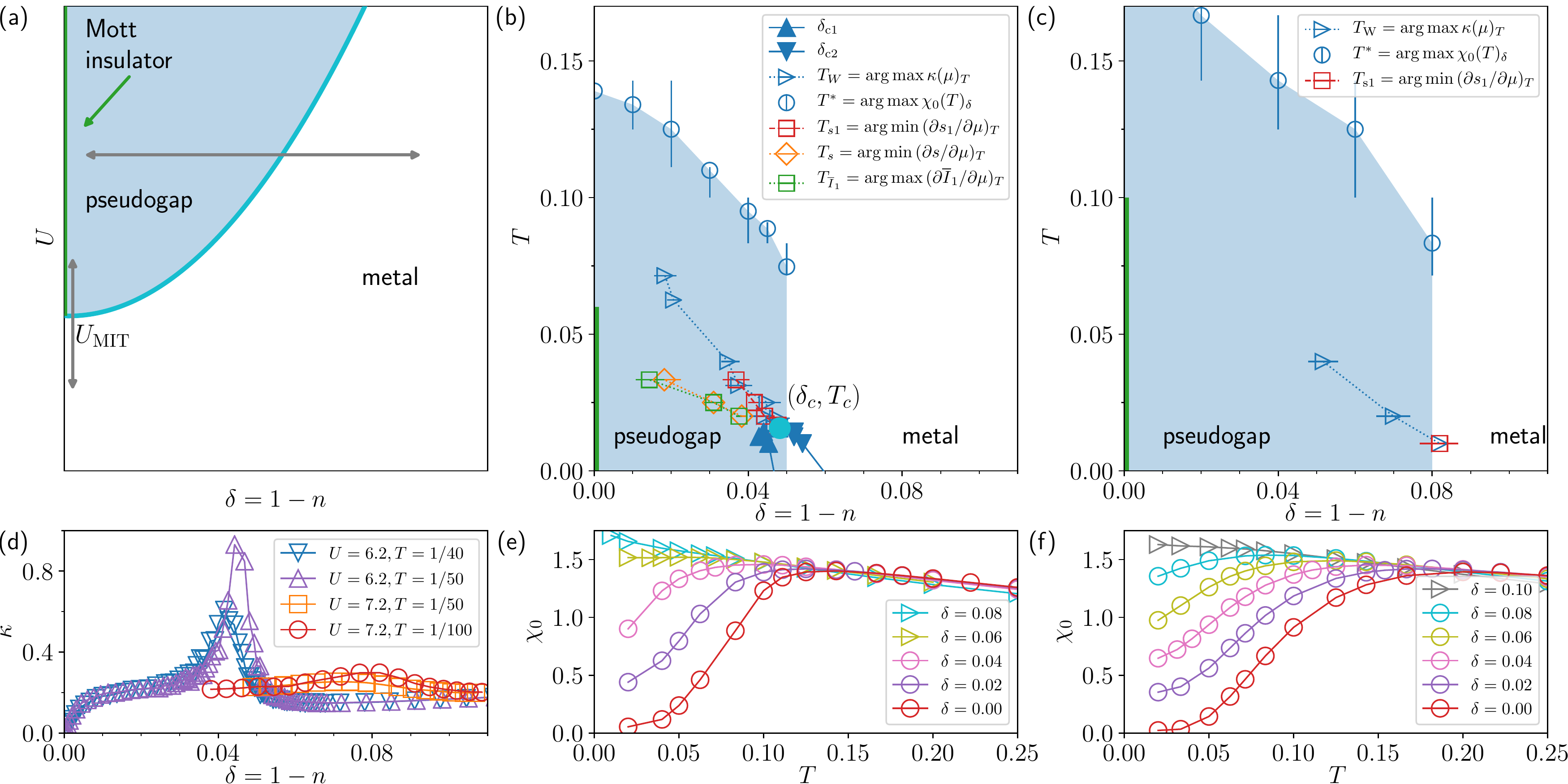}
}
\caption{(a) $U-\delta$ phase diagram at low temperature, in the normal state. Cyan line indicates a first-order transition. It separates a metal from a Mott insulator (green line) at half filling ($\delta=0$), and a strongly correlated pseudogap (shaded blue area) from a correlated metal at fixed $U>U_{\rm MIT}$. The vertical and horizontal double arrows indicate the $U$-driven and $\delta$-driven transitions respectively. The focus of this work is on the entanglement-related properties of the $\delta$-driven transition.
(b) $T-\delta$ phase diagram for $U=6.2>U_{\rm MIT}$. The shaded blue region shows the pseudogap, and the vertical green line shows the Mott insulator at $\delta=0$. The first-order pseudogap-metal transition in (b) is bounded by the spinodal lines $\delta_{c1}$ and $\delta_{c2}$, which are denoted by filled up and down triangles respectively. The first-order transition terminates in a second-order critical endpoint $(\delta_c,T_c)$, indicated by the filled cyan circle. The Widom line $T_{W}(\delta)$, a sharp crossover emanating from the critical endpoint, is shown with blue open triangles and defined here by the maxima in the isothermal charge compressibility $\kappa(\delta)$ as shown in panel (d).
Blue open circles indicate the high temperature precursor to the Widom line, known as the pseudogap temperature $T^*$, defined by the drop in the spin susceptibility $\chi_0(T)$ shown in panel (e).
Open red squares show the crossover line $T_{s_1}$ obtained from the positions of the inflection in $s_{1}(\mu)_T$, shown in Fig.~\ref{fig4} as a function of doping, which lies close to the Widom line.
Orange diamonds denote the crossover line $T_{s}$ arising from tracking the doping level of the inflection point in $s(\mu)_T$. Green squares denote the crossover line $T_{\overline{I}_1}$ arising from tracking the doping level of the inflection point showing the sharpest change in $\overline{I}_1(\mu)_T$. Key findings of this work are the inflections leading to $T_{s_1}$, $T_{s}$, and $T_{\overline{I}_1}$. 
(c) $T-\delta$ phase diagram for $U=7.2$. Symbols are the same as in (b). 
(d) Isothermal charge compressibility $\kappa = 1/n^2 (\partial n / \partial \mu)_T$ versus $\delta$ for $U=6.2, 7.2$ at low temperatures. Enhancement in the form of a peak is the signature of the Widom line. The loci of the maxima correspond to the blue open triangles in panels (b) and (c).
(e), (f) Spin susceptibility $\chi_0(T)=\int_0^{1/T} \langle S_z(\tau) S_z(0) \rangle d\tau$ for $U=6.2$ and $U=7.2$ for several dopings. $S_z$ is the projection of the total spin of the plaquette along $z$. The loci of the temperatures where the spin susceptibility drops on reducing $T$ defines the pseudogap temperature $T^*$, denoted by blue open circles in panels (b) and (c).  
}
\label{fig2}
\end{figure*}
The phase diagram of the two-dimensional Hubbard model is spanned by $U$, $\delta$ (or equivalently, $\mu$), and $T$. As a result of intense work~\cite{sht,sht2,ssht}, the main thermodynamic features of the plaquette CDMFT solution for the normal state are known. In contrast, the entanglement properties are largely unexplored. Let us briefly review the key thermodynamic features~\cite{sht,sht2,ssht,Giovanni:PRBcv}. As sketched in the $U-\delta$ low temperature cross-section of Fig.~\ref{fig2}a, the system features a first-order phase transition (see thick cyan line) originating at half filling $\delta=0$ and at a critical interaction $U_{\rm MIT}$ and extending to progressively higher doping with increasing $U$~\cite{sht,sht2}. The plaquette here constrains antiferromagnetic fluctuations to be short-range, as they would be in a frustrated lattice~\citep{reymbaut2020crossovers}. 

At $\delta=0$ and on varying $U$, this phase transition is between a metal and a Mott insulator~\cite{CaitlinSb}. This is the $U$-driven Mott metal-insulator transition, and is indicated by a vertical double arrow in Fig.~\ref{fig2}a. At constant $U>U_{\rm MIT}$ and on varying $\delta$, this first-order phase transition is between two metallic phases: a strongly correlated pseudogap at low $\delta$ and a correlated metal at high $\delta$~\cite{ssht}. These two phases share the same symmetry and are distinguished by their electronic density $n$ at the phase transition. Therefore, on varying $\delta$ for $U$ larger than the threshold $U_{\rm MIT}$ for the Mott transition at $\delta=0$, the system undergoes a continuous change between a Mott state and a pseudogap, and an abrupt (first-order) change between a pseudogap and a metal~\cite{ssht}. With further increasing doping, this metal progressively becomes less correlated, and eventually the system becomes a band insulator at $\delta=1$ ($n=0$, i.e. no particles). This sequence of phases realizes the $\delta$-driven Mott metal-insulator transition, and is indicated by a horizontal double arrow in Fig.~\ref{fig2}a. 

By adding the temperature axis, the first-order transition line in the $U-\delta$ phase diagram evolves into a first-order surface in the $U-\delta-T$ phase diagram. On increasing $T$, this transition surface is interrupted at a second-order critical line. Beyond this critical temperature, sharp crossovers emerge from the critical line. This is best shown by taking cross-sections of the $U-\delta-T$ phase diagram at constant $U$, for $U>U_{\rm MIT}$, as shown in Figures~\ref{fig2}b and \ref{fig2}c, which show our plaquette CDMFT results for $U=6.2$ and $U=7.2$. The system features a first-order transition bounded by spinodal lines $\delta_{c1}$ and $\delta_{c2}$, where the strongly correlated pseudogap and correlated metal disappear respectively. This transition ends in a second-order critical point at $(\delta_c, T_c)$. 
Beyond this, a sharp crossover emerges -- this is $T_{W}(\delta)$, the so-called Widom line (shown as open blue triangles). It is defined as the locus of the maxima of the correlation length~\cite{water1, supercritical, ssht}. Extrema of thermodynamic response functions converge to the Widom line upon asymptotically approaching the endpoint. It has been shown that charge compressibility~\cite{sht,sht2,ssht}, thermodynamic and nonlocal density fluctuations~\cite{CaitlinOpalescence}, and specific heat~\cite{Giovanni:PRBcv} all show anomalous enhancement upon crossing the Widom line. Operationally, we define the Widom line using the doping level at which the isothermal compressibility $\kappa(\delta)$ peaks~\cite{ssht}, see Fig.~\ref{fig2}d. 

Note that the second-order critical point moves to progressively larger doping and lower temperature with increasing $U$~\cite{sht,sht2}: at $U=7.2$ it is below the lowest temperature we can access because of the sign problem, but the Widom line is clearly visible. This is where the Widom line proves as a useful indicator: when the endpoint and the underlying first-order transition lie below some inaccessible region, the Widom line is a high temperature predictor of these phenomena. The low temperature region may be inaccessible because the sign problem prevents us from probing low temperatures, but also experimentally it could be inaccessible because ordered phases hide the transition, for example. 

The endpoint and the Widom line in the supercritical region have a high temperature precursor, the so-called $T^*(\delta)$ line (open blue circles), which occurs for $\delta \le \delta_c$ only. This crossover can be interpreted as the pseudogap temperature~\cite{sshtRHO, Alexis:2019}. Operationally it can be defined as the loci of the temperatures where, on reducing $T$, the spin susceptibility $\chi_0(T)_\delta$ drops, or $c$-axis resistivity increases~\cite{sshtRHO}. Since the finite doping first-order transition is continuously connected to the Mott transition at half filling, the pseudogap that arises because of the associated Widom line is a consequence of the Mott transition at half filling. 

Overall the phase diagram of the two-dimensional Hubbard model shows many similarities with the phase diagram of hole-doped cuprates~\cite{Alloul2013, ssht,Giovanni:PRBcv}. 

In Ref.~\cite{Caitlin:PRL2019} we showed that the local entropy and total mutual information detect the $U$-driven Mott transition of the half filled model, by showing characteristic inflections as a function of $U$. Now we generalize the study of these entanglement-related properties to the metallic phases at nonzero doping.

\section{Local entropy} 
\label{sec:s1}

We shall show that the local entropy $s_1$ defined in Sec.~\ref{sec:EntProp} detects the first-order transition between pseudogap and metal, the second-order endpoint, and the crossover emerging from it.

\subsection{Doping dependence of $s_1$}

\begin{figure}[ht!]
\centering{
\includegraphics[width=0.999\linewidth]{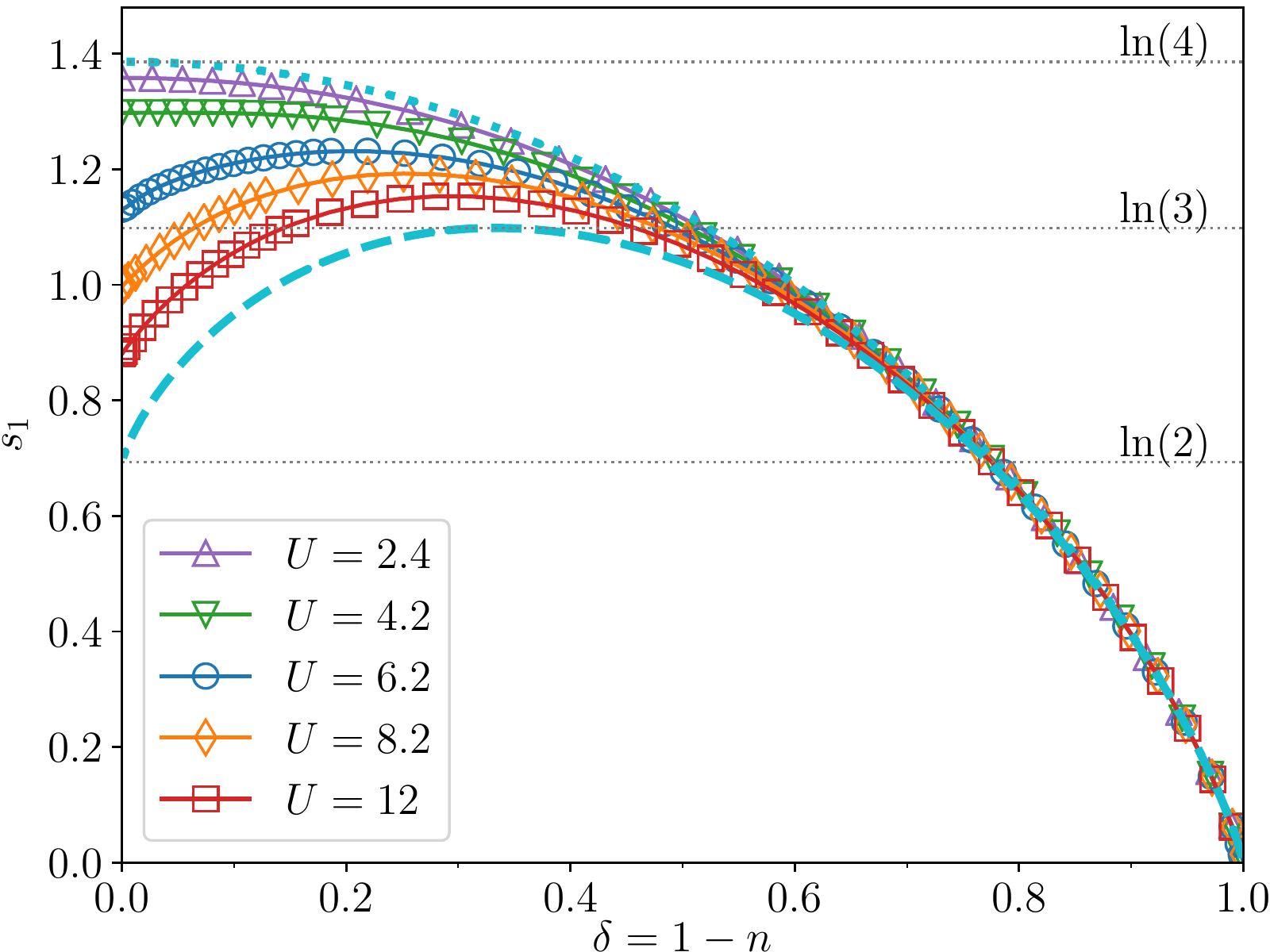}
}
\caption{Local entropy $s_1$ versus $\delta$ at $T=1/10$ for several values of $U$, bounded by the limiting cases $U=0$ (cyan dotted line) and $U=\infty$ (cyan dashed line). Grey dotted lines indicate $\ln(2)$, $\ln(3)$, and $\ln(4)$. 
}
\label{fig3}
\end{figure}
Figure~\ref{fig3} shows the local entropy $s_1$ as a function of $\delta=1-n$ for different values of $U$, at the intermediate temperature $T=1/10$. To understand the behavior of $s_1(\delta)$ it is useful to consider the $U=0$ (cyan dotted line) and $U=\infty$ limits (cyan dashed line). 

For $U=0$, the probability of double occupancy $p_{\uparrow\downarrow}=n^2/4$, and thus $s_1$ monotonically decreases with increasing $\delta$ from $\ln(4)$ to $0$. Physically, less states become available with increasing $\delta$. In information theory language, the uncertainty about the site occupation decreases and hence its entropy decreases. 

In contrast, for $U=\infty$, $D=0$ and  $s_1$ is non-monotonic with $\delta$. Exactly at half filling, the limit of $s_1$ is $\ln(2)$. This is the entropy of a free spin. Its value coincides with the entanglement entropy of a singlet. Then $s_1$ increases with increasing doping to $\ln(3)$, and subsequently decreases to zero. Indeed, $U$ suppresses the double occupancy and thus the number of available states on a single site and thus $s_1$, especially close to the Mott insulator at $\delta=0$ $(n=1)$. In information theory language, the suppression of double occupancy leads to less uncertainty about the site occupation, and hence to a decrease of the local entropy upon approaching $\delta=0$.

Therefore the behavior of the CDMFT data for $s_1(\delta)_T$ results from the competition between Fermi statistics and Mott localization. For $\delta \gtrsim 0.5$ CDMFT data are not very dependent on $U$, suggesting that Fermi statistics dominate at large $\delta$. At low $\delta$, for $U$ larger than the critical interaction, Mott localization leads to a decrease in $s_1$ with decreasing $\delta$.

\subsection{$s_1$ detects the first-order transition between strongly correlated pseudogap and correlated metal}

\begin{figure*}[ht!]
\centering{
\includegraphics[width=0.999\linewidth]{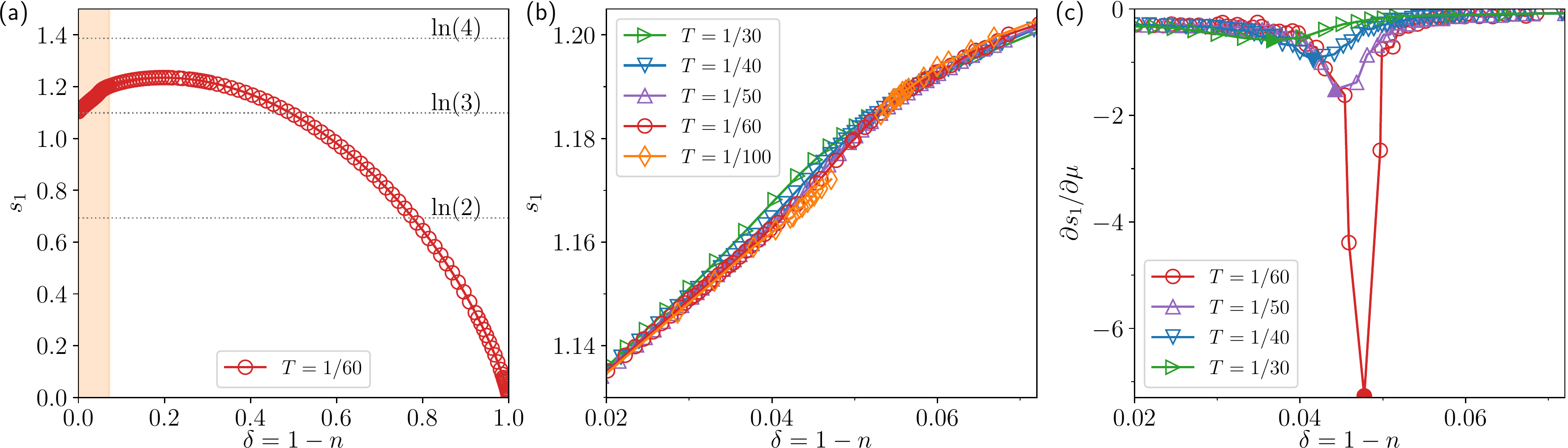}
}
\caption{(a) Local entropy $s_1$ versus $\delta$ at $T=1/60$ for $U=6.2>U_{\rm MIT}$. 
(b) $s_1$ versus $\delta$ for $U=6.2$ and low temperatures in the narrow doping interval shaded in (a). For $T=1/100 < T_c$, $s_1(\delta)$ is discontinuous. 
(c) $\partial s_1 / \partial \mu$ versus $\delta$ at $U=6.2$ for different temperatures above $T_c$. Upon approaching the endpoint $(\delta_c,T_c)$, the peak in $\partial s_1 / \partial \mu$ sharpens and narrows, and diverges at the endpoint - a signature of the Widom line. The loci of the minima in $\partial s_1 / \partial \mu$ for each temperature (see filled symbols) define the crossover line $T_{s_1}$ in Fig.~\ref{fig2}b. 
}
\label{fig4}
\end{figure*}
Next we investigate whether the pseudogap to correlated metal transition and its associated crossovers in the thermodynamic properties leave signatures in the local entropy. Figures~\ref{fig4}a, b show the local entropy $s_1(\delta)_T$ across the pseudogap to correlated metal transition at $U=6.2$ and for different temperatures --in other words, we are taking fixed $T$ cross sections of the phase diagram in Fig.~\ref{fig2}b.

For $T< T_c$, the transformation between two correlated metals -a strongly correlated pseudogap and a correlated metal- is first-order, with an abrupt jump in the density $n(\mu)$~\cite{sht,sht2}. This thermodynamic change is imprinted on the local entropy $s_1(\delta)_T$, which is also discontinuous: at $\delta_{c1}$ the pseudogap disappears, and at $\delta_{c2}$ the correlated metal disappears. The local entropy $s_1$ shows hysteresis as a function of the chemical potential $\mu$, as seen in Fig.~\ref{figSMmu62}b of Appendix~\ref{AppendixA}. We found $(s_1)_{\rm pseudogap} < (s_1)_{\textrm{correlated metal}}$, since the double occupancy $D$ is smaller in the pseudogap (this result about $D$ can be obtained from Clausius-Clapeyron equation~\cite{sht2}). 

Physically, in the strongly correlated pseudogap phase, doped holes (holons) move in a background of short-ranged singlet states caused by the superexchange mechanism. This is shown by the drop in the spin susceptibility $\chi_0(T)_\delta$ (see Figs.~\ref{fig2}e, f) and by analyzing the statistical weight of the plaquette eigenstates~\cite{hauleDOPING,sht2, ssht}. At larger doping, in the correlated metal, the carriers are quasiparticles. 
In summary, the sudden change in the many-body state between two correlated metals is captured by the discontinuous change in the local entropy $s_1$ across the transition.

\subsection{$s_1$ detects the critical endpoint and its Widom line}

The first-order transition between two metals ends in a second-order endpoint at $(\delta_c, T_c)$, where thermodynamic response functions such as charge compressibility~\cite{sht,sht2,ssht}, density fluctuations~\cite{CaitlinOpalescence}, and specific heat~\cite{Giovanni:PRBcv} diverge. The local entropy $s_1(\mu)_T$ detects the endpoint by showing an inflection as a function of $\mu$ with a vertical tangent. 
This can be understood by writing $s_1(\mu)_T$ using the chain rule as a function of $n$, as   $(\partial s_1/\partial n) (\partial n /\partial \mu)$, where $(\partial s_1/\partial n)$ is regular and $(\partial n /\partial \mu)$ is singular. The charge compressibility $\kappa$ is proportional to $\partial n / \partial \mu$ and indeed diverges at $(\delta_c,T_c)$~\cite{ssht,CaitlinOpalescence}, and hence so does $s_1$.  

For $T > T_c$, the singular behavior of the thermodynamic response functions is replaced by a sharp crossover, known as the Widom line. This thermodynamic signature also affects the local entropy $s_1$, which shows inflections versus $\mu$ (that are not visible in $s_1$ versus $\delta$). The position of the inflections can be located by taking the derivative of $s_1(\mu)_T$ with respect to $\mu$. Fig.~\ref{fig4}c shows $\partial s_1 / \partial \mu$ as a function of $\delta$ for several $T$ above $T_c$. The peaks narrow and sharpen as $T\rightarrow T_c$ from above. This is the distinctive feature of the Widom line (see Figs.~\ref{figSMinflections}a, d of Appendix~\ref{AppendixD} for $s_1$ and $\partial s_1 / \partial \mu$ versus $\mu$). 

The loci of inflections of $s_1(\mu)_T$ and of $\kappa(\mu)_T$ are close and converge at $T_c$ (see filled symbols in Fig.~\ref{fig4}c showing the inflection points). Their location is represented as open red squares in the phase diagram of Fig.~\ref{fig2}b. Similar results are obtained for $U=7.2$ (see Fig.~\ref{fig2}c for the $T-\delta$ phase diagram at $U=7.2$ and Fig.~\ref{figSMu72} of Appendix~\ref{AppendixA} for the parallel to Fig.~\ref{fig4}). 
Hence $s_1$ detects the critical endpoint of the first-order pseudogap-metal transition and its associated Widom line.

\section{Thermodynamic entropy} 
\label{sec:s}

The thermodynamic entropy per site is $s=-\Tr(\rho \ln \rho)/N$, where $\rho$ is the density matrix. To find $s$, we follow the protocol previously described in Refs.~\cite{CaitlinSb,Cocchi:PRX2017}. 
It exploits the Gibbs-Duhem relation
\begin{align}
sdT -adP +nd\mu & =0, 
\end{align}
where $a$ is the area and $P$ the pressure. At constant $T$ and $U$, it simplifies to $nd\mu = a dP$. Therefore we extract $s$ by following three steps (as illustrated in Fig.~\ref{figSM-s} of Appendix~\ref{AppendixB}). First we calculate the occupation $n(\mu)_T$, then we integrate it over $\mu$ to obtain the pressure $p(\mu)_T$: 
\begin{align}
p(\mu)_T & = \frac{1}{a} \int_{-\infty}^\mu n(\mu')_T d\mu' .
\end{align}
Finally, by differentiating the pressure with respect to temperature, we extract the thermodynamic entropy $s(\mu)_T=a(dP/dT)_\mu$.

\subsection{Doping dependence and maximum of $s$}

\begin{figure}
\centering{
\includegraphics[width=0.999\linewidth]{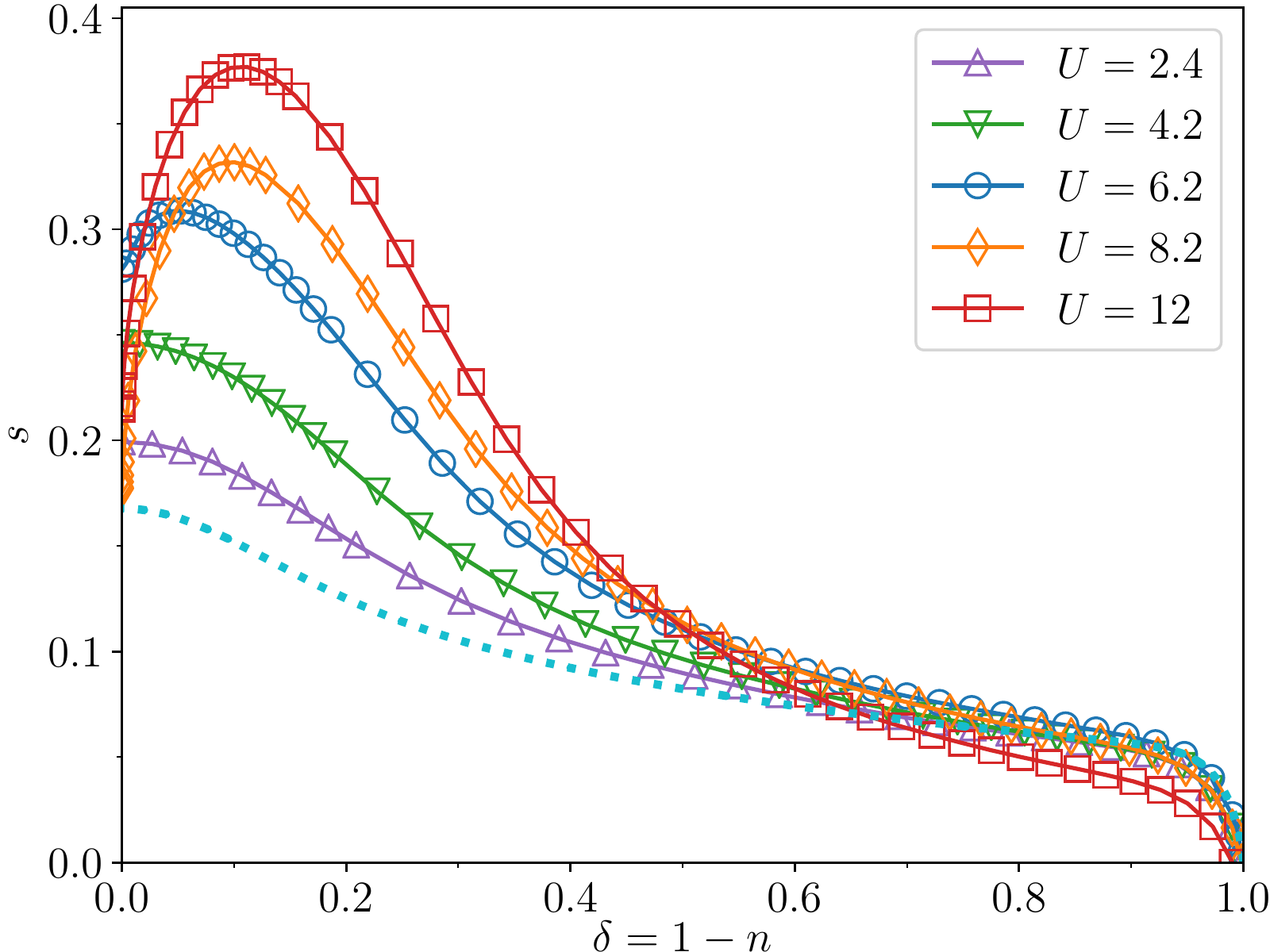}
}
\caption{Thermodynamic entropy per site $s$ versus $\delta$ at $T=1/10$ for several values of the interaction strength $U$. We estimate $s$ at $T=1/10$ by taking finite differences between pressures curves at $T=1/8$ and $T=1/10$. The limiting case for $U=0$ at $T=1/10$ is shown by the cyan dotted line. 
}
\label{fig5}
\end{figure}
Figure~\ref{fig5} shows the entropy per site $s$ as a function of $\delta$ at $T=1/10$ for several values of $U$. 
At small $U$, $s(\delta)$ is monotonically decreasing with increasing $\delta$. In contrast, for large values of $U$ (i.e. upon doping the Mott insulator), $s(\delta)$ increases until it reaches a maximum, and then decreases to $0$ as $\delta \rightarrow 1$ ($n \rightarrow 0$). The increase of $s$ with increasing $\delta$ at large $U$ is due to the fact that in the Mott insulator charge fluctuations are suppressed and doping releases them, thereby increasing $s$. 
On the other hand, upon further increasing doping, $s$ decreases with increasing doping. 
This is because the effect of interactions becomes negligible at large doping where the band is almost empty, as can be seen by comparison with the non-interacting case (cyan dotted line). Note that the interactions almost always increase the entropy compared with the non-interacting case. Also, interactions lead to a maximum in entropy in a region of doping where, in cuprates, ordered states are found at low temperature.

Previous work with CDMFT~\cite{sht,sht2, Alexis:2019} and other methods (such as the dynamical cluster approximation~\cite{MikelsonThermodynamics:2009}) revealed the maximum in $s(\delta)$, which can also be found via Maxwell relations by the crossing of the isotherms $(\partial n /\partial T)_\mu = 0$. The maximum is also captured by the atomic limit, and is another manifestation of localisation/delocalisation physics.

\subsection{$s$ detects the critical endpoint of the first-order transition and its Widom line}

\begin{figure*}
\centering{
\includegraphics[width=0.999\linewidth]{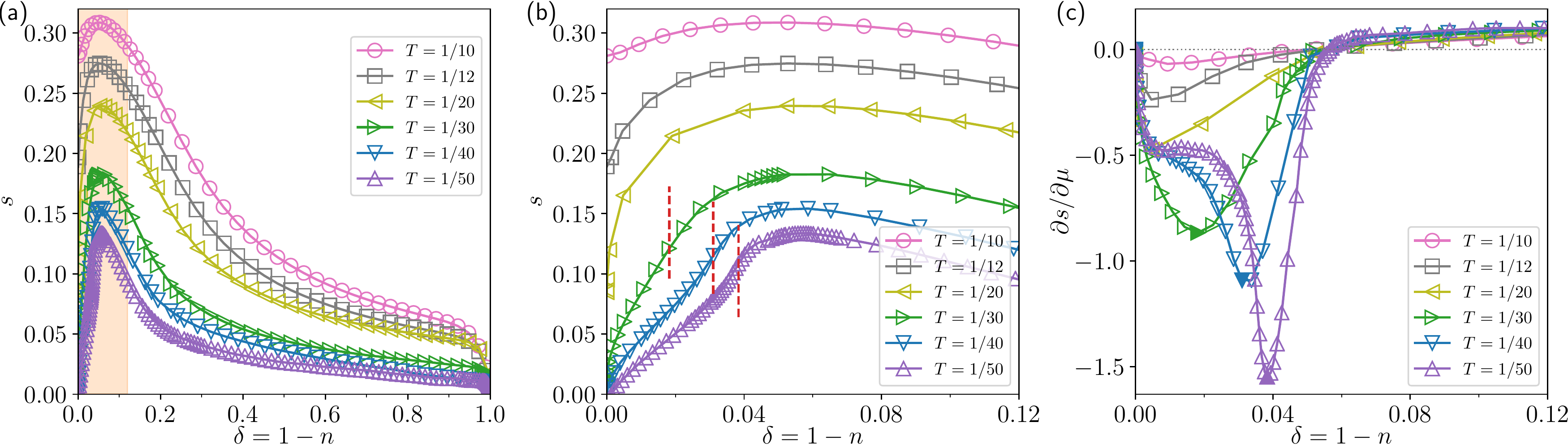}
}
\caption{(a) Thermodynamic entropy per site $s$ versus $\delta$ at $U=6.2>U_{\rm MIT}$ for different temperatures. We estimate $s$ at $T=1/12, 1/20, 1/30, 1/40, 1/50$ by taking finite differences between pressures curves at $T=1/10$ and $T=1/12$, at $1/12$ and $1/20$, at $1/20$ and $1/30$, at $1/30$ and $1/40$, and at $1/40$ and $1/50$, respectively. 
(b) The same as (a) for the doping interval shaded in (a), and highlighting the positions of the inflections that can be seen in $s(\mu)_{T}$ (which are not visible as a function of $\delta$) with vertical dashed lines. 
(c) $\partial s / \partial \mu$ versus $\delta$ for different temperatures at $U=6.2$, showing the minima which become sharper and more negative on approaching $T_c$. We have used the loci of these minima (see filled symbols) to obtain the crossover line $T_s$ in the $T-\delta$ phase diagram of Fig.~\ref{fig2}b.
}
\label{fig6}
\end{figure*}

Fig.~\ref{fig6}a shows $s(\delta)_T$ for $U=6.2$ and different temperatures. Because the procedure of extracting $s$ is computationally expensive and both the sign problem and the error propagation in calculating $s$ degrade with lowering $T$, we restrict our analysis to  $U=6.2$ only, and for $T\ge 1/50$. 

As noticed in earlier work~\cite{sht2, Alexis:2019}, the maximum in $s(\delta)$ occurs at a doping larger than the transition $\delta_c$. The central feature that we reveal here is the {\it inflection} in $s$ as a function of $\mu$ (see vertical bar in Fig.~\ref{fig6}b), where the entropy shows the largest variation with $\mu$. 
To locate the inflections, we perform the numerical derivative of $s$ with respect to $\mu$. Fig.~\ref{fig6}c shows the result $\partial s /\partial\mu$ plotted as a function of $\delta$. Note that $s$ and $\partial s /\partial\mu$ are shown as functions of $\mu$ in Figs.~\ref{figSMinflections}b, e of Appendix~\ref{AppendixD}. The position of the inflection in $s(\mu)_T$ is marked with a dashed vertical line in Fig.~\ref{fig6}b as the inflection itself is not visible as a function of doping. 
The slope becomes steeper on approaching $T_c$ (see $\partial s /\partial \mu$ curves in Fig.~\ref{fig6}c, where the minima become sharper and more negative). Thus we expect that the inflection becomes a vertical slope, hence $s$ shows singular behavior. 
 
By tracking the doping level of the inflection point of $s(\delta)_T$ at different temperatures we obtain a crossover line in the $T-\delta$ diagram. This crossover is denoted by $T_s$ (orange diamonds in Fig.~\ref{fig2}b). 
Although the precise location of $T_s$ in the $T-\delta$ diagram is affected by the uncertainty introduced by our method of extracting $s$, the important result we obtain is a crossover which extends into the supercritical region.

\subsection{$s$ detects the first-order transition between strongly correlated pseudogap and correlated metal} 

For $T<T_c$, the entropy will be discontinuous at the pseudogap to correlated metal transition. Using the Clausius-Clapeyron relation along the first-order transition, Refs.~\cite{sht,sht2} inferred that the entropy of the pseudogap is smaller than that of the correlated metal, $(s_{\rm pseudogap}) < (s_{\textrm{correlated metal}})$. The error propagation prevents us from obtaining $s$ below $T_c$.

\section{Total mutual information} 
\label{sec:I1}

Next we turn to the total mutual information between a single site and its environment, which measures classical and quantum correlations. It is simply $\overline{I}_1 = s_1 - s$. Figure~\ref{figSM-I1} of Appendix~\ref{AppendixC} illustrates graphically the construction of $\overline{I}_1$.

\subsection{Doping dependence of $\overline{I}_1$}
\begin{figure}[ht!]
\centering{
\includegraphics[width=0.999\linewidth]{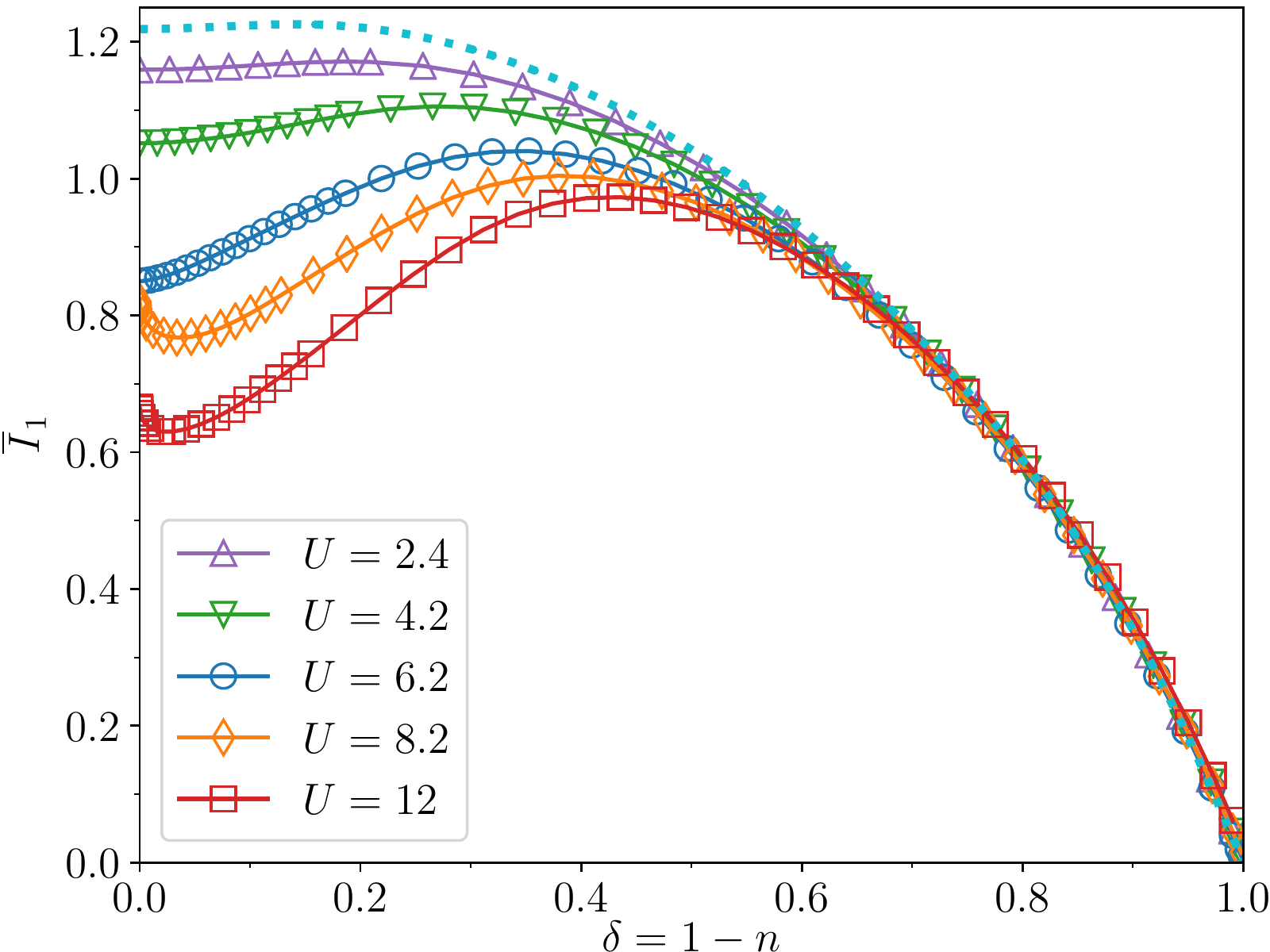}
}
\caption{Total mutual information $\overline{I}_1=s_1-s$ versus $\delta$ at $T=1/10$ for several values of the interaction strength $U$. The limiting case for $U=0$ at $T=1/10$ is shown by the cyan dotted line. 
}
\label{fig7}
\end{figure}
Figure~\ref{fig7} shows $\overline{I}_1$ as a function of doping $\delta$ for several values of $U$ at $T=1/10$. 
For large doping $\delta \gtrsim 0.5$, $\overline{I}_1$ is governed by $s_1$, does not depend much on $U$, and decreases with increasing $\delta$. Physically, this reflects the decrease of quantum and classical correlations on approaching the full insulating hole band, or equivalently the empty electron band at $\delta=1$. The dilution leads to a mutual information that does not depend on the interaction $U$. 

For $\delta \lesssim 0.5$, $\overline{I}_1$ develops a local minimum (at $\delta=0$ or at finite $\delta$ depending on the interaction $U$) that occurs from the mismatch between the positions of the peaks in $s_1(\delta)$ and in $s(\delta)$. Physically, at large $U$ the minimum comes from the competing spin and charge correlations. On the right side of the minimum, charge correlations win because states become more extended, hence the density matrix does not factor in position space, and hence correlations encoded in $\overline{I}_1$ increase with increasing $\delta$. In contrast, on the left side of the minimum, spin correlations win over charge correlations. This is because at large $U$ the superexchange coupling grows with decreasing $\delta$, hence $\overline{I}_1$ increases on approaching the Mott insulator at $\delta=0$. 

Note that $\overline{I}_1$ is non-zero in the Mott state at $\delta=0$, implying non-zero correlations between a single site and its surroundings~\cite{Cocchi:PRX2017, Caitlin:PRL2019}. Physically, this is due to the superexchange that locks the spins into singlet states. 

Note that, again, at large doping the total mutual information returns to its non-interacting value (cyan dotted line). In addition, interactions decrease the total mutual information compared with the non-interacting case because they tend to make many-body states more localized, hence reducing the correlations between a site and its environment.

\subsection{$\overline{I}_1$ detects the critical endpoint of the first-order transition and its Widom line}

\begin{figure*}
\centering{
\includegraphics[width=0.999\linewidth]{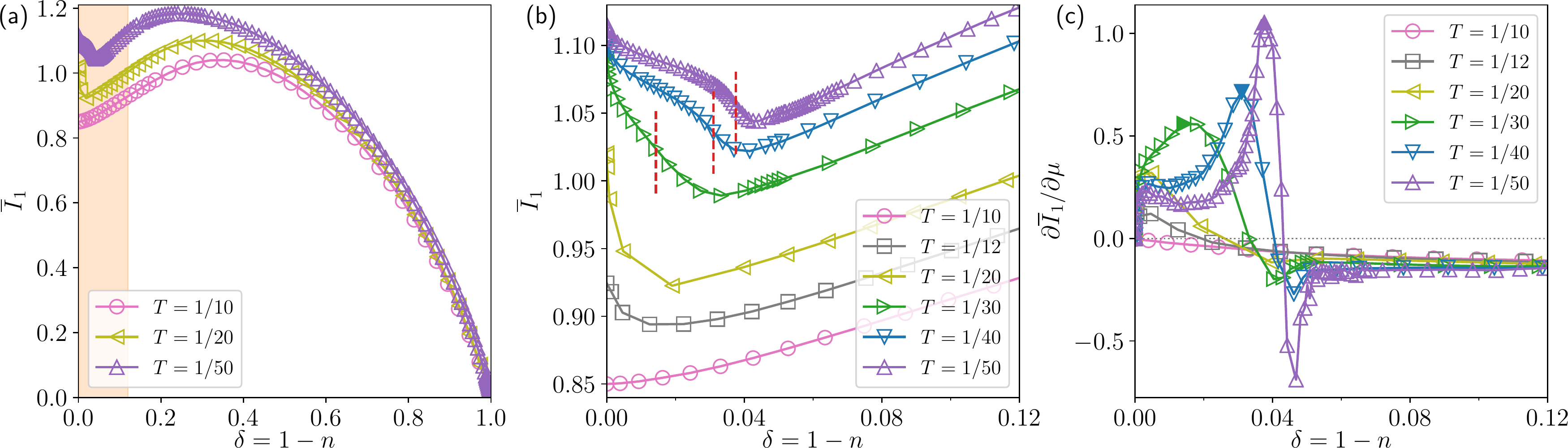}
}
\caption{(a) Total mutual information $\overline{I}_1$ versus $\delta$ for $U=6.2>U_{\rm MIT}$ and different temperatures. 
(b) The same as (a) for the doping interval shaded in (a), and highlighting the positions of the inflections that can be seen in $\overline{I}_{1}(\mu)_{T}$ (which are not visible as a function of $\delta$) with vertical dashed lines. 
(c) $\partial \overline{I}_1 / \partial \mu$ versus $\delta$ for few temperatures at $U=6.2$, showing the peaks which become more pronounced on approaching $T_c$. We have used the loci of the inflections that show the sharpest change in $\overline{I}_1(\mu)$ to obtain the crossover line $T_{\overline{I}_1}$ in the $T-\delta$ phase diagram of Fig.~\ref{fig2}b. 
}
\label{fig8}
\end{figure*}

Fig.~\ref{fig8}a shows $\overline{I}_1$ for $U=6.2>U_{\rm MIT}$ at different temperatures. Overall, $\overline{I}_1$ increases with decreasing $T$. Physically, this means that the correlations between a single site and its surroundings increase with decreasing $T$. 
The temperature dependence is stronger close to the Mott insulator at $\delta=0$ and weakens at larger $\delta$, where electrons are more diluted in the lattice. This reveals that superexchange physics comes into play upon lowering $T$ close to the Mott state. 

Next we turn to the behavior of  $\overline{I}_1(\delta)$ in the supercritical region emerging from the endpoint at $(\delta_c, T_c)$, see Fig.~\ref{fig8}b. The position of the local minimum in $\overline{I}_1(\delta)_T$ shifts to higher doping upon approaching the endpoint from above. Between the Mott insulator at $\delta=0$ and the locus of the minimum, $\overline{I}_1(\mu)_T$ develops an inflection as a function of the chemical potential $\mu$. The dashed vertical line in Fig.~\ref{fig8}b indicates the doping level of the inflection in $\overline{I}_1(\mu)_T$, as the inflection itself does not occur in $\overline{I}_1(\delta)_T$. The inflection marks a rapid change of $\overline{I}_1(\mu)_T$ in the vicinity of the endpoint and the Widom line. This rapid variation reflects the slopes in $s_1(\mu)$ and $s(\mu)$, characterised by inflections, that become infinite slopes at $(\delta_c, T_c)$, as discussed in Fig.~\ref{fig4}c and Fig.~\ref{fig6}c. 

To locate the inflections in $\overline{I}_1(\mu)_T$, we perform the numerical derivative with respect to $\mu$ -- see $\partial \overline{I}_1/\partial\mu$ plotted as a function of $\delta$ in Fig.~\ref{fig8}c and as a function of $\mu$ in Fig.~\ref{figSMinflections}f of Appendix~\ref{AppendixD}. Note that among the different inflections in $\overline{I}_1(\mu)_T$ (or equivalently peaks in $(\partial \overline{I}_1/\partial\mu)_T$), we have used the one showing the sharpest change in $\mu$ (i.e. the upward peak in Fig.~\ref{fig8}c). 
Although the precise determination of the inflections suffers from the uncertainty in our evaluation of $s$, this choice for the inflection is also consistent with our analysis at half filling in Ref.~\cite{Caitlin:PRL2019}.

Taking into account these limitations, the loci of these inflections marking the sharpest change of $\overline{I}_1(\mu)_T$ at different temperatures define the crossover line $T_{\overline{I}_1}$ in the $T-\delta$ phase diagram in Fig.~\ref{fig2}b (green squares). 
The loci of infinite slopes in $s_1(\mu)$ and $s(\mu)$ coincide at the endpoint $(\delta_c,T_c)$, so that $\overline{I}_1(\mu)$ also has an infinite slope at the endpoint.
Therefore $\overline{I}_1$ detects the pseudogap to correlated metal crossover emerging from the critical endpoint. 

For $T < T_c$, $\overline{I}_1(\delta)$ will be discontinuous. However, since we cannot reliably extract $s$ below $T_c$, we cannot conclude on whether $(\overline{I}_1)_{\rm pseudogap} \gtrless (\overline{I}_1)_{\textrm{correlated metal}}$. 
A hint may come from the study of $\overline{I}_1$ across the $U$-driven Mott transition as in Ref.~\cite{Caitlin:PRL2019}. At half filling, at the metal to Mott insulator transition we found $(\overline{I}_1)_{\rm metal} < (\overline{I}_1)_{\textrm{Mott insulator}}$. Since the pseudogap to metal transition is connected to the $U$-driven Mott transition (see cyan line in the $U-\delta$ diagram of Fig.~\ref{fig2}a), we can speculate that $(\overline{I}_1)_{\textrm{correlated metal}} <  (\overline{I}_1)_{\rm pseudogap}$, although further work is needed to investigate this.

\subsection{Comparing the local entropy with and without tunneling to neighboring sites}

\begin{figure}
\centering{
\includegraphics[width=0.999\linewidth]{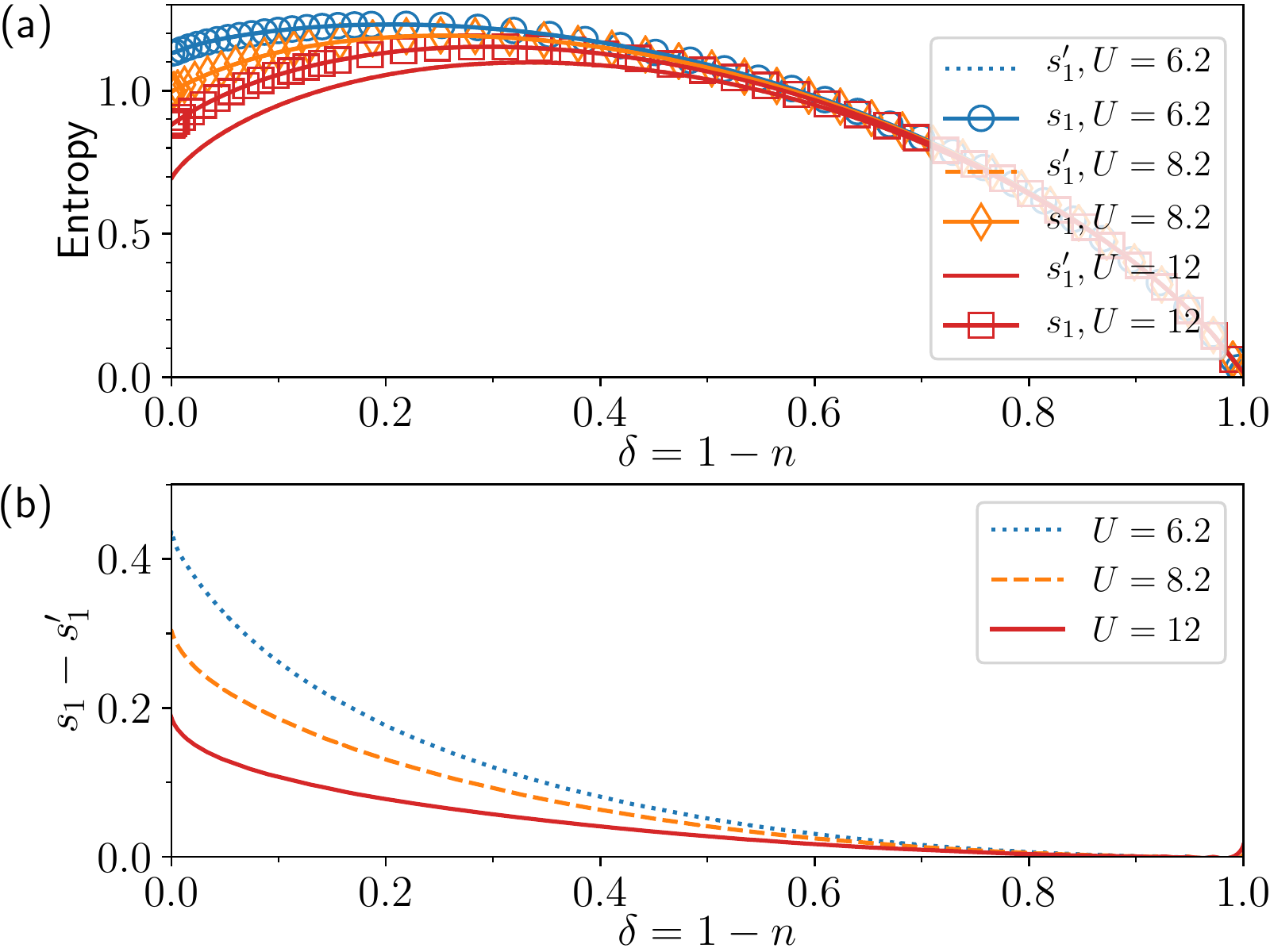}
}
\caption{(a) Entropies $s_1$ and $s_1'$ at $T=1/10$ for different values of $U$. $s_1'$ is the entropy of an isolated site in the grand-canonical ensemble, which at this temperature and for this interaction range, is essentially independent of $U$. (b) $s_1-s_1'$ for the same values as panel (a). }
\label{fig11}
\end{figure}

We can compute a different kind of mutual information, by calculating $s_1 - s_1'$, where $s_1'$ is the entropy of an isolated site in the grand canonical ensemble, see Figure~\ref{fig11}. 
An advantage of calculating mutual information in this way is that $s_1'$ captures only thermal contributions without any entanglement coming from quantum mechanical tunneling between the site and its neighbors. The vastly different behaviour of the two measures of correlation (cf Fig.~\ref{fig7} and Fig.~\ref{fig11}b) demonstrates the importance of quantum mechanical correlations. In particular, note that tunneling is never turned off in the calculations of total mutual information in Fig.~\ref{fig7}. That quantity is thus systematically larger than $s_1 - s_1'$ which compares the entropy of a single site with and without tunneling to the neighbors.   

\section{Conclusions and outlook} 
\label{sec:Conclusions}

Our work brings a quantum information perspective to a central unresolved problem in quantum many-body systems: what is the role of entanglement and classical correlations at a key phase transition in interacting fermions, the doping-driven Mott transition.

Specifically we studied two entanglement-related properties, the local entropy and the total mutual information, in the normal state of the two-dimensional Hubbard model solved with plaquette CDMFT. 
We considered a wide range of interaction strength, doping, and temperature. This allowed us to reveal a complex interplay of quantum and classical correlations. 

We focused on the doping-driven Mott transition that contains a finite-$T$ and finite-$\delta$ endpoint between two compressible (metallic) phases: a strongly correlated pseudogap and a correlated metal~\cite{sht, sht2, ssht}. 
Upon doping the Mott insulator, we showed that the local entropy $s_1$, thermodynamic entropy $s$, and total mutual information $\overline{I}_1$ all detect the pseudogap to correlated metal transition, its endpoint at finite temperature and finite doping, and its associated Widom line. Discontinuity below $T_c$ evolves into inflections (versus $\mu$) with a vertical tangent at $T_c$ and with a decreasing slope for $T>T_c$, up in the supercritical region. 
These sharp variations of correlations across the pseudogap to correlated metal transition indicate a  dramatic electronic reorganisation, persisting up to high temperature. 

Furthermore, our study provides a simple approach that is generically and immediately applicable to other computational methods, as it requires only the knowledge of the occupation and the double occupation. Our approach could also be generalized to other correlated many-body quantum systems, thereby paving the way for a deeper understanding of the signatures of entanglement and classical correlations in the properties of these systems. 

On the quantum information theory side, our work is thus a contribution to the understanding of the behavior of entanglement-related properties across a first-order transition ending in a finite temperature critical endpoint. Probing entanglement at this type of phase transition~\cite{eeQCD2017, Caitlin:PRL2019} has received less attention than at transitions across quantum critical points~\cite{amicoRMP2008, Anfossi_Giorda_Montorsi_Traversa_2005,Amico_Patan_2007,larssonPRA2006,LarssonScalingHubbard:2005, Amico_Patan_2007, Gabbrielli:2019, Frerot:2019} and finite temperature continuous transitions~\cite{Melko:2010, Singh:2011, Kallin:2011, Wilms_Troyer_Verstraete_2011, Wilms:2012, Iaconis:2013, Wald:JSM2020}, although it may have an impact on our understanding of critical behavior of correlated many-body quantum systems. 

Indeed, contrary to zero temperature where the study of phase transitions through the lenses of quantum information theory is well developed~\cite{amicoRMP2008}, at finite temperature the competition of quantum and classical correlations at phase transitions is less clear and remains one of the open challenges in this field and one that is receiving growing attention~\cite{Gabbrielli:2019, Frerot:2019, Wald:JSM2020, Grover:PRB2019, Grover:arXiv2019}. Our work contributes to unravelling this challenge. Note that at finite temperature the entanglement entropy is contaminated by thermodynamic entropy, and thus the quantities $s_1$ and $\overline{I}_1$ in our study cannot separate the quantum and classical contributions. It is of great recent interest to try to separate pure quantum correlations from classical correlations at finite temperature phase transitions~\cite{Grover:PRB2019, Grover:arXiv2019, Wald:JSM2020}. In light of the results of our work it would be an interesting open direction to understand this in all three of these regimes (low temperature phase transition, finite temperature endpoint, and crossover in the supercritical region). 

Further work is needed to characterise the structure of the entanglement by varying the subsystem size. Here we considered only the entanglement between a single site and the rest of the lattice. Yet, we showed that $s_1$ and $\overline{I}_{1}$ are a simple practical probe to characterise electronic phases in the Hubbard model. When the subsystem $A$ is larger than one site, one of the difficulties is that the reduced density matrix $\rho_A$ is no longer diagonal, and thus the procedure described in Sec.~\ref{sec:EntProp} cannot be directly applied. A hint on how to overcome this problem could come from Ref.~\cite{Udagawa_Motome:2015} where an algorithm is presented to extract the entanglement spectrum with cluster methods. Ref.~\cite{Phillips:PRD2019} provides a scaling of the one-site and two-site entropies close to a Mott transition in a related model.  

Going beyond this work and probing the entanglement entropy and the mutual information for different sizes of the subsystem $A$ could give new insights on the structure of the entanglement in the Mott insulator, the pseudogap, and the metallic phases. Such characterisation of the distribution of the entanglement could also be done with ultracold atom experiments, for instance see the pioneering results of Ref.~\cite{greinerNat2015} in a Bose Hubbard model.

Our results establish a substantial link between quantum information theory and experiments with cold gases. This is because our results have the advantage that they could be tested with ultracold atom realisations of the two-dimensional fermionic Hubbard model~\cite{GrossScience2017}. This is an advantage because measurements of entanglement-related properties have often remained elusive because of the lack of experimental probes. Recent groundbreaking experiments with ultracold atoms changed this~\cite{greinerNat2015, Kaufman:Science2016, Cocchi:PRX2017, Lukin:Science2019, Brydges:Science2019}. Our work thus provides a theoretical framework and new predictions for such experiments. Specifically, the high-temperature doping dependence of $s_1$, $s$, and $\overline{I}_1$ are compatible with the experimental findings of Ref.~\cite{Cocchi:PRX2017}, and could be further explored with current experiments. On the other hand, at lower temperatures, a quantitative comparison may not be possible because some of the low temperature features revealed here may be hidden by long range orders like superconductivity and antiferromagnetism. However moderate frustration can reduce the ordered temperatures, hence qualitative comparison with experiments could be possible~\cite{reymbaut2020crossovers}.  

Finally, we comment on the implication of our analysis for the pseudogap problem in hole-doped cuprates. The driving mechanism of the pseudogap remains a central puzzle~\cite{keimerRev}. In the plaquette CDMFT solution of the two-dimensional Hubbard model, the pseudogap originates from Mott physics plus short-range correlations locking electrons into singlets~\cite{kyung, hauleDOPING, ssht}. Hence the pseudogap to metal transition is a purely electronic transition. Broken symmetry states~\cite{Zhao:2017} may also occur but they are not necessary for the appearance of a pseudogap at finite temperature. 

Our work shows that the electronic rearrangement at the onset of the pseudogap as a function of doping is characterized by sharp variations in the local entropy and total mutual information (discontinuities at low temperatures and inflections at and above the finite-temperature endpoint). Growing research interest is devoted to studying the role of the entanglement close to the pseudogap in cuprates~\cite{Scheurer:PNAS2019, Zaanen:SciPost2019, Bagrov:arXiv2019}. Our findings suggest that changes in the entanglement-related properties - and not ordered phases - could be the key to understanding the finite temperature pseudogap in hole-doped cuprates.


\section*{Acknowledgments}
This work has been supported by the Natural Sciences and Engineering Research Council of Canada (NSERC) under grants RGPIN-2019-05312, the Canada First Research Excellence Fund and by the Research Chair in the Theory of Quantum Materials. PS work was supported by the U.S. Department of Energy, Office of Science, Basic Energy Sciences as a part of the Computational Materials Science Program. Simulations were performed on computers provided by the Canadian Foundation for Innovation, the Minist\`ere de l'\'Education des Loisirs et du Sport (Qu\'ebec), Calcul Qu\'ebec, and Compute Canada.

\appendix

\section{Further analysis of the local entropy $s_1$}
\label{AppendixA}

In this appendix, we show further figures of $s_1$ to expand on the behavior discussed in the main text.

\begin{figure}
\centering{
\includegraphics[width=0.999\linewidth]{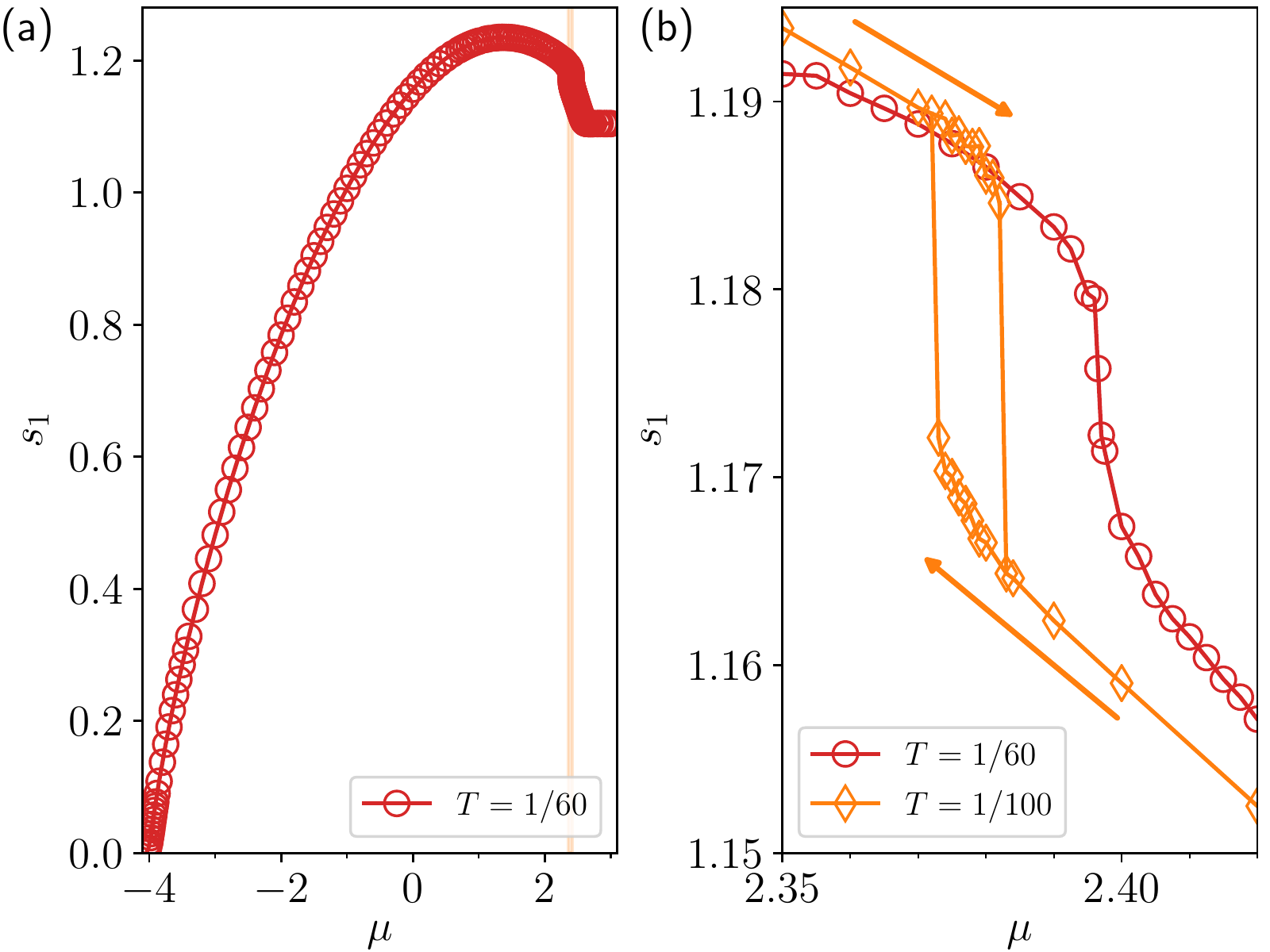}
}
\caption{(a) Local entropy $s_1$ versus chemical potential $\mu$ at $T=1/60$ for $U=6.2$. Figure~\ref{fig4}a of the main text shows the same data as a function of doping $\delta$. 
(b) $s_1$ versus $\mu$ for $U=6.2$ and low temperatures in the narrow chemical potential interval shaded in (a). $T=1/60$ is close to $T_c$, and $s_1(\mu)$ shows a drop marked by an inflection. $T=1/100$ is below $T_c$ and the coexistence loop in $s_1(\mu)$ is visible. Arrows indicate the sweep direction. We have found $(s_{1}) _{\textrm{pseudogap}} < (s_{1} )_{\textrm{correlated~metal}}$. 
}
\label{figSMmu62}
\end{figure}

Figure~\ref{figSMmu62} shows $s_1$ vs chemical potential $\mu$ for $U=6.2$. This figure is a direct parallel of the results of $s_1$ versus $\delta$ in Figure~\ref{fig4} of the main text, where the symbols have the same meaning, but is displayed here as a function of $\mu$ in order to see the hysteresis and inflection.

\begin{figure*}
\centering{
\includegraphics[width=0.999\linewidth]{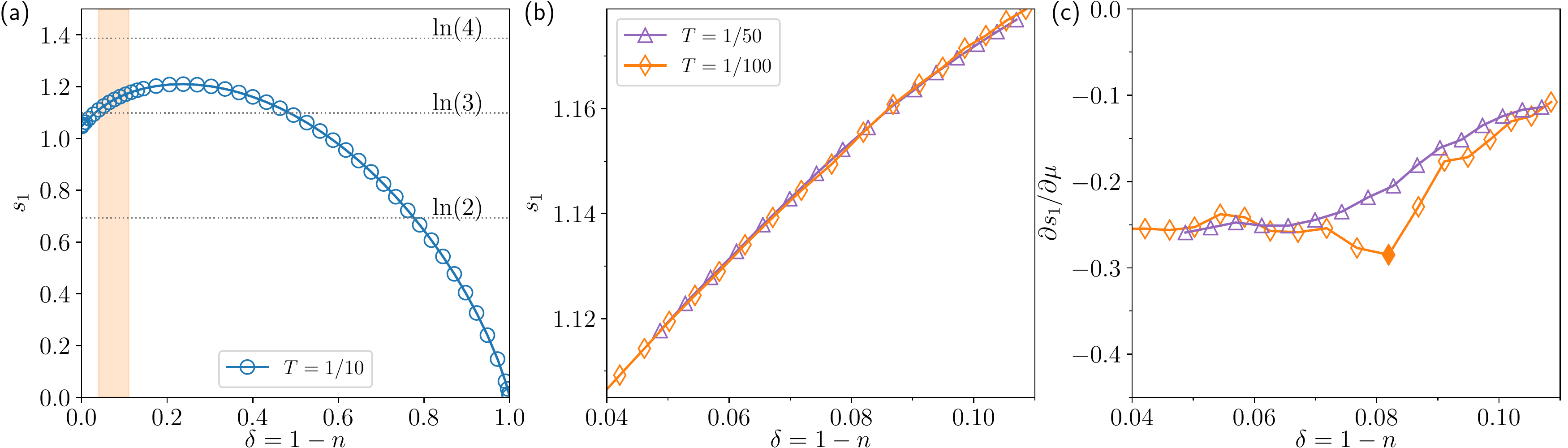}
}
\caption{The analog of Fig.~\ref{fig4} of the main text, but for $U=7.2$. 
(a) $s_1$ versus $\delta$ at $T=1/10$ for $U=7.2$. 
(b) $s_1$ versus $\delta$ for $U=7.2$ and low temperatures in the doping interval shaded in (a). 
(c) $\partial s_1 / \partial \mu$ versus $\delta$ at $U=7.2$. Upon approaching $(\delta_c,T_c)$, the peak in $\partial s_1 / \partial \mu$ sharpens and narrows. The sign problem prevents us from accessing temperatures smaller than $T\approx 1/100$. The locus of the minimum of $\partial s_1 / \partial \mu$ at $T=1/100$ defines the crossover $T_{s_1}$ in Fig.~\ref{fig2}c of the main text, which lies close to the Widom line. 
}
\label{figSMu72}
\end{figure*}

Figure~\ref{figSMu72} is the equivalent of Figure~\ref{fig4} of the main text but for $U=7.2$. It shows $s_1$ and $\partial s_1/\partial \mu$ as a function of doping.

\section{Construction of the thermodynamic entropy $s$}
\label{AppendixB}

\begin{figure*}
\centering{
\includegraphics[width=0.999\linewidth]{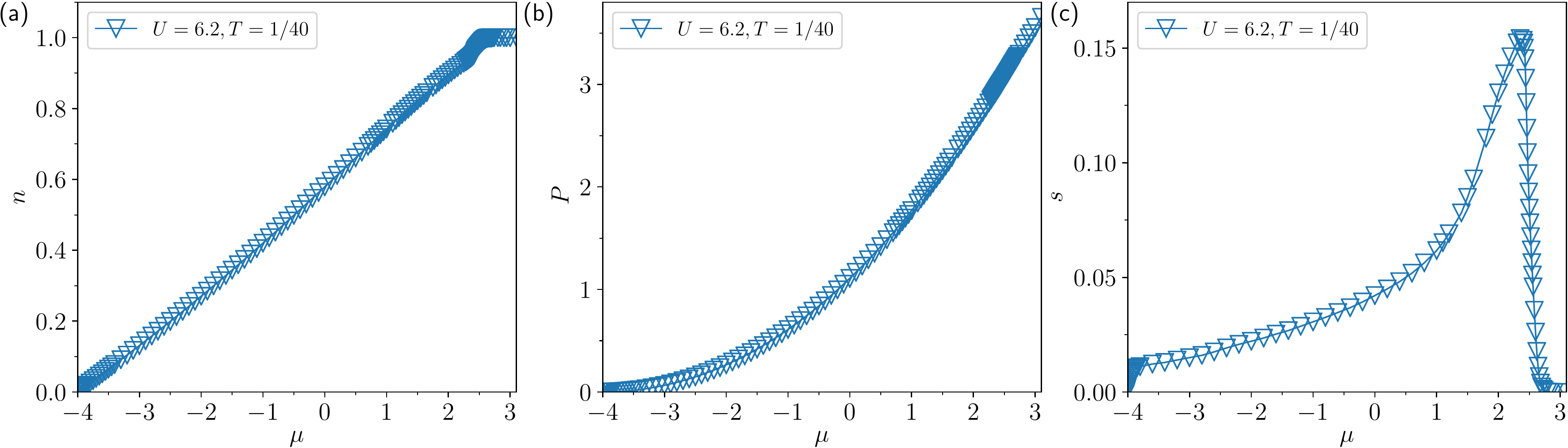}
}
\caption{Illustrating the three steps in the construction of the thermodynamic entropy per site $s$. (a) Occupation $n$ versus $\mu$ at constant temperature. 
(b) Pressure $P$ versus $\mu$ at constant temperature, calculated by integrating $n$ over $\mu$. 
(c) Thermodynamic entropy $s$ versus $\mu$ at constant temperature, obtained by differentiating the pressure with respect to temperature. 
Data are for $U=6.2$ and $T=1/40$. The calculation exploiting the Gibbs-Duhem relation is shown in the main text, where we follow the procedure of our work of Ref.~\cite{CaitlinSb}.
}
\label{figSM-s}
\end{figure*}

In Section~\ref{sec:s} of the main text, we discussed how to extract the pressure $P$ and the thermodynamic entropy $s$ from the occupation $n$ versus $\mu$ curves, using the Gibbs-Duhem relation. In this appendix we illustrate this procedure for $U=6.2$ and $T=1/40$. Figure~\ref{figSM-s} shows the occupation $n$, the pressure $P$, and the thermodynamic entropy $s$ versus $\mu$ for these values. 

\section{Construction of the total mutual information $\overline{I}_1$}
\label{AppendixC}

\begin{figure}
\centering{
\includegraphics[width=0.999\linewidth]{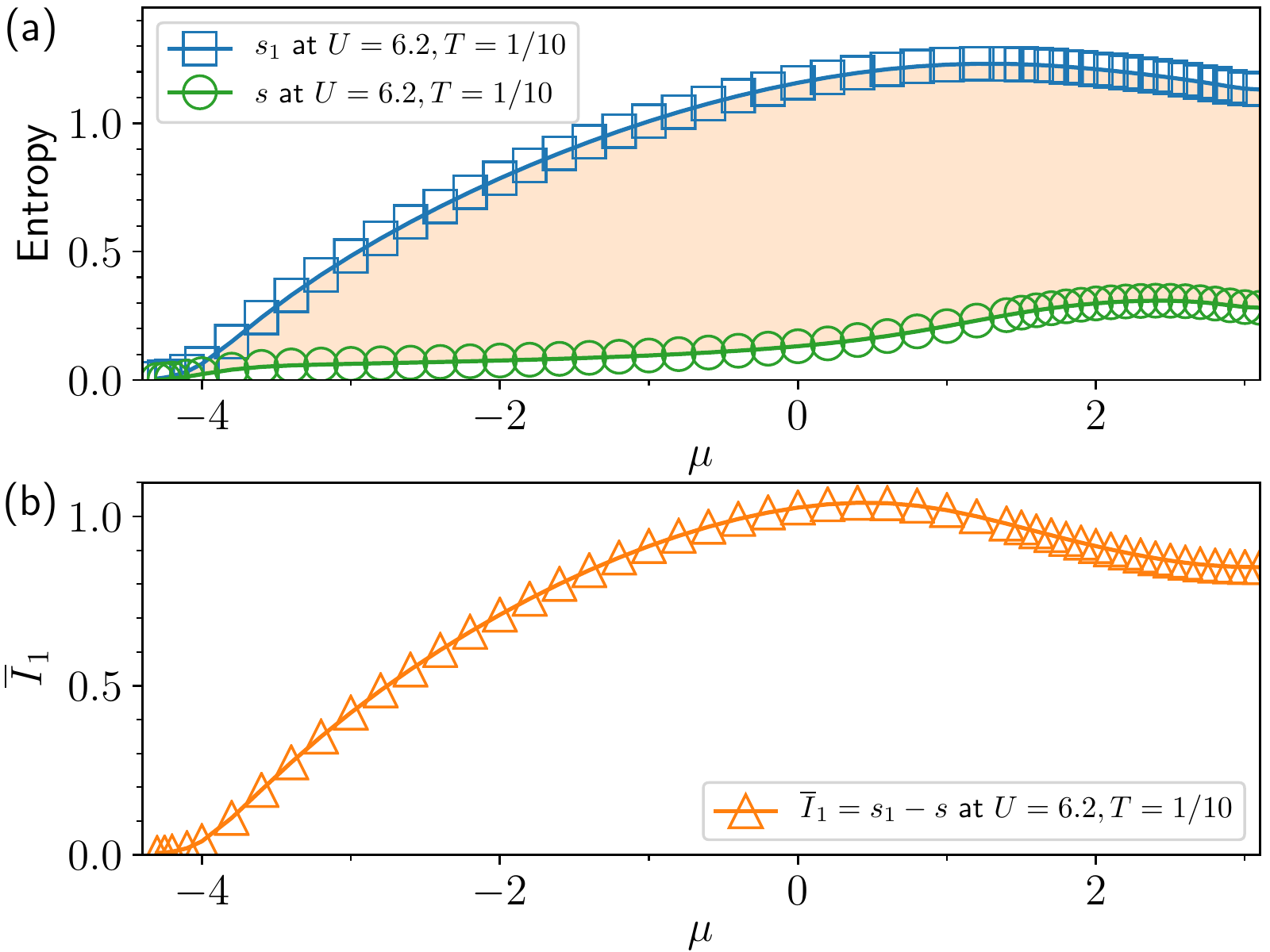}
}
\caption{(a) Local entropy $s_1$ (blue squares) and thermodynamic entropy $s$ (green circles) as a function of the chemical potential $\mu$ for $U=6.2$ and $T=1/10$. The shaded area indicates the total mutual information $\overline{I}_1 = s_1 - s$, which is also shown in panel (b).
(b) Total mutual information $\overline{I}_{1}$ as a function of $\mu$ for U=6.2 and T=1/10.
}
\label{figSM-I1}
\end{figure}

In this appendix we show the entropy and total mutual information $\overline{I}_1$ as a function of $\mu$ in order to show visually how $\overline{I}_1$ is constructed as the difference between $s_1$ and $s$. This is shown for $U=6.2$ and $T=1/10$ in Fig.~\ref{figSM-I1}, which complements the set of data shown versus $\delta$ in Figure~\ref{fig8} of the main text. In Figure~\ref{figSMexp} of the main text, we show a similar figure in which we compare our results at $U=8.2$ with experimental data obtained at high temperatures.

\section{Finding inflections}
\label{AppendixD}

\begin{figure*}
\centering{
\includegraphics[width=0.999\linewidth]{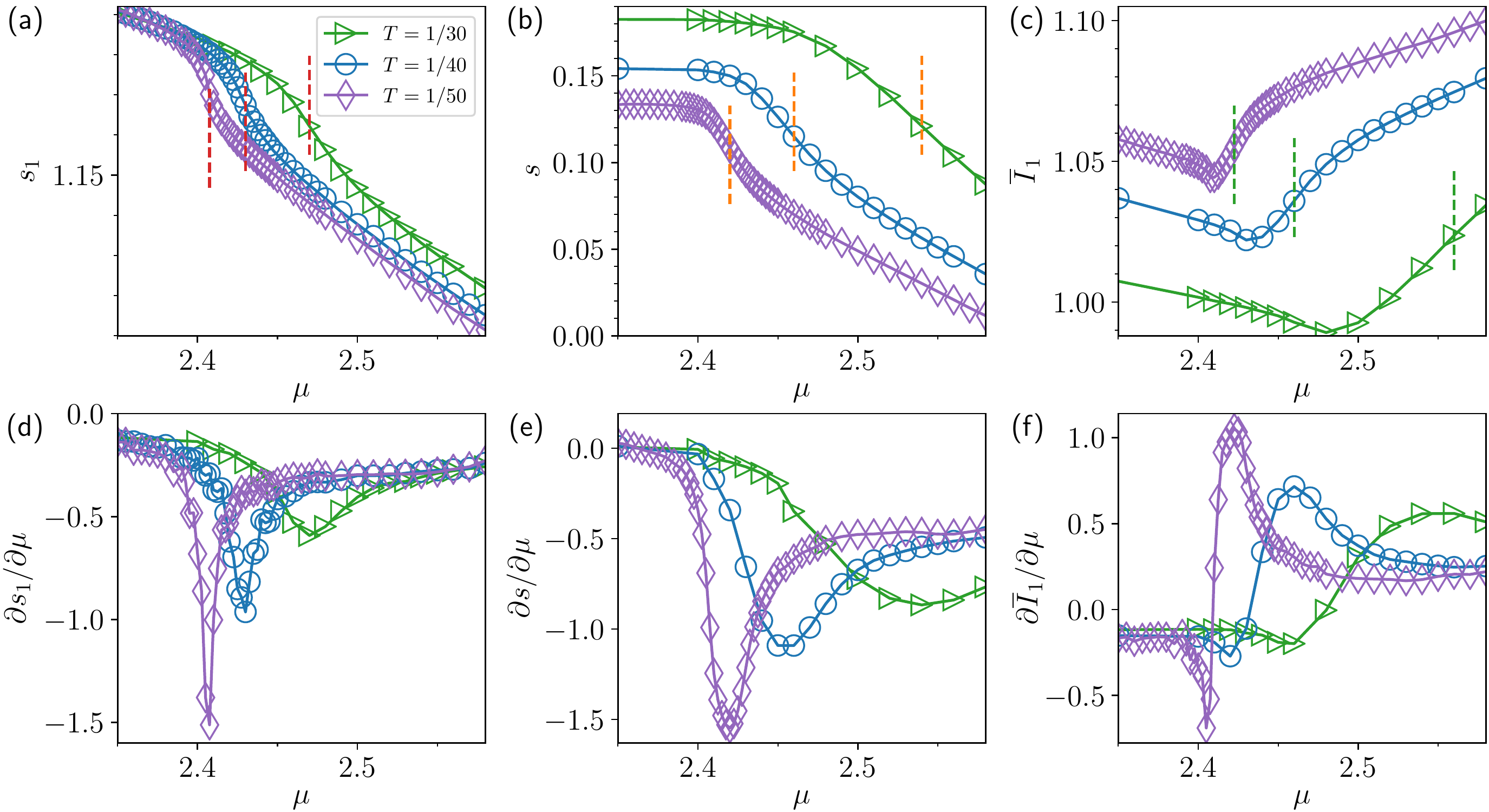}
}
\caption{(a) Local entropy $s_1$ versus $\mu$ for $U=6.2$ and several temperatures. (b) Thermodynamic entropy $s$ versus $\mu$. (c) Total mutual information $\overline I_1 = s_1 - s$ versus $\mu$. The position of the inflections are marked by dashed lines. 
(d), (e), (f) Numerical derivatives with respect to $\mu$ of the quantities in (a), (b), (c). We use the position of the peaks to determine the inflection points in (a), (b), (c). The peaks become narrower and the magnitude increases upon approaching the critical temperature $T_{c}$. The loci of the inflection versus $\mu$ giving the sharpest change in $\overline{I}_1$ was chosen to define a crossover line $T_{\overline{I}_1}$ in the phase diagram of Fig.~\ref{fig2}b in the main text. 
All panels are for $U=6.2$ and the same temperatures and color scheme as in (a).
}
\label{figSMinflections}
\end{figure*}

Inflections in the local entropy $s_1$, the thermodynamic entropy $s$, and the total mutual information $\overline{I}_{1}$ occur as a function of $\mu$. In Figures~\ref{fig4}, \ref{fig6}, and \ref{fig8} of the main text, we have shown these quantities as a function of doping where the position of the inflection vs $\mu$ is marked with a dashed vertical line. In this appendix, we illustrate these inflections as a function of $\mu$. Figure~\ref{figSMinflections} shows these quantities along with their numerical derivatives with respect to $\mu$ for $U=6.2$ and different temperatures.
%


\begin{thebibliography}{98}%
\makeatletter
\providecommand \@ifxundefined [1]{%
 \@ifx{#1\undefined}
}%
\providecommand \@ifnum [1]{%
 \ifnum #1\expandafter \@firstoftwo
 \else \expandafter \@secondoftwo
 \fi
}%
\providecommand \@ifx [1]{%
 \ifx #1\expandafter \@firstoftwo
 \else \expandafter \@secondoftwo
 \fi
}%
\providecommand \natexlab [1]{#1}%
\providecommand \enquote  [1]{``#1''}%
\providecommand \bibnamefont  [1]{#1}%
\providecommand \bibfnamefont [1]{#1}%
\providecommand \citenamefont [1]{#1}%
\providecommand \href@noop [0]{\@secondoftwo}%
\providecommand \href [0]{\begingroup \@sanitize@url \@href}%
\providecommand \@href[1]{\@@startlink{#1}\@@href}%
\providecommand \@@href[1]{\endgroup#1\@@endlink}%
\providecommand \@sanitize@url [0]{\catcode `\\12\catcode `\$12\catcode
  `\&12\catcode `\#12\catcode `\^12\catcode `\_12\catcode `\%12\relax}%
\providecommand \@@startlink[1]{}%
\providecommand \@@endlink[0]{}%
\providecommand \url  [0]{\begingroup\@sanitize@url \@url }%
\providecommand \@url [1]{\endgroup\@href {#1}{\urlprefix }}%
\providecommand \urlprefix  [0]{URL }%
\providecommand \Eprint [0]{\href }%
\providecommand \doibase [0]{http://dx.doi.org/}%
\providecommand \selectlanguage [0]{\@gobble}%
\providecommand \bibinfo  [0]{\@secondoftwo}%
\providecommand \bibfield  [0]{\@secondoftwo}%
\providecommand \translation [1]{[#1]}%
\providecommand \BibitemOpen [0]{}%
\providecommand \bibitemStop [0]{}%
\providecommand \bibitemNoStop [0]{.\EOS\space}%
\providecommand \EOS [0]{\spacefactor3000\relax}%
\providecommand \BibitemShut  [1]{\csname bibitem#1\endcsname}%
\let\auto@bib@innerbib\@empty
\bibitem [{\citenamefont {Cover}\ and\ \citenamefont
  {Thomas}(2006)}]{CoverInformation}%
  \BibitemOpen
  \bibfield  {author} {\bibinfo {author} {\bibfnamefont {Thomas~M.}\
  \bibnamefont {Cover}}\ and\ \bibinfo {author} {\bibfnamefont {Joy~A.}\
  \bibnamefont {Thomas}},\ }\href@noop {} {\emph {\bibinfo {title} {{Elements
  of Information Theory (Wiley Series in Telecommunications and Signal
  Processing)}}}}\ (\bibinfo  {publisher} {Wiley-Interscience},\ \bibinfo
  {address} {New York, NY, USA},\ \bibinfo {year} {2006})\BibitemShut {NoStop}%
\bibitem [{\citenamefont {Watrous}(2018)}]{watrous2018}%
  \BibitemOpen
  \bibfield  {author} {\bibinfo {author} {\bibfnamefont {John}\ \bibnamefont
  {Watrous}},\ }\href@noop {} {\emph {\bibinfo {title} {The Theory of Quantum
  Information}}}\ (\bibinfo  {publisher} {Cambridge University Press},\
  \bibinfo {year} {2018})\BibitemShut {NoStop}%
\bibitem [{\citenamefont {Amico}\ \emph {et~al.}(2008)\citenamefont {Amico},
  \citenamefont {Fazio}, \citenamefont {Osterloh},\ and\ \citenamefont
  {Vedral}}]{amicoRMP2008}%
  \BibitemOpen
  \bibfield  {author} {\bibinfo {author} {\bibfnamefont {Luigi}\ \bibnamefont
  {Amico}}, \bibinfo {author} {\bibfnamefont {Rosario}\ \bibnamefont {Fazio}},
  \bibinfo {author} {\bibfnamefont {Andreas}\ \bibnamefont {Osterloh}}, \ and\
  \bibinfo {author} {\bibfnamefont {Vlatko}\ \bibnamefont {Vedral}},\
  }\bibfield  {title} {\enquote {\bibinfo {title} {Entanglement in many-body
  systems},}\ }\href {\doibase 10.1103/RevModPhys.80.517} {\bibfield  {journal}
  {\bibinfo  {journal} {Rev. Mod. Phys.}\ }\textbf {\bibinfo {volume} {80}},\
  \bibinfo {pages} {517--576} (\bibinfo {year} {2008})}\BibitemShut {NoStop}%
\bibitem [{\citenamefont {Eisert}\ \emph {et~al.}(2010)\citenamefont {Eisert},
  \citenamefont {Cramer},\ and\ \citenamefont {Plenio}}]{EisertArea2010}%
  \BibitemOpen
  \bibfield  {author} {\bibinfo {author} {\bibfnamefont {J.}~\bibnamefont
  {Eisert}}, \bibinfo {author} {\bibfnamefont {M.}~\bibnamefont {Cramer}}, \
  and\ \bibinfo {author} {\bibfnamefont {M.~B.}\ \bibnamefont {Plenio}},\
  }\bibfield  {title} {\enquote {\bibinfo {title} {Colloquium: Area laws for
  the entanglement entropy},}\ }\href {\doibase 10.1103/RevModPhys.82.277}
  {\bibfield  {journal} {\bibinfo  {journal} {Rev. Mod. Phys.}\ }\textbf
  {\bibinfo {volume} {82}},\ \bibinfo {pages} {277--306} (\bibinfo {year}
  {2010})}\BibitemShut {NoStop}%
\bibitem [{\citenamefont {Laflorencie}(2016)}]{Laflorencie:PhysRep2016}%
  \BibitemOpen
  \bibfield  {author} {\bibinfo {author} {\bibfnamefont {Nicolas}\ \bibnamefont
  {Laflorencie}},\ }\bibfield  {title} {\enquote {\bibinfo {title} {Quantum
  entanglement in condensed matter systems},}\ }\href {\doibase
  https://doi.org/10.1016/j.physrep.2016.06.008} {\bibfield  {journal}
  {\bibinfo  {journal} {Physics Reports}\ }\textbf {\bibinfo {volume} {646}},\
  \bibinfo {pages} {1--59} (\bibinfo {year} {2016})}\BibitemShut {NoStop}%
\bibitem [{\citenamefont {Zeng}\ \emph {et~al.}(2019)\citenamefont {Zeng},
  \citenamefont {Chen}, \citenamefont {Zhou},\ and\ \citenamefont
  {Wen}}]{zhengBOOK}%
  \BibitemOpen
  \bibfield  {author} {\bibinfo {author} {\bibfnamefont {B.}~\bibnamefont
  {Zeng}}, \bibinfo {author} {\bibfnamefont {X.}~\bibnamefont {Chen}}, \bibinfo
  {author} {\bibfnamefont {D.-L.}\ \bibnamefont {Zhou}}, \ and\ \bibinfo
  {author} {\bibfnamefont {X.-G.}\ \bibnamefont {Wen}},\ }\href {\doibase
  10.1007/978-1-4939-9084-9} {\emph {\bibinfo {title} {{Quantum Information
  Meets Quantum Matter}}}}\ (\bibinfo  {publisher} {Springer-Verlag},\ \bibinfo
  {address} {New York},\ \bibinfo {year} {2019})\BibitemShut {NoStop}%
\bibitem [{\citenamefont {Wen}(2017)}]{Wen:RMP2017}%
  \BibitemOpen
  \bibfield  {author} {\bibinfo {author} {\bibfnamefont {Xiao-Gang}\
  \bibnamefont {Wen}},\ }\bibfield  {title} {\enquote {\bibinfo {title}
  {Colloquium: Zoo of quantum-topological phases of matter},}\ }\href {\doibase
  10.1103/RevModPhys.89.041004} {\bibfield  {journal} {\bibinfo  {journal}
  {Rev. Mod. Phys.}\ }\textbf {\bibinfo {volume} {89}},\ \bibinfo {pages}
  {041004} (\bibinfo {year} {2017})}\BibitemShut {NoStop}%
\bibitem [{\citenamefont {Islam}\ \emph {et~al.}(2015)\citenamefont {Islam},
  \citenamefont {Ma}, \citenamefont {Preiss}, \citenamefont {Tai},
  \citenamefont {Lukin}, \citenamefont {Rispoli},\ and\ \citenamefont
  {Greiner}}]{greinerNat2015}%
  \BibitemOpen
  \bibfield  {author} {\bibinfo {author} {\bibfnamefont {Rajibul}\ \bibnamefont
  {Islam}}, \bibinfo {author} {\bibfnamefont {Ruichao}\ \bibnamefont {Ma}},
  \bibinfo {author} {\bibfnamefont {Philipp~M}\ \bibnamefont {Preiss}},
  \bibinfo {author} {\bibfnamefont {M~Eric}\ \bibnamefont {Tai}}, \bibinfo
  {author} {\bibfnamefont {Alexander}\ \bibnamefont {Lukin}}, \bibinfo {author}
  {\bibfnamefont {Matthew}\ \bibnamefont {Rispoli}}, \ and\ \bibinfo {author}
  {\bibfnamefont {Markus}\ \bibnamefont {Greiner}},\ }\bibfield  {title}
  {\enquote {\bibinfo {title} {Measuring entanglement entropy in a quantum
  many-body system},}\ }\href {\doibase 10.1038/nature15750} {\bibfield
  {journal} {\bibinfo  {journal} {Nature}\ }\textbf {\bibinfo {volume} {528}},\
  \bibinfo {pages} {77} (\bibinfo {year} {2015})}\BibitemShut {NoStop}%
\bibitem [{\citenamefont {Kaufman}\ \emph {et~al.}(2016)\citenamefont
  {Kaufman}, \citenamefont {Tai}, \citenamefont {Lukin}, \citenamefont
  {Rispoli}, \citenamefont {Schittko}, \citenamefont {Preiss},\ and\
  \citenamefont {Greiner}}]{Kaufman:Science2016}%
  \BibitemOpen
  \bibfield  {author} {\bibinfo {author} {\bibfnamefont {Adam~M.}\ \bibnamefont
  {Kaufman}}, \bibinfo {author} {\bibfnamefont {M.~Eric}\ \bibnamefont {Tai}},
  \bibinfo {author} {\bibfnamefont {Alexander}\ \bibnamefont {Lukin}}, \bibinfo
  {author} {\bibfnamefont {Matthew}\ \bibnamefont {Rispoli}}, \bibinfo {author}
  {\bibfnamefont {Robert}\ \bibnamefont {Schittko}}, \bibinfo {author}
  {\bibfnamefont {Philipp~M.}\ \bibnamefont {Preiss}}, \ and\ \bibinfo {author}
  {\bibfnamefont {Markus}\ \bibnamefont {Greiner}},\ }\bibfield  {title}
  {\enquote {\bibinfo {title} {Quantum thermalization through entanglement in
  an isolated many-body system},}\ }\href {\doibase 10.1126/science.aaf6725}
  {\bibfield  {journal} {\bibinfo  {journal} {Science}\ }\textbf {\bibinfo
  {volume} {353}},\ \bibinfo {pages} {794--800} (\bibinfo {year}
  {2016})}\BibitemShut {NoStop}%
\bibitem [{\citenamefont {Cocchi}\ \emph {et~al.}(2017)\citenamefont {Cocchi},
  \citenamefont {Miller}, \citenamefont {Drewes}, \citenamefont {Chan},
  \citenamefont {Pertot}, \citenamefont {Brennecke},\ and\ \citenamefont
  {K\"ohl}}]{Cocchi:PRX2017}%
  \BibitemOpen
  \bibfield  {author} {\bibinfo {author} {\bibfnamefont {E.}~\bibnamefont
  {Cocchi}}, \bibinfo {author} {\bibfnamefont {L.~A.}\ \bibnamefont {Miller}},
  \bibinfo {author} {\bibfnamefont {J.~H.}\ \bibnamefont {Drewes}}, \bibinfo
  {author} {\bibfnamefont {C.~F.}\ \bibnamefont {Chan}}, \bibinfo {author}
  {\bibfnamefont {D.}~\bibnamefont {Pertot}}, \bibinfo {author} {\bibfnamefont
  {F.}~\bibnamefont {Brennecke}}, \ and\ \bibinfo {author} {\bibfnamefont
  {M.}~\bibnamefont {K\"ohl}},\ }\bibfield  {title} {\enquote {\bibinfo {title}
  {{Measuring Entropy and Short-Range Correlations in the Two-Dimensional
  Hubbard Model}},}\ }\href {\doibase 10.1103/PhysRevX.7.031025} {\bibfield
  {journal} {\bibinfo  {journal} {Phys. Rev. X}\ }\textbf {\bibinfo {volume}
  {7}},\ \bibinfo {pages} {031025} (\bibinfo {year} {2017})}\BibitemShut
  {NoStop}%
\bibitem [{\citenamefont {Lukin}\ \emph {et~al.}(2019)\citenamefont {Lukin},
  \citenamefont {Rispoli}, \citenamefont {Schittko}, \citenamefont {Tai},
  \citenamefont {Kaufman}, \citenamefont {Choi}, \citenamefont {Khemani},
  \citenamefont {L{\'e}onard},\ and\ \citenamefont
  {Greiner}}]{Lukin:Science2019}%
  \BibitemOpen
  \bibfield  {author} {\bibinfo {author} {\bibfnamefont {Alexander}\
  \bibnamefont {Lukin}}, \bibinfo {author} {\bibfnamefont {Matthew}\
  \bibnamefont {Rispoli}}, \bibinfo {author} {\bibfnamefont {Robert}\
  \bibnamefont {Schittko}}, \bibinfo {author} {\bibfnamefont {M.~Eric}\
  \bibnamefont {Tai}}, \bibinfo {author} {\bibfnamefont {Adam~M.}\ \bibnamefont
  {Kaufman}}, \bibinfo {author} {\bibfnamefont {Soonwon}\ \bibnamefont {Choi}},
  \bibinfo {author} {\bibfnamefont {Vedika}\ \bibnamefont {Khemani}}, \bibinfo
  {author} {\bibfnamefont {Julian}\ \bibnamefont {L{\'e}onard}}, \ and\
  \bibinfo {author} {\bibfnamefont {Markus}\ \bibnamefont {Greiner}},\
  }\bibfield  {title} {\enquote {\bibinfo {title} {Probing entanglement in a
  many-body{\textendash}localized system},}\ }\href {\doibase
  10.1126/science.aau0818} {\bibfield  {journal} {\bibinfo  {journal}
  {Science}\ }\textbf {\bibinfo {volume} {364}},\ \bibinfo {pages} {256--260}
  (\bibinfo {year} {2019})}\BibitemShut {NoStop}%
\bibitem [{\citenamefont {Brydges}\ \emph {et~al.}(2019)\citenamefont
  {Brydges}, \citenamefont {Elben}, \citenamefont {Jurcevic}, \citenamefont
  {Vermersch}, \citenamefont {Maier}, \citenamefont {Lanyon}, \citenamefont
  {Zoller}, \citenamefont {Blatt},\ and\ \citenamefont
  {Roos}}]{Brydges:Science2019}%
  \BibitemOpen
  \bibfield  {author} {\bibinfo {author} {\bibfnamefont {Tiff}\ \bibnamefont
  {Brydges}}, \bibinfo {author} {\bibfnamefont {Andreas}\ \bibnamefont
  {Elben}}, \bibinfo {author} {\bibfnamefont {Petar}\ \bibnamefont {Jurcevic}},
  \bibinfo {author} {\bibfnamefont {Beno{\^\i}t}\ \bibnamefont {Vermersch}},
  \bibinfo {author} {\bibfnamefont {Christine}\ \bibnamefont {Maier}}, \bibinfo
  {author} {\bibfnamefont {Ben~P.}\ \bibnamefont {Lanyon}}, \bibinfo {author}
  {\bibfnamefont {Peter}\ \bibnamefont {Zoller}}, \bibinfo {author}
  {\bibfnamefont {Rainer}\ \bibnamefont {Blatt}}, \ and\ \bibinfo {author}
  {\bibfnamefont {Christian~F.}\ \bibnamefont {Roos}},\ }\bibfield  {title}
  {\enquote {\bibinfo {title} {Probing r{\'e}nyi entanglement entropy via
  randomized measurements},}\ }\href {\doibase 10.1126/science.aau4963}
  {\bibfield  {journal} {\bibinfo  {journal} {Science}\ }\textbf {\bibinfo
  {volume} {364}},\ \bibinfo {pages} {260--263} (\bibinfo {year}
  {2019})}\BibitemShut {NoStop}%
\bibitem [{\citenamefont {Jaksch}\ and\ \citenamefont {Zoller}(2005)}]{jzHM}%
  \BibitemOpen
  \bibfield  {author} {\bibinfo {author} {\bibfnamefont {D.}~\bibnamefont
  {Jaksch}}\ and\ \bibinfo {author} {\bibfnamefont {P.}~\bibnamefont
  {Zoller}},\ }\bibfield  {title} {\enquote {\bibinfo {title} {The cold atom
  hubbard toolbox},}\ }\href@noop {} {\bibfield  {journal} {\bibinfo  {journal}
  {Ann. Phys.}\ }\textbf {\bibinfo {volume} {315}},\ \bibinfo {eid} {52}
  (\bibinfo {year} {2005})}\BibitemShut {NoStop}%
\bibitem [{\citenamefont {Esslinger}(2010)}]{Esslinger:2010}%
  \BibitemOpen
  \bibfield  {author} {\bibinfo {author} {\bibfnamefont {Tilman}\ \bibnamefont
  {Esslinger}},\ }\bibfield  {title} {\enquote {\bibinfo {title}
  {{Fermi-Hubbard Physics with Atoms in an Optical Lattice}},}\ }\href
  {\doibase 10.1146/annurev-conmatphys-070909-104059} {\bibfield  {journal}
  {\bibinfo  {journal} {Annual Review of Condensed Matter Physics}\ }\textbf
  {\bibinfo {volume} {1}},\ \bibinfo {pages} {129--152} (\bibinfo {year}
  {2010})}\BibitemShut {NoStop}%
\bibitem [{\citenamefont {Jordens}\ \emph {et~al.}(2008)\citenamefont
  {Jordens}, \citenamefont {Strohmaier}, \citenamefont {Gunter}, \citenamefont
  {Moritz},\ and\ \citenamefont {Esslinger}}]{Jordens:2008}%
  \BibitemOpen
  \bibfield  {author} {\bibinfo {author} {\bibfnamefont {Robert}\ \bibnamefont
  {Jordens}}, \bibinfo {author} {\bibfnamefont {Niels}\ \bibnamefont
  {Strohmaier}}, \bibinfo {author} {\bibfnamefont {Kenneth}\ \bibnamefont
  {Gunter}}, \bibinfo {author} {\bibfnamefont {Henning}\ \bibnamefont
  {Moritz}}, \ and\ \bibinfo {author} {\bibfnamefont {Tilman}\ \bibnamefont
  {Esslinger}},\ }\bibfield  {title} {\enquote {\bibinfo {title} {{A Mott
  insulator of fermionic atoms in an optical lattice}},}\ }\href {\doibase
  http://dx.doi.org/10.1038/nature07244} {\bibfield  {journal} {\bibinfo
  {journal} {Nature}\ }\textbf {\bibinfo {volume} {455}},\ \bibinfo {pages}
  {204--207} (\bibinfo {year} {2008})}\BibitemShut {NoStop}%
\bibitem [{\citenamefont {Schneider}\ \emph {et~al.}(2008)\citenamefont
  {Schneider}, \citenamefont {Hackerm{\"u}ller}, \citenamefont {Will},
  \citenamefont {Best}, \citenamefont {Bloch}, \citenamefont {Costi},
  \citenamefont {Helmes}, \citenamefont {Rasch},\ and\ \citenamefont
  {Rosch}}]{Schneider:2008}%
  \BibitemOpen
  \bibfield  {author} {\bibinfo {author} {\bibfnamefont {U.}~\bibnamefont
  {Schneider}}, \bibinfo {author} {\bibfnamefont {L.}~\bibnamefont
  {Hackerm{\"u}ller}}, \bibinfo {author} {\bibfnamefont {S.}~\bibnamefont
  {Will}}, \bibinfo {author} {\bibfnamefont {Th.}\ \bibnamefont {Best}},
  \bibinfo {author} {\bibfnamefont {I.}~\bibnamefont {Bloch}}, \bibinfo
  {author} {\bibfnamefont {T.~A.}\ \bibnamefont {Costi}}, \bibinfo {author}
  {\bibfnamefont {R.~W.}\ \bibnamefont {Helmes}}, \bibinfo {author}
  {\bibfnamefont {D.}~\bibnamefont {Rasch}}, \ and\ \bibinfo {author}
  {\bibfnamefont {A.}~\bibnamefont {Rosch}},\ }\bibfield  {title} {\enquote
  {\bibinfo {title} {{Metallic and Insulating Phases of Repulsively Interacting
  Fermions in a 3D Optical Lattice}},}\ }\href {\doibase
  10.1126/science.1165449} {\bibfield  {journal} {\bibinfo  {journal}
  {Science}\ }\textbf {\bibinfo {volume} {322}},\ \bibinfo {pages} {1520--1525}
  (\bibinfo {year} {2008})}\BibitemShut {NoStop}%
\bibitem [{\citenamefont {Hofrichter}\ \emph {et~al.}(2016)\citenamefont
  {Hofrichter}, \citenamefont {Riegger}, \citenamefont {Scazza}, \citenamefont
  {H\"ofer}, \citenamefont {Fernandes}, \citenamefont {Bloch},\ and\
  \citenamefont {F\"olling}}]{Hofrichter:PRX2016}%
  \BibitemOpen
  \bibfield  {author} {\bibinfo {author} {\bibfnamefont {Christian}\
  \bibnamefont {Hofrichter}}, \bibinfo {author} {\bibfnamefont {Luis}\
  \bibnamefont {Riegger}}, \bibinfo {author} {\bibfnamefont {Francesco}\
  \bibnamefont {Scazza}}, \bibinfo {author} {\bibfnamefont {Moritz}\
  \bibnamefont {H\"ofer}}, \bibinfo {author} {\bibfnamefont {Diogo~Rio}\
  \bibnamefont {Fernandes}}, \bibinfo {author} {\bibfnamefont {Immanuel}\
  \bibnamefont {Bloch}}, \ and\ \bibinfo {author} {\bibfnamefont {Simon}\
  \bibnamefont {F\"olling}},\ }\bibfield  {title} {\enquote {\bibinfo {title}
  {{Direct Probing of the Mott Crossover in the $\mathrm{SU}(N)$ Fermi-Hubbard
  Model}},}\ }\href {\doibase 10.1103/PhysRevX.6.021030} {\bibfield  {journal}
  {\bibinfo  {journal} {Phys. Rev. X}\ }\textbf {\bibinfo {volume} {6}},\
  \bibinfo {pages} {021030} (\bibinfo {year} {2016})}\BibitemShut {NoStop}%
\bibitem [{\citenamefont {Cocchi}\ \emph {et~al.}(2016)\citenamefont {Cocchi},
  \citenamefont {Miller}, \citenamefont {Drewes}, \citenamefont {Koschorreck},
  \citenamefont {Pertot}, \citenamefont {Brennecke},\ and\ \citenamefont
  {K\"ohl}}]{Cocchi:PRL2016}%
  \BibitemOpen
  \bibfield  {author} {\bibinfo {author} {\bibfnamefont {Eugenio}\ \bibnamefont
  {Cocchi}}, \bibinfo {author} {\bibfnamefont {Luke~A.}\ \bibnamefont
  {Miller}}, \bibinfo {author} {\bibfnamefont {Jan~H.}\ \bibnamefont {Drewes}},
  \bibinfo {author} {\bibfnamefont {Marco}\ \bibnamefont {Koschorreck}},
  \bibinfo {author} {\bibfnamefont {Daniel}\ \bibnamefont {Pertot}}, \bibinfo
  {author} {\bibfnamefont {Ferdinand}\ \bibnamefont {Brennecke}}, \ and\
  \bibinfo {author} {\bibfnamefont {Michael}\ \bibnamefont {K\"ohl}},\
  }\bibfield  {title} {\enquote {\bibinfo {title} {{Equation of State of the
  Two-Dimensional Hubbard Model}},}\ }\href {\doibase
  10.1103/PhysRevLett.116.175301} {\bibfield  {journal} {\bibinfo  {journal}
  {Phys. Rev. Lett.}\ }\textbf {\bibinfo {volume} {116}},\ \bibinfo {pages}
  {175301} (\bibinfo {year} {2016})}\BibitemShut {NoStop}%
\bibitem [{\citenamefont {Cheuk}\ \emph {et~al.}(2016)\citenamefont {Cheuk},
  \citenamefont {Nichols}, \citenamefont {Lawrence}, \citenamefont {Okan},
  \citenamefont {Zhang}, \citenamefont {Khatami}, \citenamefont {Trivedi},
  \citenamefont {Paiva}, \citenamefont {Rigol},\ and\ \citenamefont
  {Zwierlein}}]{Cheuk:2016}%
  \BibitemOpen
  \bibfield  {author} {\bibinfo {author} {\bibfnamefont {Lawrence~W.}\
  \bibnamefont {Cheuk}}, \bibinfo {author} {\bibfnamefont {Matthew~A.}\
  \bibnamefont {Nichols}}, \bibinfo {author} {\bibfnamefont {Katherine~R.}\
  \bibnamefont {Lawrence}}, \bibinfo {author} {\bibfnamefont {Melih}\
  \bibnamefont {Okan}}, \bibinfo {author} {\bibfnamefont {Hao}\ \bibnamefont
  {Zhang}}, \bibinfo {author} {\bibfnamefont {Ehsan}\ \bibnamefont {Khatami}},
  \bibinfo {author} {\bibfnamefont {Nandini}\ \bibnamefont {Trivedi}}, \bibinfo
  {author} {\bibfnamefont {Thereza}\ \bibnamefont {Paiva}}, \bibinfo {author}
  {\bibfnamefont {Marcos}\ \bibnamefont {Rigol}}, \ and\ \bibinfo {author}
  {\bibfnamefont {Martin~W.}\ \bibnamefont {Zwierlein}},\ }\bibfield  {title}
  {\enquote {\bibinfo {title} {{Observation of spatial charge and spin
  correlations in the 2D Fermi-Hubbard model}},}\ }\href {\doibase
  10.1126/science.aag3349} {\bibfield  {journal} {\bibinfo  {journal}
  {Science}\ }\textbf {\bibinfo {volume} {353}},\ \bibinfo {pages} {1260--1264}
  (\bibinfo {year} {2016})}\BibitemShut {NoStop}%
\bibitem [{\citenamefont {Parsons}\ \emph {et~al.}(2016)\citenamefont
  {Parsons}, \citenamefont {Mazurenko}, \citenamefont {Chiu}, \citenamefont
  {Ji}, \citenamefont {Greif},\ and\ \citenamefont {Greiner}}]{Parsons:2016}%
  \BibitemOpen
  \bibfield  {author} {\bibinfo {author} {\bibfnamefont {Maxwell~F.}\
  \bibnamefont {Parsons}}, \bibinfo {author} {\bibfnamefont {Anton}\
  \bibnamefont {Mazurenko}}, \bibinfo {author} {\bibfnamefont {Christie~S.}\
  \bibnamefont {Chiu}}, \bibinfo {author} {\bibfnamefont {Geoffrey}\
  \bibnamefont {Ji}}, \bibinfo {author} {\bibfnamefont {Daniel}\ \bibnamefont
  {Greif}}, \ and\ \bibinfo {author} {\bibfnamefont {Markus}\ \bibnamefont
  {Greiner}},\ }\bibfield  {title} {\enquote {\bibinfo {title} {{Site-resolved
  measurement of the spin-correlation function in the Fermi-Hubbard model}},}\
  }\href {\doibase 10.1126/science.aag1430} {\bibfield  {journal} {\bibinfo
  {journal} {Science}\ }\textbf {\bibinfo {volume} {353}},\ \bibinfo {pages}
  {1253--1256} (\bibinfo {year} {2016})}\BibitemShut {NoStop}%
\bibitem [{\citenamefont {Boll}\ \emph {et~al.}(2016)\citenamefont {Boll},
  \citenamefont {Hilker}, \citenamefont {Salomon}, \citenamefont {Omran},
  \citenamefont {Nespolo}, \citenamefont {Pollet}, \citenamefont {Bloch},\ and\
  \citenamefont {Gross}}]{Boll:2016}%
  \BibitemOpen
  \bibfield  {author} {\bibinfo {author} {\bibfnamefont {Martin}\ \bibnamefont
  {Boll}}, \bibinfo {author} {\bibfnamefont {Timon~A.}\ \bibnamefont {Hilker}},
  \bibinfo {author} {\bibfnamefont {Guillaume}\ \bibnamefont {Salomon}},
  \bibinfo {author} {\bibfnamefont {Ahmed}\ \bibnamefont {Omran}}, \bibinfo
  {author} {\bibfnamefont {Jacopo}\ \bibnamefont {Nespolo}}, \bibinfo {author}
  {\bibfnamefont {Lode}\ \bibnamefont {Pollet}}, \bibinfo {author}
  {\bibfnamefont {Immanuel}\ \bibnamefont {Bloch}}, \ and\ \bibinfo {author}
  {\bibfnamefont {Christian}\ \bibnamefont {Gross}},\ }\bibfield  {title}
  {\enquote {\bibinfo {title} {{Spin- and density-resolved microscopy of
  antiferromagnetic correlations in Fermi-Hubbard chains}},}\ }\href {\doibase
  10.1126/science.aag1635} {\bibfield  {journal} {\bibinfo  {journal}
  {Science}\ }\textbf {\bibinfo {volume} {353}},\ \bibinfo {pages} {1257--1260}
  (\bibinfo {year} {2016})}\BibitemShut {NoStop}%
\bibitem [{\citenamefont {Drewes}\ \emph {et~al.}(2016)\citenamefont {Drewes},
  \citenamefont {Cocchi}, \citenamefont {Miller}, \citenamefont {Chan},
  \citenamefont {Pertot}, \citenamefont {Brennecke},\ and\ \citenamefont
  {K\"ohl}}]{DrewesPRL2016}%
  \BibitemOpen
  \bibfield  {author} {\bibinfo {author} {\bibfnamefont {J.~H.}\ \bibnamefont
  {Drewes}}, \bibinfo {author} {\bibfnamefont {E.}~\bibnamefont {Cocchi}},
  \bibinfo {author} {\bibfnamefont {L.~A.}\ \bibnamefont {Miller}}, \bibinfo
  {author} {\bibfnamefont {C.~F.}\ \bibnamefont {Chan}}, \bibinfo {author}
  {\bibfnamefont {D.}~\bibnamefont {Pertot}}, \bibinfo {author} {\bibfnamefont
  {F.}~\bibnamefont {Brennecke}}, \ and\ \bibinfo {author} {\bibfnamefont
  {M.}~\bibnamefont {K\"ohl}},\ }\bibfield  {title} {\enquote {\bibinfo {title}
  {{Thermodynamics versus Local Density Fluctuations in the
  Metal--Mott-Insulator Crossover}},}\ }\href {\doibase
  10.1103/PhysRevLett.117.135301} {\bibfield  {journal} {\bibinfo  {journal}
  {Phys. Rev. Lett.}\ }\textbf {\bibinfo {volume} {117}},\ \bibinfo {pages}
  {135301} (\bibinfo {year} {2016})}\BibitemShut {NoStop}%
\bibitem [{\citenamefont {Drewes}\ \emph {et~al.}(2017)\citenamefont {Drewes},
  \citenamefont {Miller}, \citenamefont {Cocchi}, \citenamefont {Chan},
  \citenamefont {Wurz}, \citenamefont {Gall}, \citenamefont {Pertot},
  \citenamefont {Brennecke},\ and\ \citenamefont {K\"ohl}}]{drewesPRL2017}%
  \BibitemOpen
  \bibfield  {author} {\bibinfo {author} {\bibfnamefont {J.~H.}\ \bibnamefont
  {Drewes}}, \bibinfo {author} {\bibfnamefont {L.~A.}\ \bibnamefont {Miller}},
  \bibinfo {author} {\bibfnamefont {E.}~\bibnamefont {Cocchi}}, \bibinfo
  {author} {\bibfnamefont {C.~F.}\ \bibnamefont {Chan}}, \bibinfo {author}
  {\bibfnamefont {N.}~\bibnamefont {Wurz}}, \bibinfo {author} {\bibfnamefont
  {M.}~\bibnamefont {Gall}}, \bibinfo {author} {\bibfnamefont {D.}~\bibnamefont
  {Pertot}}, \bibinfo {author} {\bibfnamefont {F.}~\bibnamefont {Brennecke}}, \
  and\ \bibinfo {author} {\bibfnamefont {M.}~\bibnamefont {K\"ohl}},\
  }\bibfield  {title} {\enquote {\bibinfo {title} {Antiferromagnetic
  correlations in two-dimensional fermionic mott-insulating and metallic
  phases},}\ }\href {\doibase 10.1103/PhysRevLett.118.170401} {\bibfield
  {journal} {\bibinfo  {journal} {Phys. Rev. Lett.}\ }\textbf {\bibinfo
  {volume} {118}},\ \bibinfo {pages} {170401} (\bibinfo {year}
  {2017})}\BibitemShut {NoStop}%
\bibitem [{\citenamefont {Nichols}\ \emph {et~al.}(2018)\citenamefont
  {Nichols}, \citenamefont {Cheuk}, \citenamefont {Okan}, \citenamefont
  {Hartke}, \citenamefont {Mendez}, \citenamefont {Senthil}, \citenamefont
  {Khatami}, \citenamefont {Zhang},\ and\ \citenamefont
  {Zwierlein}}]{Nichols:2018}%
  \BibitemOpen
  \bibfield  {author} {\bibinfo {author} {\bibfnamefont {Matthew~A.}\
  \bibnamefont {Nichols}}, \bibinfo {author} {\bibfnamefont {Lawrence~W.}\
  \bibnamefont {Cheuk}}, \bibinfo {author} {\bibfnamefont {Melih}\ \bibnamefont
  {Okan}}, \bibinfo {author} {\bibfnamefont {Thomas~R.}\ \bibnamefont
  {Hartke}}, \bibinfo {author} {\bibfnamefont {Enrique}\ \bibnamefont
  {Mendez}}, \bibinfo {author} {\bibfnamefont {T.}~\bibnamefont {Senthil}},
  \bibinfo {author} {\bibfnamefont {Ehsan}\ \bibnamefont {Khatami}}, \bibinfo
  {author} {\bibfnamefont {Hao}\ \bibnamefont {Zhang}}, \ and\ \bibinfo
  {author} {\bibfnamefont {Martin~W.}\ \bibnamefont {Zwierlein}},\ }\bibfield
  {title} {\enquote {\bibinfo {title} {Spin transport in a mott insulator of
  ultracold fermions},}\ }\href {\doibase 10.1126/science.aat4387} {\bibfield
  {journal} {\bibinfo  {journal} {Science}\ } (\bibinfo {year} {2018}),\
  10.1126/science.aat4387}\BibitemShut {NoStop}%
\bibitem [{\citenamefont {Brown}\ \emph {et~al.}(2019)\citenamefont {Brown},
  \citenamefont {Mitra}, \citenamefont {Guardado-Sanchez}, \citenamefont
  {Nourafkan}, \citenamefont {Reymbaut}, \citenamefont {H{\'e}bert},
  \citenamefont {Bergeron}, \citenamefont {Tremblay}, \citenamefont {Kokalj},
  \citenamefont {Huse}, \citenamefont {Schau{\ss}},\ and\ \citenamefont
  {Bakr}}]{Brown:Science2019}%
  \BibitemOpen
  \bibfield  {author} {\bibinfo {author} {\bibfnamefont {Peter~T.}\
  \bibnamefont {Brown}}, \bibinfo {author} {\bibfnamefont {Debayan}\
  \bibnamefont {Mitra}}, \bibinfo {author} {\bibfnamefont {Elmer}\ \bibnamefont
  {Guardado-Sanchez}}, \bibinfo {author} {\bibfnamefont {Reza}\ \bibnamefont
  {Nourafkan}}, \bibinfo {author} {\bibfnamefont {Alexis}\ \bibnamefont
  {Reymbaut}}, \bibinfo {author} {\bibfnamefont {Charles-David}\ \bibnamefont
  {H{\'e}bert}}, \bibinfo {author} {\bibfnamefont {Simon}\ \bibnamefont
  {Bergeron}}, \bibinfo {author} {\bibfnamefont {A.-M.~S.}\ \bibnamefont
  {Tremblay}}, \bibinfo {author} {\bibfnamefont {Jure}\ \bibnamefont {Kokalj}},
  \bibinfo {author} {\bibfnamefont {David~A.}\ \bibnamefont {Huse}}, \bibinfo
  {author} {\bibfnamefont {Peter}\ \bibnamefont {Schau{\ss}}}, \ and\ \bibinfo
  {author} {\bibfnamefont {Waseem~S.}\ \bibnamefont {Bakr}},\ }\bibfield
  {title} {\enquote {\bibinfo {title} {Bad metallic transport in a cold atom
  fermi-hubbard system},}\ }\href {\doibase 10.1126/science.aat4134} {\bibfield
   {journal} {\bibinfo  {journal} {Science}\ }\textbf {\bibinfo {volume}
  {363}},\ \bibinfo {pages} {379--382} (\bibinfo {year} {2019})}\BibitemShut
  {NoStop}%
\bibitem [{\citenamefont {Anderson}(1987)}]{Anderson:1987}%
  \BibitemOpen
  \bibfield  {author} {\bibinfo {author} {\bibfnamefont {P.~W.}\ \bibnamefont
  {Anderson}},\ }\bibfield  {title} {\enquote {\bibinfo {title} {{The
  resonating valence bond state in La$_2$CuO$_4$ and superconductivity}},}\
  }\href {\doibase 10.1126/science.235.4793.1196} {\bibfield  {journal}
  {\bibinfo  {journal} {Science}\ }\textbf {\bibinfo {volume} {235}},\ \bibinfo
  {pages} {1196--1198} (\bibinfo {year} {1987})}\BibitemShut {NoStop}%
\bibitem [{\citenamefont {Walsh}\ \emph
  {et~al.}(2019{\natexlab{a}})\citenamefont {Walsh}, \citenamefont {S\'emon},
  \citenamefont {Poulin}, \citenamefont {Sordi},\ and\ \citenamefont
  {Tremblay}}]{Caitlin:PRL2019}%
  \BibitemOpen
  \bibfield  {author} {\bibinfo {author} {\bibfnamefont {C.}~\bibnamefont
  {Walsh}}, \bibinfo {author} {\bibfnamefont {P.}~\bibnamefont {S\'emon}},
  \bibinfo {author} {\bibfnamefont {D.}~\bibnamefont {Poulin}}, \bibinfo
  {author} {\bibfnamefont {G.}~\bibnamefont {Sordi}}, \ and\ \bibinfo {author}
  {\bibfnamefont {A.-M.~S.}\ \bibnamefont {Tremblay}},\ }\bibfield  {title}
  {\enquote {\bibinfo {title} {Local entanglement entropy and mutual
  information across the mott transition in the two-dimensional hubbard
  model},}\ }\href {\doibase 10.1103/PhysRevLett.122.067203} {\bibfield
  {journal} {\bibinfo  {journal} {Phys. Rev. Lett.}\ }\textbf {\bibinfo
  {volume} {122}},\ \bibinfo {pages} {067203} (\bibinfo {year}
  {2019}{\natexlab{a}})}\BibitemShut {NoStop}%
\bibitem [{\citenamefont {Groisman}\ \emph {et~al.}(2005)\citenamefont
  {Groisman}, \citenamefont {Popescu},\ and\ \citenamefont
  {Winter}}]{groisman2005}%
  \BibitemOpen
  \bibfield  {author} {\bibinfo {author} {\bibfnamefont {Berry}\ \bibnamefont
  {Groisman}}, \bibinfo {author} {\bibfnamefont {Sandu}\ \bibnamefont
  {Popescu}}, \ and\ \bibinfo {author} {\bibfnamefont {Andreas}\ \bibnamefont
  {Winter}},\ }\bibfield  {title} {\enquote {\bibinfo {title} {Quantum,
  classical, and total amount of correlations in a quantum state},}\ }\href
  {\doibase 10.1103/PhysRevA.72.032317} {\bibfield  {journal} {\bibinfo
  {journal} {Phys. Rev. A}\ }\textbf {\bibinfo {volume} {72}},\ \bibinfo
  {pages} {032317} (\bibinfo {year} {2005})}\BibitemShut {NoStop}%
\bibitem [{\citenamefont {Lee}\ \emph {et~al.}(2006)\citenamefont {Lee},
  \citenamefont {Nagaosa},\ and\ \citenamefont {Wen}}]{LeeRMP:2006}%
  \BibitemOpen
  \bibfield  {author} {\bibinfo {author} {\bibfnamefont {Patrick~A.}\
  \bibnamefont {Lee}}, \bibinfo {author} {\bibfnamefont {Naoto}\ \bibnamefont
  {Nagaosa}}, \ and\ \bibinfo {author} {\bibfnamefont {Xiao-Gang}\ \bibnamefont
  {Wen}},\ }\bibfield  {title} {\enquote {\bibinfo {title} {Doping a mott
  insulator: Physics of high-temperature superconductivity},}\ }\href {\doibase
  10.1103/RevModPhys.78.17} {\bibfield  {journal} {\bibinfo  {journal} {Rev.
  Mod. Phys.}\ }\textbf {\bibinfo {volume} {78}},\ \bibinfo {pages} {17--85}
  (\bibinfo {year} {2006})}\BibitemShut {NoStop}%
\bibitem [{\citenamefont {Tremblay}\ \emph {et~al.}(2006)\citenamefont
  {Tremblay}, \citenamefont {Kyung},\ and\ \citenamefont
  {S\'{e}n\'{e}chal}}]{tremblayR}%
  \BibitemOpen
  \bibfield  {author} {\bibinfo {author} {\bibfnamefont {A.-M.~S.}\
  \bibnamefont {Tremblay}}, \bibinfo {author} {\bibfnamefont {B.}~\bibnamefont
  {Kyung}}, \ and\ \bibinfo {author} {\bibfnamefont {D.}~\bibnamefont
  {S\'{e}n\'{e}chal}},\ }\bibfield  {title} {\enquote {\bibinfo {title}
  {{Pseudogap and high-temperature superconductivity from weak to strong
  coupling. Towards a quantitative theory}},}\ }\href {\doibase
  10.1063/1.2199446} {\bibfield  {journal} {\bibinfo  {journal} {Low Temp.
  Phys.}\ }\textbf {\bibinfo {volume} {32}},\ \bibinfo {pages} {424} (\bibinfo
  {year} {2006})}\BibitemShut {NoStop}%
\bibitem [{\citenamefont {Sordi}\ \emph {et~al.}(2010)\citenamefont {Sordi},
  \citenamefont {Haule},\ and\ \citenamefont {Tremblay}}]{sht}%
  \BibitemOpen
  \bibfield  {author} {\bibinfo {author} {\bibfnamefont {G.}~\bibnamefont
  {Sordi}}, \bibinfo {author} {\bibfnamefont {K.}~\bibnamefont {Haule}}, \ and\
  \bibinfo {author} {\bibfnamefont {A.-M.~S.}\ \bibnamefont {Tremblay}},\
  }\bibfield  {title} {\enquote {\bibinfo {title} {{Finite Doping Signatures of
  the Mott Transition in the Two-Dimensional Hubbard Model}},}\ }\href
  {\doibase 10.1103/PhysRevLett.104.226402} {\bibfield  {journal} {\bibinfo
  {journal} {Phys. Rev. Lett.}\ }\textbf {\bibinfo {volume} {104}},\ \bibinfo
  {pages} {226402} (\bibinfo {year} {2010})}\BibitemShut {NoStop}%
\bibitem [{\citenamefont {Wolf}\ \emph {et~al.}(2008)\citenamefont {Wolf},
  \citenamefont {Verstraete}, \citenamefont {Hastings},\ and\ \citenamefont
  {Cirac}}]{wolfPRL2008}%
  \BibitemOpen
  \bibfield  {author} {\bibinfo {author} {\bibfnamefont {Michael~M.}\
  \bibnamefont {Wolf}}, \bibinfo {author} {\bibfnamefont {Frank}\ \bibnamefont
  {Verstraete}}, \bibinfo {author} {\bibfnamefont {Matthew~B.}\ \bibnamefont
  {Hastings}}, \ and\ \bibinfo {author} {\bibfnamefont {J.~Ignacio}\
  \bibnamefont {Cirac}},\ }\bibfield  {title} {\enquote {\bibinfo {title} {Area
  laws in quantum systems: Mutual information and correlations},}\ }\href
  {\doibase 10.1103/PhysRevLett.100.070502} {\bibfield  {journal} {\bibinfo
  {journal} {Phys. Rev. Lett.}\ }\textbf {\bibinfo {volume} {100}},\ \bibinfo
  {pages} {070502} (\bibinfo {year} {2008})}\BibitemShut {NoStop}%
\bibitem [{\citenamefont {Amico}\ and\ \citenamefont
  {Patan\`e}(2007)}]{Amico_Patan_2007}%
  \BibitemOpen
  \bibfield  {author} {\bibinfo {author} {\bibfnamefont {L.}~\bibnamefont
  {Amico}}\ and\ \bibinfo {author} {\bibfnamefont {D.}~\bibnamefont
  {Patan\`e}},\ }\bibfield  {title} {\enquote {\bibinfo {title} {Entanglement
  crossover close to a quantum critical point},}\ }\href {\doibase
  10.1209/0295-5075/77/17001} {\bibfield  {journal} {\bibinfo  {journal} {EPL
  (Europhysics Letters)}\ }\textbf {\bibinfo {volume} {77}},\ \bibinfo {pages}
  {17001} (\bibinfo {year} {2007})}\BibitemShut {NoStop}%
\bibitem [{\citenamefont {Melko}\ \emph {et~al.}(2010)\citenamefont {Melko},
  \citenamefont {Kallin},\ and\ \citenamefont {Hastings}}]{Melko:2010}%
  \BibitemOpen
  \bibfield  {author} {\bibinfo {author} {\bibfnamefont {Roger~G.}\
  \bibnamefont {Melko}}, \bibinfo {author} {\bibfnamefont {Ann~B.}\
  \bibnamefont {Kallin}}, \ and\ \bibinfo {author} {\bibfnamefont {Matthew~B.}\
  \bibnamefont {Hastings}},\ }\bibfield  {title} {\enquote {\bibinfo {title}
  {{Finite-size scaling of mutual information in Monte Carlo simulations:
  Application to the spin-$\frac{1}{2}$ $XXZ$ model}},}\ }\href {\doibase
  10.1103/PhysRevB.82.100409} {\bibfield  {journal} {\bibinfo  {journal} {Phys.
  Rev. B}\ }\textbf {\bibinfo {volume} {82}},\ \bibinfo {pages} {100409}
  (\bibinfo {year} {2010})}\BibitemShut {NoStop}%
\bibitem [{\citenamefont {Singh}\ \emph {et~al.}(2011)\citenamefont {Singh},
  \citenamefont {Hastings}, \citenamefont {Kallin},\ and\ \citenamefont
  {Melko}}]{Singh:2011}%
  \BibitemOpen
  \bibfield  {author} {\bibinfo {author} {\bibfnamefont {Rajiv R.~P.}\
  \bibnamefont {Singh}}, \bibinfo {author} {\bibfnamefont {Matthew~B.}\
  \bibnamefont {Hastings}}, \bibinfo {author} {\bibfnamefont {Ann~B.}\
  \bibnamefont {Kallin}}, \ and\ \bibinfo {author} {\bibfnamefont {Roger~G.}\
  \bibnamefont {Melko}},\ }\bibfield  {title} {\enquote {\bibinfo {title}
  {Finite-temperature critical behavior of mutual information},}\ }\href
  {\doibase 10.1103/PhysRevLett.106.135701} {\bibfield  {journal} {\bibinfo
  {journal} {Phys. Rev. Lett.}\ }\textbf {\bibinfo {volume} {106}},\ \bibinfo
  {pages} {135701} (\bibinfo {year} {2011})}\BibitemShut {NoStop}%
\bibitem [{\citenamefont {Kallin}\ \emph {et~al.}(2011)\citenamefont {Kallin},
  \citenamefont {Hastings}, \citenamefont {Melko},\ and\ \citenamefont
  {Singh}}]{Kallin:2011}%
  \BibitemOpen
  \bibfield  {author} {\bibinfo {author} {\bibfnamefont {Ann~B.}\ \bibnamefont
  {Kallin}}, \bibinfo {author} {\bibfnamefont {Matthew~B.}\ \bibnamefont
  {Hastings}}, \bibinfo {author} {\bibfnamefont {Roger~G.}\ \bibnamefont
  {Melko}}, \ and\ \bibinfo {author} {\bibfnamefont {Rajiv R.~P.}\ \bibnamefont
  {Singh}},\ }\bibfield  {title} {\enquote {\bibinfo {title} {{Anomalies in the
  entanglement properties of the square-lattice Heisenberg model}},}\ }\href
  {\doibase 10.1103/PhysRevB.84.165134} {\bibfield  {journal} {\bibinfo
  {journal} {Phys. Rev. B}\ }\textbf {\bibinfo {volume} {84}},\ \bibinfo
  {pages} {165134} (\bibinfo {year} {2011})}\BibitemShut {NoStop}%
\bibitem [{\citenamefont {Wilms}\ \emph {et~al.}(2011)\citenamefont {Wilms},
  \citenamefont {Troyer},\ and\ \citenamefont
  {Verstraete}}]{Wilms_Troyer_Verstraete_2011}%
  \BibitemOpen
  \bibfield  {author} {\bibinfo {author} {\bibfnamefont {Johannes}\
  \bibnamefont {Wilms}}, \bibinfo {author} {\bibfnamefont {Matthias}\
  \bibnamefont {Troyer}}, \ and\ \bibinfo {author} {\bibfnamefont {Frank}\
  \bibnamefont {Verstraete}},\ }\bibfield  {title} {\enquote {\bibinfo {title}
  {Mutual information in classical spin models},}\ }\href {\doibase
  10.1088/1742-5468/2011/10/P10011} {\bibfield  {journal} {\bibinfo  {journal}
  {Journal of Statistical Mechanics: Theory and Experiment}\ }\textbf {\bibinfo
  {volume} {2011}},\ \bibinfo {pages} {P10011} (\bibinfo {year}
  {2011})}\BibitemShut {NoStop}%
\bibitem [{\citenamefont {Wilms}\ \emph {et~al.}(2012)\citenamefont {Wilms},
  \citenamefont {Vidal}, \citenamefont {Verstraete},\ and\ \citenamefont
  {Dusuel}}]{Wilms:2012}%
  \BibitemOpen
  \bibfield  {author} {\bibinfo {author} {\bibfnamefont {Johannes}\
  \bibnamefont {Wilms}}, \bibinfo {author} {\bibfnamefont {Julien}\
  \bibnamefont {Vidal}}, \bibinfo {author} {\bibfnamefont {Frank}\ \bibnamefont
  {Verstraete}}, \ and\ \bibinfo {author} {\bibfnamefont {S\'ebastien}\
  \bibnamefont {Dusuel}},\ }\bibfield  {title} {\enquote {\bibinfo {title}
  {Finite-temperature mutual information in a simple phase transition},}\
  }\href {http://stacks.iop.org/1742-5468/2012/i=01/a=P01023} {\bibfield
  {journal} {\bibinfo  {journal} {Journal of Statistical Mechanics: Theory and
  Experiment}\ }\textbf {\bibinfo {volume} {2012}},\ \bibinfo {pages} {P01023}
  (\bibinfo {year} {2012})}\BibitemShut {NoStop}%
\bibitem [{\citenamefont {Iaconis}\ \emph {et~al.}(2013)\citenamefont
  {Iaconis}, \citenamefont {Inglis}, \citenamefont {Kallin},\ and\
  \citenamefont {Melko}}]{Iaconis:2013}%
  \BibitemOpen
  \bibfield  {author} {\bibinfo {author} {\bibfnamefont {Jason}\ \bibnamefont
  {Iaconis}}, \bibinfo {author} {\bibfnamefont {Stephen}\ \bibnamefont
  {Inglis}}, \bibinfo {author} {\bibfnamefont {Ann~B.}\ \bibnamefont {Kallin}},
  \ and\ \bibinfo {author} {\bibfnamefont {Roger~G.}\ \bibnamefont {Melko}},\
  }\bibfield  {title} {\enquote {\bibinfo {title} {{Detecting classical phase
  transitions with Renyi mutual information}},}\ }\href {\doibase
  10.1103/PhysRevB.87.195134} {\bibfield  {journal} {\bibinfo  {journal} {Phys.
  Rev. B}\ }\textbf {\bibinfo {volume} {87}},\ \bibinfo {pages} {195134}
  (\bibinfo {year} {2013})}\BibitemShut {NoStop}%
\bibitem [{\citenamefont {{Gabbrielli}}\ \emph {et~al.}(2018)\citenamefont
  {{Gabbrielli}}, \citenamefont {{Smerzi}},\ and\ \citenamefont
  {{Pezz{\`e}}}}]{Gabbrielli:2019}%
  \BibitemOpen
  \bibfield  {author} {\bibinfo {author} {\bibfnamefont {Marco}\ \bibnamefont
  {{Gabbrielli}}}, \bibinfo {author} {\bibfnamefont {Augusto}\ \bibnamefont
  {{Smerzi}}}, \ and\ \bibinfo {author} {\bibfnamefont {Luca}\ \bibnamefont
  {{Pezz{\`e}}}},\ }\bibfield  {title} {\enquote {\bibinfo {title}
  {{Multipartite Entanglement at Finite Temperature}},}\ }\href {\doibase
  10.1038/s41598-018-31761-3} {\bibfield  {journal} {\bibinfo  {journal}
  {Scientific Reports}\ }\textbf {\bibinfo {volume} {8}},\ \bibinfo {eid}
  {15663} (\bibinfo {year} {2018})}\BibitemShut {NoStop}%
\bibitem [{\citenamefont {{Fr{\'e}rot}}\ and\ \citenamefont
  {{Roscilde}}(2019)}]{Frerot:2019}%
  \BibitemOpen
  \bibfield  {author} {\bibinfo {author} {\bibfnamefont {Ir{\'e}n{\'e}e}\
  \bibnamefont {{Fr{\'e}rot}}}\ and\ \bibinfo {author} {\bibfnamefont
  {Tommaso}\ \bibnamefont {{Roscilde}}},\ }\bibfield  {title} {\enquote
  {\bibinfo {title} {{Reconstructing the quantum critical fan of strongly
  correlated systems using quantum correlations}},}\ }\href {\doibase
  10.1038/s41467-019-08324-9} {\bibfield  {journal} {\bibinfo  {journal}
  {Nature Communications}\ }\textbf {\bibinfo {volume} {10}},\ \bibinfo {eid}
  {577} (\bibinfo {year} {2019})}\BibitemShut {NoStop}%
\bibitem [{\citenamefont {Wald}\ \emph {et~al.}(2020)\citenamefont {Wald},
  \citenamefont {Arias},\ and\ \citenamefont {Alba}}]{Wald:JSM2020}%
  \BibitemOpen
  \bibfield  {author} {\bibinfo {author} {\bibfnamefont {Sascha}\ \bibnamefont
  {Wald}}, \bibinfo {author} {\bibfnamefont {Ra{\'{u}}l}\ \bibnamefont
  {Arias}}, \ and\ \bibinfo {author} {\bibfnamefont {Vincenzo}\ \bibnamefont
  {Alba}},\ }\bibfield  {title} {\enquote {\bibinfo {title} {Entanglement and
  classical fluctuations at finite-temperature critical points},}\ }\href
  {\doibase 10.1088/1742-5468/ab6b19} {\bibfield  {journal} {\bibinfo
  {journal} {Journal of Statistical Mechanics: Theory and Experiment}\ }\textbf
  {\bibinfo {volume} {2020}},\ \bibinfo {pages} {033105} (\bibinfo {year}
  {2020})}\BibitemShut {NoStop}%
\bibitem [{\citenamefont {Sordi}\ \emph {et~al.}(2011)\citenamefont {Sordi},
  \citenamefont {Haule},\ and\ \citenamefont {Tremblay}}]{sht2}%
  \BibitemOpen
  \bibfield  {author} {\bibinfo {author} {\bibfnamefont {G.}~\bibnamefont
  {Sordi}}, \bibinfo {author} {\bibfnamefont {K.}~\bibnamefont {Haule}}, \ and\
  \bibinfo {author} {\bibfnamefont {A.-M.~S.}\ \bibnamefont {Tremblay}},\
  }\bibfield  {title} {\enquote {\bibinfo {title} {{Mott physics and
  first-order transition between two metals in the normal-state phase diagram
  of the two-dimensional Hubbard model}},}\ }\href {\doibase
  10.1103/PhysRevB.84.075161} {\bibfield  {journal} {\bibinfo  {journal} {Phys.
  Rev. B}\ }\textbf {\bibinfo {volume} {84}},\ \bibinfo {pages} {075161}
  (\bibinfo {year} {2011})}\BibitemShut {NoStop}%
\bibitem [{\citenamefont {Sordi}\ \emph {et~al.}(2012)\citenamefont {Sordi},
  \citenamefont {S\'emon}, \citenamefont {Haule},\ and\ \citenamefont
  {Tremblay}}]{ssht}%
  \BibitemOpen
  \bibfield  {author} {\bibinfo {author} {\bibfnamefont {G.}~\bibnamefont
  {Sordi}}, \bibinfo {author} {\bibfnamefont {P.}~\bibnamefont {S\'emon}},
  \bibinfo {author} {\bibfnamefont {K.}~\bibnamefont {Haule}}, \ and\ \bibinfo
  {author} {\bibfnamefont {A.-M.~S.}\ \bibnamefont {Tremblay}},\ }\bibfield
  {title} {\enquote {\bibinfo {title} {{Pseudogap temperature as a Widom line
  in doped Mott insulators}},}\ }\href {\doibase doi:10.1038/srep00547}
  {\bibfield  {journal} {\bibinfo  {journal} {Sci. Rep.}\ }\textbf {\bibinfo
  {volume} {2}},\ \bibinfo {pages} {547} (\bibinfo {year} {2012})}\BibitemShut
  {NoStop}%
\bibitem [{\citenamefont {Sordi}\ \emph {et~al.}(2013)\citenamefont {Sordi},
  \citenamefont {S\'emon}, \citenamefont {Haule},\ and\ \citenamefont
  {Tremblay}}]{sshtRHO}%
  \BibitemOpen
  \bibfield  {author} {\bibinfo {author} {\bibfnamefont {G.}~\bibnamefont
  {Sordi}}, \bibinfo {author} {\bibfnamefont {P.}~\bibnamefont {S\'emon}},
  \bibinfo {author} {\bibfnamefont {K.}~\bibnamefont {Haule}}, \ and\ \bibinfo
  {author} {\bibfnamefont {A.-M.~S.}\ \bibnamefont {Tremblay}},\ }\bibfield
  {title} {\enquote {\bibinfo {title} {{$c$-axis resistivity, pseudogap,
  superconductivity, and Widom line in doped Mott insulators }},}\ }\href
  {\doibase 10.1103/PhysRevB.87.041101} {\bibfield  {journal} {\bibinfo
  {journal} {Phys. Rev. B}\ }\textbf {\bibinfo {volume} {87}},\ \bibinfo
  {pages} {041101} (\bibinfo {year} {2013})}\BibitemShut {NoStop}%
\bibitem [{\citenamefont {Reymbaut}\ \emph {et~al.}(2019)\citenamefont
  {Reymbaut}, \citenamefont {Bergeron}, \citenamefont {Garioud}, \citenamefont
  {Th\'enault}, \citenamefont {Charlebois}, \citenamefont {S\'emon},\ and\
  \citenamefont {Tremblay}}]{Alexis:2019}%
  \BibitemOpen
  \bibfield  {author} {\bibinfo {author} {\bibfnamefont {A.}~\bibnamefont
  {Reymbaut}}, \bibinfo {author} {\bibfnamefont {S.}~\bibnamefont {Bergeron}},
  \bibinfo {author} {\bibfnamefont {R.}~\bibnamefont {Garioud}}, \bibinfo
  {author} {\bibfnamefont {M.}~\bibnamefont {Th\'enault}}, \bibinfo {author}
  {\bibfnamefont {M.}~\bibnamefont {Charlebois}}, \bibinfo {author}
  {\bibfnamefont {P.}~\bibnamefont {S\'emon}}, \ and\ \bibinfo {author}
  {\bibfnamefont {A.-M.~S.}\ \bibnamefont {Tremblay}},\ }\bibfield  {title}
  {\enquote {\bibinfo {title} {Pseudogap, van hove singularity, maximum in
  entropy, and specific heat for hole-doped mott insulators},}\ }\href
  {\doibase 10.1103/PhysRevResearch.1.023015} {\bibfield  {journal} {\bibinfo
  {journal} {Phys. Rev. Research}\ }\textbf {\bibinfo {volume} {1}},\ \bibinfo
  {pages} {023015} (\bibinfo {year} {2019})}\BibitemShut {NoStop}%
\bibitem [{\citenamefont {Alloul}\ \emph {et~al.}(1989)\citenamefont {Alloul},
  \citenamefont {Ohno},\ and\ \citenamefont {Mendels}}]{Alloul:1989}%
  \BibitemOpen
  \bibfield  {author} {\bibinfo {author} {\bibfnamefont {H.}~\bibnamefont
  {Alloul}}, \bibinfo {author} {\bibfnamefont {T.}~\bibnamefont {Ohno}}, \ and\
  \bibinfo {author} {\bibfnamefont {P.}~\bibnamefont {Mendels}},\ }\bibfield
  {title} {\enquote {\bibinfo {title} {{$^{89}\mathrm{Y}$ NMR evidence for a
  fermi-liquid behavior in
  ${\mathrm{YBa}}_{2}$${\mathrm{Cu}}_{3}$${\mathrm{O}}_{6+\mathrm{x}}$}},}\
  }\href {\doibase 10.1103/PhysRevLett.63.1700} {\bibfield  {journal} {\bibinfo
   {journal} {Phys. Rev. Lett.}\ }\textbf {\bibinfo {volume} {63}},\ \bibinfo
  {pages} {1700--1703} (\bibinfo {year} {1989})}\BibitemShut {NoStop}%
\bibitem [{\citenamefont {Reymbaut}\ \emph {et~al.}(2020)\citenamefont
  {Reymbaut}, \citenamefont {Boulay}, \citenamefont {Fratino}, \citenamefont
  {Sémon}, \citenamefont {Wu}, \citenamefont {Sordi},\ and\ \citenamefont
  {Tremblay}}]{reymbaut2020crossovers}%
  \BibitemOpen
  \bibfield  {author} {\bibinfo {author} {\bibfnamefont {A.}~\bibnamefont
  {Reymbaut}}, \bibinfo {author} {\bibfnamefont {M.}~\bibnamefont {Boulay}},
  \bibinfo {author} {\bibfnamefont {L.}~\bibnamefont {Fratino}}, \bibinfo
  {author} {\bibfnamefont {P.}~\bibnamefont {Sémon}}, \bibinfo {author}
  {\bibfnamefont {Wei}\ \bibnamefont {Wu}}, \bibinfo {author} {\bibfnamefont
  {G.}~\bibnamefont {Sordi}}, \ and\ \bibinfo {author} {\bibfnamefont
  {A.~M.~S.}\ \bibnamefont {Tremblay}},\ }\href@noop {} {\enquote {\bibinfo
  {title} {Mott transition and high-temperature crossovers at half-filling},}\
  } (\bibinfo {year} {2020}),\ \Eprint {http://arxiv.org/abs/2004.02302}
  {arXiv:2004.02302 [cond-mat.str-el]} \BibitemShut {NoStop}%
\bibitem [{\citenamefont {Maier}\ \emph {et~al.}(2005)\citenamefont {Maier},
  \citenamefont {Jarrell}, \citenamefont {Pruschke},\ and\ \citenamefont
  {Hettler}}]{maier}%
  \BibitemOpen
  \bibfield  {author} {\bibinfo {author} {\bibfnamefont {Thomas}\ \bibnamefont
  {Maier}}, \bibinfo {author} {\bibfnamefont {Mark}\ \bibnamefont {Jarrell}},
  \bibinfo {author} {\bibfnamefont {Thomas}\ \bibnamefont {Pruschke}}, \ and\
  \bibinfo {author} {\bibfnamefont {Matthias~H.}\ \bibnamefont {Hettler}},\
  }\bibfield  {title} {\enquote {\bibinfo {title} {Quantum cluster theories},}\
  }\href {\doibase 10.1103/RevModPhys.77.1027} {\bibfield  {journal} {\bibinfo
  {journal} {Rev. Mod. Phys.}\ }\textbf {\bibinfo {volume} {77}},\ \bibinfo
  {pages} {1027--1080} (\bibinfo {year} {2005})}\BibitemShut {NoStop}%
\bibitem [{\citenamefont {Kotliar}\ \emph {et~al.}(2006)\citenamefont
  {Kotliar}, \citenamefont {Savrasov}, \citenamefont {Haule}, \citenamefont
  {Oudovenko}, \citenamefont {Parcollet},\ and\ \citenamefont
  {Marianetti}}]{kotliarRMP}%
  \BibitemOpen
  \bibfield  {author} {\bibinfo {author} {\bibfnamefont {G.}~\bibnamefont
  {Kotliar}}, \bibinfo {author} {\bibfnamefont {S.~Y.}\ \bibnamefont
  {Savrasov}}, \bibinfo {author} {\bibfnamefont {K.}~\bibnamefont {Haule}},
  \bibinfo {author} {\bibfnamefont {V.~S.}\ \bibnamefont {Oudovenko}}, \bibinfo
  {author} {\bibfnamefont {O.}~\bibnamefont {Parcollet}}, \ and\ \bibinfo
  {author} {\bibfnamefont {C.~A.}\ \bibnamefont {Marianetti}},\ }\bibfield
  {title} {\enquote {\bibinfo {title} {{Electronic structure calculations with
  dynamical mean-field theory}},}\ }\href {\doibase 10.1103/RevModPhys.78.865}
  {\bibfield  {journal} {\bibinfo  {journal} {Rev. Mod. Phys.}\ }\textbf
  {\bibinfo {volume} {78}},\ \bibinfo {eid} {865} (\bibinfo {year}
  {2006})}\BibitemShut {NoStop}%
\bibitem [{\citenamefont {Georges}\ \emph {et~al.}(1996)\citenamefont
  {Georges}, \citenamefont {Kotliar}, \citenamefont {Krauth},\ and\
  \citenamefont {Rozenberg}}]{rmp}%
  \BibitemOpen
  \bibfield  {author} {\bibinfo {author} {\bibfnamefont {Antoine}\ \bibnamefont
  {Georges}}, \bibinfo {author} {\bibfnamefont {Gabriel}\ \bibnamefont
  {Kotliar}}, \bibinfo {author} {\bibfnamefont {Werner}\ \bibnamefont
  {Krauth}}, \ and\ \bibinfo {author} {\bibfnamefont {Marcelo~J.}\ \bibnamefont
  {Rozenberg}},\ }\bibfield  {title} {\enquote {\bibinfo {title} {{Dynamical
  mean-field theory of strongly correlated fermion systems and the limit of
  infinite dimensions}},}\ }\href {\doibase 10.1103/RevModPhys.68.13}
  {\bibfield  {journal} {\bibinfo  {journal} {Rev. Mod. Phys.}\ }\textbf
  {\bibinfo {volume} {68}},\ \bibinfo {pages} {13} (\bibinfo {year}
  {1996})}\BibitemShut {NoStop}%
\bibitem [{\citenamefont {Gull}\ \emph {et~al.}(2011)\citenamefont {Gull},
  \citenamefont {Millis}, \citenamefont {Lichtenstein}, \citenamefont
  {Rubtsov}, \citenamefont {Troyer},\ and\ \citenamefont {Werner}}]{millisRMP}%
  \BibitemOpen
  \bibfield  {author} {\bibinfo {author} {\bibfnamefont {Emanuel}\ \bibnamefont
  {Gull}}, \bibinfo {author} {\bibfnamefont {Andrew~J.}\ \bibnamefont
  {Millis}}, \bibinfo {author} {\bibfnamefont {Alexander~I.}\ \bibnamefont
  {Lichtenstein}}, \bibinfo {author} {\bibfnamefont {Alexey~N.}\ \bibnamefont
  {Rubtsov}}, \bibinfo {author} {\bibfnamefont {Matthias}\ \bibnamefont
  {Troyer}}, \ and\ \bibinfo {author} {\bibfnamefont {Philipp}\ \bibnamefont
  {Werner}},\ }\bibfield  {title} {\enquote {\bibinfo {title} {{Continuous-time
  Monte~Carlo methods for quantum impurity models}},}\ }\href {\doibase
  10.1103/RevModPhys.83.349} {\bibfield  {journal} {\bibinfo  {journal} {Rev.
  Mod. Phys.}\ }\textbf {\bibinfo {volume} {83}},\ \bibinfo {pages} {349--404}
  (\bibinfo {year} {2011})}\BibitemShut {NoStop}%
\bibitem [{\citenamefont {S\'emon}\ \emph {et~al.}(2014)\citenamefont
  {S\'emon}, \citenamefont {Yee}, \citenamefont {Haule},\ and\ \citenamefont
  {Tremblay}}]{patrickSkipList}%
  \BibitemOpen
  \bibfield  {author} {\bibinfo {author} {\bibfnamefont {P.}~\bibnamefont
  {S\'emon}}, \bibinfo {author} {\bibfnamefont {Chuck-Hou}\ \bibnamefont
  {Yee}}, \bibinfo {author} {\bibfnamefont {Kristjan}\ \bibnamefont {Haule}}, \
  and\ \bibinfo {author} {\bibfnamefont {A.-M.~S.}\ \bibnamefont {Tremblay}},\
  }\bibfield  {title} {\enquote {\bibinfo {title} {{Lazy skip-lists: An
  algorithm for fast hybridization-expansion quantum Monte Carlo}},}\ }\href
  {\doibase 10.1103/PhysRevB.90.075149} {\bibfield  {journal} {\bibinfo
  {journal} {Phys. Rev. B}\ }\textbf {\bibinfo {volume} {90}},\ \bibinfo
  {pages} {075149} (\bibinfo {year} {2014})}\BibitemShut {NoStop}%
\bibitem [{\citenamefont {Walsh}\ \emph
  {et~al.}(2019{\natexlab{b}})\citenamefont {Walsh}, \citenamefont {S\'emon},
  \citenamefont {Poulin}, \citenamefont {Sordi},\ and\ \citenamefont
  {Tremblay}}]{CaitlinSb}%
  \BibitemOpen
  \bibfield  {author} {\bibinfo {author} {\bibfnamefont {C.}~\bibnamefont
  {Walsh}}, \bibinfo {author} {\bibfnamefont {P.}~\bibnamefont {S\'emon}},
  \bibinfo {author} {\bibfnamefont {D.}~\bibnamefont {Poulin}}, \bibinfo
  {author} {\bibfnamefont {G.}~\bibnamefont {Sordi}}, \ and\ \bibinfo {author}
  {\bibfnamefont {A.-M.~S.}\ \bibnamefont {Tremblay}},\ }\bibfield  {title}
  {\enquote {\bibinfo {title} {{Thermodynamic and information-theoretic
  description of the Mott transition in the two-dimensional Hubbard model}},}\
  }\href {\doibase 10.1103/PhysRevB.99.075122} {\bibfield  {journal} {\bibinfo
  {journal} {Phys. Rev. B}\ }\textbf {\bibinfo {volume} {99}},\ \bibinfo
  {pages} {075122} (\bibinfo {year} {2019}{\natexlab{b}})}\BibitemShut
  {NoStop}%
\bibitem [{\citenamefont {Sch\"afer}\ \emph {et~al.}(2015)\citenamefont
  {Sch\"afer}, \citenamefont {Geles}, \citenamefont {Rost}, \citenamefont
  {Rohringer}, \citenamefont {Arrigoni}, \citenamefont {Held}, \citenamefont
  {Bl\"umer}, \citenamefont {Aichhorn},\ and\ \citenamefont
  {Toschi}}]{Shafer2015}%
  \BibitemOpen
  \bibfield  {author} {\bibinfo {author} {\bibfnamefont {T.}~\bibnamefont
  {Sch\"afer}}, \bibinfo {author} {\bibfnamefont {F.}~\bibnamefont {Geles}},
  \bibinfo {author} {\bibfnamefont {D.}~\bibnamefont {Rost}}, \bibinfo {author}
  {\bibfnamefont {G.}~\bibnamefont {Rohringer}}, \bibinfo {author}
  {\bibfnamefont {E.}~\bibnamefont {Arrigoni}}, \bibinfo {author}
  {\bibfnamefont {K.}~\bibnamefont {Held}}, \bibinfo {author} {\bibfnamefont
  {N.}~\bibnamefont {Bl\"umer}}, \bibinfo {author} {\bibfnamefont
  {M.}~\bibnamefont {Aichhorn}}, \ and\ \bibinfo {author} {\bibfnamefont
  {A.}~\bibnamefont {Toschi}},\ }\bibfield  {title} {\enquote {\bibinfo {title}
  {{Fate of the false Mott-Hubbard transition in two dimensions}},}\ }\href
  {\doibase 10.1103/PhysRevB.91.125109} {\bibfield  {journal} {\bibinfo
  {journal} {Phys. Rev. B}\ }\textbf {\bibinfo {volume} {91}},\ \bibinfo
  {pages} {125109} (\bibinfo {year} {2015})}\BibitemShut {NoStop}%
\bibitem [{\citenamefont {Biroli}\ and\ \citenamefont
  {Kotliar}(2005)}]{BiroliExp2005}%
  \BibitemOpen
  \bibfield  {author} {\bibinfo {author} {\bibfnamefont {G.}~\bibnamefont
  {Biroli}}\ and\ \bibinfo {author} {\bibfnamefont {G.}~\bibnamefont
  {Kotliar}},\ }\bibfield  {title} {\enquote {\bibinfo {title} {Reply to
  ``comment on `cluster methods for strongly correlated electron systems'
  ''},}\ }\href {\doibase 10.1103/PhysRevB.71.037102} {\bibfield  {journal}
  {\bibinfo  {journal} {Phys. Rev. B}\ }\textbf {\bibinfo {volume} {71}},\
  \bibinfo {pages} {037102} (\bibinfo {year} {2005})}\BibitemShut {NoStop}%
\bibitem [{\citenamefont {Schäfer}\ \emph {et~al.}(2020)\citenamefont
  {Schäfer}, \citenamefont {Wentzell}, \citenamefont {Šimkovic IV},
  \citenamefont {He}, \citenamefont {Hille}, \citenamefont {Klett},
  \citenamefont {Eckhardt}, \citenamefont {Arzhang}, \citenamefont {Harkov},
  \citenamefont {Régent}, \citenamefont {Kirsch}, \citenamefont {Wang},
  \citenamefont {Kim}, \citenamefont {Kozik}, \citenamefont {Stepanov},
  \citenamefont {Kauch}, \citenamefont {Andergassen}, \citenamefont {Hansmann},
  \citenamefont {Rohe}, \citenamefont {Vilk}, \citenamefont {LeBlanc},
  \citenamefont {Zhang}, \citenamefont {Tremblay}, \citenamefont {Ferrero},
  \citenamefont {Parcollet},\ and\ \citenamefont
  {Georges}}]{Schafer:arXiv2020}%
  \BibitemOpen
  \bibfield  {author} {\bibinfo {author} {\bibfnamefont {Thomas}\ \bibnamefont
  {Schäfer}}, \bibinfo {author} {\bibfnamefont {Nils}\ \bibnamefont
  {Wentzell}}, \bibinfo {author} {\bibfnamefont {Fedor}\ \bibnamefont
  {Šimkovic IV}}, \bibinfo {author} {\bibfnamefont {Yuan-Yao}\ \bibnamefont
  {He}}, \bibinfo {author} {\bibfnamefont {Cornelia}\ \bibnamefont {Hille}},
  \bibinfo {author} {\bibfnamefont {Marcel}\ \bibnamefont {Klett}}, \bibinfo
  {author} {\bibfnamefont {Christian~J.}\ \bibnamefont {Eckhardt}}, \bibinfo
  {author} {\bibfnamefont {Behnam}\ \bibnamefont {Arzhang}}, \bibinfo {author}
  {\bibfnamefont {Viktor}\ \bibnamefont {Harkov}}, \bibinfo {author}
  {\bibfnamefont {François-Marie~Le}\ \bibnamefont {Régent}}, \bibinfo
  {author} {\bibfnamefont {Alfred}\ \bibnamefont {Kirsch}}, \bibinfo {author}
  {\bibfnamefont {Yan}\ \bibnamefont {Wang}}, \bibinfo {author} {\bibfnamefont
  {Aaram~J.}\ \bibnamefont {Kim}}, \bibinfo {author} {\bibfnamefont {Evgeny}\
  \bibnamefont {Kozik}}, \bibinfo {author} {\bibfnamefont {Evgeny~A.}\
  \bibnamefont {Stepanov}}, \bibinfo {author} {\bibfnamefont {Anna}\
  \bibnamefont {Kauch}}, \bibinfo {author} {\bibfnamefont {Sabine}\
  \bibnamefont {Andergassen}}, \bibinfo {author} {\bibfnamefont {Philipp}\
  \bibnamefont {Hansmann}}, \bibinfo {author} {\bibfnamefont {Daniel}\
  \bibnamefont {Rohe}}, \bibinfo {author} {\bibfnamefont {Yuri~M.}\
  \bibnamefont {Vilk}}, \bibinfo {author} {\bibfnamefont {James P.~F.}\
  \bibnamefont {LeBlanc}}, \bibinfo {author} {\bibfnamefont {Shiwei}\
  \bibnamefont {Zhang}}, \bibinfo {author} {\bibfnamefont {A.~M.~S.}\
  \bibnamefont {Tremblay}}, \bibinfo {author} {\bibfnamefont {Michel}\
  \bibnamefont {Ferrero}}, \bibinfo {author} {\bibfnamefont {Olivier}\
  \bibnamefont {Parcollet}}, \ and\ \bibinfo {author} {\bibfnamefont {Antoine}\
  \bibnamefont {Georges}},\ }\href@noop {} {\enquote {\bibinfo {title}
  {Tracking the footprints of spin fluctuations: A multi-method,
  multi-messenger study of the two-dimensional hubbard model},}\ } (\bibinfo
  {year} {2020}),\ \Eprint {http://arxiv.org/abs/2006.10769} {arXiv:2006.10769
  [cond-mat.str-el]} \BibitemShut {NoStop}%
\bibitem [{\citenamefont {Fratino}\ \emph {et~al.}(2017)\citenamefont
  {Fratino}, \citenamefont {S\'emon}, \citenamefont {Charlebois}, \citenamefont
  {Sordi},\ and\ \citenamefont {Tremblay}}]{LorenzoAF}%
  \BibitemOpen
  \bibfield  {author} {\bibinfo {author} {\bibfnamefont {L.}~\bibnamefont
  {Fratino}}, \bibinfo {author} {\bibfnamefont {P.}~\bibnamefont {S\'emon}},
  \bibinfo {author} {\bibfnamefont {M.}~\bibnamefont {Charlebois}}, \bibinfo
  {author} {\bibfnamefont {G.}~\bibnamefont {Sordi}}, \ and\ \bibinfo {author}
  {\bibfnamefont {A.-M.~S.}\ \bibnamefont {Tremblay}},\ }\bibfield  {title}
  {\enquote {\bibinfo {title} {{Signatures of the Mott transition in the
  antiferromagnetic state of the two-dimensional Hubbard model}},}\ }\href
  {\doibase 10.1103/PhysRevB.95.235109} {\bibfield  {journal} {\bibinfo
  {journal} {Phys. Rev. B}\ }\textbf {\bibinfo {volume} {95}},\ \bibinfo
  {pages} {235109} (\bibinfo {year} {2017})}\BibitemShut {NoStop}%
\bibitem [{\citenamefont {LeBlanc}\ \emph {et~al.}(2015)\citenamefont
  {LeBlanc}, \citenamefont {Antipov}, \citenamefont {Becca}, \citenamefont
  {Bulik}, \citenamefont {Chan}, \citenamefont {Chung}, \citenamefont {Deng},
  \citenamefont {Ferrero}, \citenamefont {Henderson}, \citenamefont
  {Jim\'enez-Hoyos}, \citenamefont {Kozik}, \citenamefont {Liu}, \citenamefont
  {Millis}, \citenamefont {Prokof'ev}, \citenamefont {Qin}, \citenamefont
  {Scuseria}, \citenamefont {Shi}, \citenamefont {Svistunov}, \citenamefont
  {Tocchio}, \citenamefont {Tupitsyn}, \citenamefont {White}, \citenamefont
  {Zhang}, \citenamefont {Zheng}, \citenamefont {Zhu},\ and\ \citenamefont
  {Gull}}]{LeBlancPRX}%
  \BibitemOpen
  \bibfield  {author} {\bibinfo {author} {\bibfnamefont {J.~P.~F.}\
  \bibnamefont {LeBlanc}}, \bibinfo {author} {\bibfnamefont {Andrey~E.}\
  \bibnamefont {Antipov}}, \bibinfo {author} {\bibfnamefont {Federico}\
  \bibnamefont {Becca}}, \bibinfo {author} {\bibfnamefont {Ireneusz~W.}\
  \bibnamefont {Bulik}}, \bibinfo {author} {\bibfnamefont {Garnet Kin-Lic}\
  \bibnamefont {Chan}}, \bibinfo {author} {\bibfnamefont {Chia-Min}\
  \bibnamefont {Chung}}, \bibinfo {author} {\bibfnamefont {Youjin}\
  \bibnamefont {Deng}}, \bibinfo {author} {\bibfnamefont {Michel}\ \bibnamefont
  {Ferrero}}, \bibinfo {author} {\bibfnamefont {Thomas~M.}\ \bibnamefont
  {Henderson}}, \bibinfo {author} {\bibfnamefont {Carlos~A.}\ \bibnamefont
  {Jim\'enez-Hoyos}}, \bibinfo {author} {\bibfnamefont {E.}~\bibnamefont
  {Kozik}}, \bibinfo {author} {\bibfnamefont {Xuan-Wen}\ \bibnamefont {Liu}},
  \bibinfo {author} {\bibfnamefont {Andrew~J.}\ \bibnamefont {Millis}},
  \bibinfo {author} {\bibfnamefont {N.~V.}\ \bibnamefont {Prokof'ev}}, \bibinfo
  {author} {\bibfnamefont {Mingpu}\ \bibnamefont {Qin}}, \bibinfo {author}
  {\bibfnamefont {Gustavo~E.}\ \bibnamefont {Scuseria}}, \bibinfo {author}
  {\bibfnamefont {Hao}\ \bibnamefont {Shi}}, \bibinfo {author} {\bibfnamefont
  {B.~V.}\ \bibnamefont {Svistunov}}, \bibinfo {author} {\bibfnamefont
  {Luca~F.}\ \bibnamefont {Tocchio}}, \bibinfo {author} {\bibfnamefont {I.~S.}\
  \bibnamefont {Tupitsyn}}, \bibinfo {author} {\bibfnamefont {Steven~R.}\
  \bibnamefont {White}}, \bibinfo {author} {\bibfnamefont {Shiwei}\
  \bibnamefont {Zhang}}, \bibinfo {author} {\bibfnamefont {Bo-Xiao}\
  \bibnamefont {Zheng}}, \bibinfo {author} {\bibfnamefont {Zhenyue}\
  \bibnamefont {Zhu}}, \ and\ \bibinfo {author} {\bibfnamefont {Emanuel}\
  \bibnamefont {Gull}} (\bibinfo {collaboration} {Simons Collaboration on the
  Many-Electron Problem}),\ }\bibfield  {title} {\enquote {\bibinfo {title}
  {{Solutions of the Two-Dimensional Hubbard Model: Benchmarks and Results from
  a Wide Range of Numerical Algorithms}},}\ }\href {\doibase
  10.1103/PhysRevX.5.041041} {\bibfield  {journal} {\bibinfo  {journal} {Phys.
  Rev. X}\ }\textbf {\bibinfo {volume} {5}},\ \bibinfo {pages} {041041}
  (\bibinfo {year} {2015})}\BibitemShut {NoStop}%
\bibitem [{\citenamefont {Cardy}\ and\ \citenamefont
  {Herzog}(2014)}]{Cardy:2017}%
  \BibitemOpen
  \bibfield  {author} {\bibinfo {author} {\bibfnamefont {John}\ \bibnamefont
  {Cardy}}\ and\ \bibinfo {author} {\bibfnamefont {Christopher~P.}\
  \bibnamefont {Herzog}},\ }\bibfield  {title} {\enquote {\bibinfo {title}
  {Universal thermal corrections to single interval entanglement entropy for
  two dimensional conformal field theories},}\ }\href {\doibase
  10.1103/PhysRevLett.112.171603} {\bibfield  {journal} {\bibinfo  {journal}
  {Phys. Rev. Lett.}\ }\textbf {\bibinfo {volume} {112}},\ \bibinfo {pages}
  {171603} (\bibinfo {year} {2014})}\BibitemShut {NoStop}%
\bibitem [{\citenamefont {Vedral}(2004)}]{Vedral:T2004}%
  \BibitemOpen
  \bibfield  {author} {\bibinfo {author} {\bibfnamefont {Vlatko}\ \bibnamefont
  {Vedral}},\ }\bibfield  {title} {\enquote {\bibinfo {title} {High-temperature
  macroscopic entanglement},}\ }\href
  {http://stacks.iop.org/1367-2630/6/i=1/a=102} {\bibfield  {journal} {\bibinfo
   {journal} {New Journal of Physics}\ }\textbf {\bibinfo {volume} {6}},\
  \bibinfo {pages} {102} (\bibinfo {year} {2004})}\BibitemShut {NoStop}%
\bibitem [{\citenamefont {Anders}\ and\ \citenamefont
  {Vedral}(2007)}]{Anders:2007}%
  \BibitemOpen
  \bibfield  {author} {\bibinfo {author} {\bibfnamefont {Janet}\ \bibnamefont
  {Anders}}\ and\ \bibinfo {author} {\bibfnamefont {Vlatko}\ \bibnamefont
  {Vedral}},\ }\bibfield  {title} {\enquote {\bibinfo {title} {Macroscopic
  entanglement and phase transitions},}\ }\href {\doibase
  10.1007/s11080-007-9034-6} {\bibfield  {journal} {\bibinfo  {journal} {Open
  Systems \& Information Dynamics}\ }\textbf {\bibinfo {volume} {14}},\
  \bibinfo {pages} {1--16} (\bibinfo {year} {2007})}\BibitemShut {NoStop}%
\bibitem [{\citenamefont {Zanardi}(2002)}]{zanardi2002}%
  \BibitemOpen
  \bibfield  {author} {\bibinfo {author} {\bibfnamefont {Paolo}\ \bibnamefont
  {Zanardi}},\ }\bibfield  {title} {\enquote {\bibinfo {title} {Quantum
  entanglement in fermionic lattices},}\ }\href {\doibase
  10.1103/PhysRevA.65.042101} {\bibfield  {journal} {\bibinfo  {journal} {Phys.
  Rev. A}\ }\textbf {\bibinfo {volume} {65}},\ \bibinfo {pages} {042101}
  (\bibinfo {year} {2002})}\BibitemShut {NoStop}%
\bibitem [{\citenamefont {Gu}\ \emph {et~al.}(2004)\citenamefont {Gu},
  \citenamefont {Deng}, \citenamefont {Li},\ and\ \citenamefont
  {Lin}}]{Gu:2004}%
  \BibitemOpen
  \bibfield  {author} {\bibinfo {author} {\bibfnamefont {Shi-Jian}\
  \bibnamefont {Gu}}, \bibinfo {author} {\bibfnamefont {Shu-Sa}\ \bibnamefont
  {Deng}}, \bibinfo {author} {\bibfnamefont {You-Quan}\ \bibnamefont {Li}}, \
  and\ \bibinfo {author} {\bibfnamefont {Hai-Qing}\ \bibnamefont {Lin}},\
  }\bibfield  {title} {\enquote {\bibinfo {title} {{Entanglement and Quantum
  Phase Transition in the Extended Hubbard Model}},}\ }\href {\doibase
  10.1103/PhysRevLett.93.086402} {\bibfield  {journal} {\bibinfo  {journal}
  {Phys. Rev. Lett.}\ }\textbf {\bibinfo {volume} {93}},\ \bibinfo {pages}
  {086402} (\bibinfo {year} {2004})}\BibitemShut {NoStop}%
\bibitem [{\citenamefont {Larsson}\ and\ \citenamefont
  {Johannesson}(2005{\natexlab{a}})}]{larssonPRL:2005}%
  \BibitemOpen
  \bibfield  {author} {\bibinfo {author} {\bibfnamefont {Daniel}\ \bibnamefont
  {Larsson}}\ and\ \bibinfo {author} {\bibfnamefont {Henrik}\ \bibnamefont
  {Johannesson}},\ }\bibfield  {title} {\enquote {\bibinfo {title}
  {{Entanglement Scaling in the One-Dimensional Hubbard Model at
  Criticality}},}\ }\href {\doibase 10.1103/PhysRevLett.95.196406} {\bibfield
  {journal} {\bibinfo  {journal} {Phys. Rev. Lett.}\ }\textbf {\bibinfo
  {volume} {95}},\ \bibinfo {pages} {196406} (\bibinfo {year}
  {2005}{\natexlab{a}})}\BibitemShut {NoStop}%
\bibitem [{\citenamefont {Larsson}\ and\ \citenamefont
  {Johannesson}(2006)}]{larssonPRA2006}%
  \BibitemOpen
  \bibfield  {author} {\bibinfo {author} {\bibfnamefont {Daniel}\ \bibnamefont
  {Larsson}}\ and\ \bibinfo {author} {\bibfnamefont {Henrik}\ \bibnamefont
  {Johannesson}},\ }\bibfield  {title} {\enquote {\bibinfo {title}
  {{Single-site entanglement of fermions at a quantum phase transition}},}\
  }\href {\doibase 10.1103/PhysRevA.73.042320} {\bibfield  {journal} {\bibinfo
  {journal} {Phys. Rev. A}\ }\textbf {\bibinfo {volume} {73}},\ \bibinfo
  {pages} {042320} (\bibinfo {year} {2006})}\BibitemShut {NoStop}%
\bibitem [{\citenamefont {Anfossi}\ \emph {et~al.}(2005)\citenamefont
  {Anfossi}, \citenamefont {Giorda}, \citenamefont {Montorsi},\ and\
  \citenamefont {Traversa}}]{Anfossi_Giorda_Montorsi_Traversa_2005}%
  \BibitemOpen
  \bibfield  {author} {\bibinfo {author} {\bibfnamefont {Alberto}\ \bibnamefont
  {Anfossi}}, \bibinfo {author} {\bibfnamefont {Paolo}\ \bibnamefont {Giorda}},
  \bibinfo {author} {\bibfnamefont {Arianna}\ \bibnamefont {Montorsi}}, \ and\
  \bibinfo {author} {\bibfnamefont {Fabio}\ \bibnamefont {Traversa}},\
  }\bibfield  {title} {\enquote {\bibinfo {title} {{Two-Point Versus
  Multipartite Entanglement in Quantum Phase Transitions}},}\ }\href {\doibase
  10.1103/PhysRevLett.95.056402} {\bibfield  {journal} {\bibinfo  {journal}
  {Phys. Rev. Lett.}\ }\textbf {\bibinfo {volume} {95}},\ \bibinfo {pages}
  {056402} (\bibinfo {year} {2005})}\BibitemShut {NoStop}%
\bibitem [{\citenamefont {Anfossi}\ \emph {et~al.}(2007)\citenamefont
  {Anfossi}, \citenamefont {Giorda},\ and\ \citenamefont
  {Montorsi}}]{Anfossi:2007}%
  \BibitemOpen
  \bibfield  {author} {\bibinfo {author} {\bibfnamefont {Alberto}\ \bibnamefont
  {Anfossi}}, \bibinfo {author} {\bibfnamefont {Paolo}\ \bibnamefont {Giorda}},
  \ and\ \bibinfo {author} {\bibfnamefont {Arianna}\ \bibnamefont {Montorsi}},\
  }\bibfield  {title} {\enquote {\bibinfo {title} {{Entanglement in extended
  Hubbard models and quantum phase transitions}},}\ }\href {\doibase
  10.1103/PhysRevB.75.165106} {\bibfield  {journal} {\bibinfo  {journal} {Phys.
  Rev. B}\ }\textbf {\bibinfo {volume} {75}},\ \bibinfo {pages} {165106}
  (\bibinfo {year} {2007})}\BibitemShut {NoStop}%
\bibitem [{\citenamefont {Byczuk}\ \emph {et~al.}(2012)\citenamefont {Byczuk},
  \citenamefont {Kune\ifmmode~\check{s}\else \v{s}\fi{}}, \citenamefont
  {Hofstetter},\ and\ \citenamefont {Vollhardt}}]{byczukPRL2012}%
  \BibitemOpen
  \bibfield  {author} {\bibinfo {author} {\bibfnamefont {Krzysztof}\
  \bibnamefont {Byczuk}}, \bibinfo {author} {\bibfnamefont {Jan}\ \bibnamefont
  {Kune\ifmmode~\check{s}\else \v{s}\fi{}}}, \bibinfo {author} {\bibfnamefont
  {Walter}\ \bibnamefont {Hofstetter}}, \ and\ \bibinfo {author} {\bibfnamefont
  {Dieter}\ \bibnamefont {Vollhardt}},\ }\bibfield  {title} {\enquote {\bibinfo
  {title} {Quantification of correlations in quantum many-particle systems},}\
  }\href {\doibase 10.1103/PhysRevLett.108.087004} {\bibfield  {journal}
  {\bibinfo  {journal} {Phys. Rev. Lett.}\ }\textbf {\bibinfo {volume} {108}},\
  \bibinfo {pages} {087004} (\bibinfo {year} {2012})}\BibitemShut {NoStop}%
\bibitem [{\citenamefont {Lanat\`a}\ \emph {et~al.}(2014)\citenamefont
  {Lanat\`a}, \citenamefont {Strand}, \citenamefont {Yao},\ and\ \citenamefont
  {Kotliar}}]{lanataEnt}%
  \BibitemOpen
  \bibfield  {author} {\bibinfo {author} {\bibfnamefont {Nicola}\ \bibnamefont
  {Lanat\`a}}, \bibinfo {author} {\bibfnamefont {Hugo U.~R.}\ \bibnamefont
  {Strand}}, \bibinfo {author} {\bibfnamefont {Yongxin}\ \bibnamefont {Yao}}, \
  and\ \bibinfo {author} {\bibfnamefont {Gabriel}\ \bibnamefont {Kotliar}},\
  }\bibfield  {title} {\enquote {\bibinfo {title} {Principle of maximum
  entanglement entropy and local physics of strongly correlated materials},}\
  }\href {\doibase 10.1103/PhysRevLett.113.036402} {\bibfield  {journal}
  {\bibinfo  {journal} {Phys. Rev. Lett.}\ }\textbf {\bibinfo {volume} {113}},\
  \bibinfo {pages} {036402} (\bibinfo {year} {2014})}\BibitemShut {NoStop}%
\bibitem [{\citenamefont {Udagawa}\ and\ \citenamefont
  {Motome}(2015)}]{Udagawa_Motome:2015}%
  \BibitemOpen
  \bibfield  {author} {\bibinfo {author} {\bibfnamefont {Masafumi}\
  \bibnamefont {Udagawa}}\ and\ \bibinfo {author} {\bibfnamefont {Yukitoshi}\
  \bibnamefont {Motome}},\ }\bibfield  {title} {\enquote {\bibinfo {title}
  {Entanglement spectrum in cluster dynamical mean-field theory},}\ }\href
  {http://stacks.iop.org/1742-5468/2015/i=1/a=P01016} {\bibfield  {journal}
  {\bibinfo  {journal} {Journal of Statistical Mechanics: Theory and
  Experiment}\ }\textbf {\bibinfo {volume} {2015}},\ \bibinfo {pages} {P01016}
  (\bibinfo {year} {2015})}\BibitemShut {NoStop}%
\bibitem [{\citenamefont {{Bakr}}\ \emph {et~al.}(2009)\citenamefont {{Bakr}},
  \citenamefont {{Gillen}}, \citenamefont {{Peng}}, \citenamefont
  {{F{\"o}lling}},\ and\ \citenamefont {{Greiner}}}]{Bakr:2009}%
  \BibitemOpen
  \bibfield  {author} {\bibinfo {author} {\bibfnamefont {Waseem~S.}\
  \bibnamefont {{Bakr}}}, \bibinfo {author} {\bibfnamefont {Jonathon~I.}\
  \bibnamefont {{Gillen}}}, \bibinfo {author} {\bibfnamefont {Amy}\
  \bibnamefont {{Peng}}}, \bibinfo {author} {\bibfnamefont {Simon}\
  \bibnamefont {{F{\"o}lling}}}, \ and\ \bibinfo {author} {\bibfnamefont
  {Markus}\ \bibnamefont {{Greiner}}},\ }\bibfield  {title} {\enquote {\bibinfo
  {title} {{A quantum gas microscope for detecting single atoms in a
  Hubbard-regime optical lattice}},}\ }\href {\doibase 10.1038/nature08482}
  {\bibfield  {journal} {\bibinfo  {journal} {\nat}\ }\textbf {\bibinfo
  {volume} {462}},\ \bibinfo {pages} {74--77} (\bibinfo {year}
  {2009})}\BibitemShut {NoStop}%
\bibitem [{\citenamefont {{Sherson}}\ \emph {et~al.}(2010)\citenamefont
  {{Sherson}}, \citenamefont {{Weitenberg}}, \citenamefont {{Endres}},
  \citenamefont {{Cheneau}}, \citenamefont {{Bloch}},\ and\ \citenamefont
  {{Kuhr}}}]{Sherson:2010}%
  \BibitemOpen
  \bibfield  {author} {\bibinfo {author} {\bibfnamefont {Jacob~F.}\
  \bibnamefont {{Sherson}}}, \bibinfo {author} {\bibfnamefont {Christof}\
  \bibnamefont {{Weitenberg}}}, \bibinfo {author} {\bibfnamefont {Manuel}\
  \bibnamefont {{Endres}}}, \bibinfo {author} {\bibfnamefont {Marc}\
  \bibnamefont {{Cheneau}}}, \bibinfo {author} {\bibfnamefont {Immanuel}\
  \bibnamefont {{Bloch}}}, \ and\ \bibinfo {author} {\bibfnamefont {Stefan}\
  \bibnamefont {{Kuhr}}},\ }\bibfield  {title} {\enquote {\bibinfo {title}
  {{Single-atom-resolved fluorescence imaging of an atomic Mott insulator}},}\
  }\href {\doibase 10.1038/nature09378} {\bibfield  {journal} {\bibinfo
  {journal} {\nat}\ }\textbf {\bibinfo {volume} {467}},\ \bibinfo {pages}
  {68--72} (\bibinfo {year} {2010})}\BibitemShut {NoStop}%
\bibitem [{\citenamefont {Cheuk}\ \emph {et~al.}(2015)\citenamefont {Cheuk},
  \citenamefont {Nichols}, \citenamefont {Okan}, \citenamefont {Gersdorf},
  \citenamefont {Ramasesh}, \citenamefont {Bakr}, \citenamefont {Lompe},\ and\
  \citenamefont {Zwierlein}}]{Cheuk:PRL2015}%
  \BibitemOpen
  \bibfield  {author} {\bibinfo {author} {\bibfnamefont {Lawrence~W.}\
  \bibnamefont {Cheuk}}, \bibinfo {author} {\bibfnamefont {Matthew~A.}\
  \bibnamefont {Nichols}}, \bibinfo {author} {\bibfnamefont {Melih}\
  \bibnamefont {Okan}}, \bibinfo {author} {\bibfnamefont {Thomas}\ \bibnamefont
  {Gersdorf}}, \bibinfo {author} {\bibfnamefont {Vinay~V.}\ \bibnamefont
  {Ramasesh}}, \bibinfo {author} {\bibfnamefont {Waseem~S.}\ \bibnamefont
  {Bakr}}, \bibinfo {author} {\bibfnamefont {Thomas}\ \bibnamefont {Lompe}}, \
  and\ \bibinfo {author} {\bibfnamefont {Martin~W.}\ \bibnamefont
  {Zwierlein}},\ }\bibfield  {title} {\enquote {\bibinfo {title} {Quantum-gas
  microscope for fermionic atoms},}\ }\href {\doibase
  10.1103/PhysRevLett.114.193001} {\bibfield  {journal} {\bibinfo  {journal}
  {Phys. Rev. Lett.}\ }\textbf {\bibinfo {volume} {114}},\ \bibinfo {pages}
  {193001} (\bibinfo {year} {2015})}\BibitemShut {NoStop}%
\bibitem [{\citenamefont {{Haller}}\ \emph {et~al.}(2015)\citenamefont
  {{Haller}}, \citenamefont {{Hudson}}, \citenamefont {{Kelly}}, \citenamefont
  {{Cotta}}, \citenamefont {{Peaudecerf}}, \citenamefont {{Bruce}},\ and\
  \citenamefont {{Kuhr}}}]{Haller:NatPhys2015}%
  \BibitemOpen
  \bibfield  {author} {\bibinfo {author} {\bibfnamefont {Elmar}\ \bibnamefont
  {{Haller}}}, \bibinfo {author} {\bibfnamefont {James}\ \bibnamefont
  {{Hudson}}}, \bibinfo {author} {\bibfnamefont {Andrew}\ \bibnamefont
  {{Kelly}}}, \bibinfo {author} {\bibfnamefont {Dylan~A.}\ \bibnamefont
  {{Cotta}}}, \bibinfo {author} {\bibfnamefont {Bruno}\ \bibnamefont
  {{Peaudecerf}}}, \bibinfo {author} {\bibfnamefont {Graham~D.}\ \bibnamefont
  {{Bruce}}}, \ and\ \bibinfo {author} {\bibfnamefont {Stefan}\ \bibnamefont
  {{Kuhr}}},\ }\bibfield  {title} {\enquote {\bibinfo {title} {{Single-atom
  imaging of fermions in a quantum-gas microscope}},}\ }\href {\doibase
  10.1038/nphys3403} {\bibfield  {journal} {\bibinfo  {journal} {Nature
  Physics}\ }\textbf {\bibinfo {volume} {11}},\ \bibinfo {pages} {738--742}
  (\bibinfo {year} {2015})}\BibitemShut {NoStop}%
\bibitem [{\citenamefont {Parsons}\ \emph {et~al.}(2015)\citenamefont
  {Parsons}, \citenamefont {Huber}, \citenamefont {Mazurenko}, \citenamefont
  {Chiu}, \citenamefont {Setiawan}, \citenamefont {Wooley-Brown}, \citenamefont
  {Blatt},\ and\ \citenamefont {Greiner}}]{Parsons:PRL2015}%
  \BibitemOpen
  \bibfield  {author} {\bibinfo {author} {\bibfnamefont {Maxwell~F.}\
  \bibnamefont {Parsons}}, \bibinfo {author} {\bibfnamefont {Florian}\
  \bibnamefont {Huber}}, \bibinfo {author} {\bibfnamefont {Anton}\ \bibnamefont
  {Mazurenko}}, \bibinfo {author} {\bibfnamefont {Christie~S.}\ \bibnamefont
  {Chiu}}, \bibinfo {author} {\bibfnamefont {Widagdo}\ \bibnamefont
  {Setiawan}}, \bibinfo {author} {\bibfnamefont {Katherine}\ \bibnamefont
  {Wooley-Brown}}, \bibinfo {author} {\bibfnamefont {Sebastian}\ \bibnamefont
  {Blatt}}, \ and\ \bibinfo {author} {\bibfnamefont {Markus}\ \bibnamefont
  {Greiner}},\ }\bibfield  {title} {\enquote {\bibinfo {title} {Site-resolved
  imaging of fermionic $^{6}\mathrm{Li}$ in an optical lattice},}\ }\href
  {\doibase 10.1103/PhysRevLett.114.213002} {\bibfield  {journal} {\bibinfo
  {journal} {Phys. Rev. Lett.}\ }\textbf {\bibinfo {volume} {114}},\ \bibinfo
  {pages} {213002} (\bibinfo {year} {2015})}\BibitemShut {NoStop}%
\bibitem [{\citenamefont {Omran}\ \emph {et~al.}(2015)\citenamefont {Omran},
  \citenamefont {Boll}, \citenamefont {Hilker}, \citenamefont {Kleinlein},
  \citenamefont {Salomon}, \citenamefont {Bloch},\ and\ \citenamefont
  {Gross}}]{Omran:PRL2015}%
  \BibitemOpen
  \bibfield  {author} {\bibinfo {author} {\bibfnamefont {Ahmed}\ \bibnamefont
  {Omran}}, \bibinfo {author} {\bibfnamefont {Martin}\ \bibnamefont {Boll}},
  \bibinfo {author} {\bibfnamefont {Timon~A.}\ \bibnamefont {Hilker}}, \bibinfo
  {author} {\bibfnamefont {Katharina}\ \bibnamefont {Kleinlein}}, \bibinfo
  {author} {\bibfnamefont {Guillaume}\ \bibnamefont {Salomon}}, \bibinfo
  {author} {\bibfnamefont {Immanuel}\ \bibnamefont {Bloch}}, \ and\ \bibinfo
  {author} {\bibfnamefont {Christian}\ \bibnamefont {Gross}},\ }\bibfield
  {title} {\enquote {\bibinfo {title} {Microscopic observation of pauli
  blocking in degenerate fermionic lattice gases},}\ }\href {\doibase
  10.1103/PhysRevLett.115.263001} {\bibfield  {journal} {\bibinfo  {journal}
  {Phys. Rev. Lett.}\ }\textbf {\bibinfo {volume} {115}},\ \bibinfo {pages}
  {263001} (\bibinfo {year} {2015})}\BibitemShut {NoStop}%
\bibitem [{\citenamefont {Edge}\ \emph {et~al.}(2015)\citenamefont {Edge},
  \citenamefont {Anderson}, \citenamefont {Jervis}, \citenamefont {McKay},
  \citenamefont {Day}, \citenamefont {Trotzky},\ and\ \citenamefont
  {Thywissen}}]{Edge:PRA2015}%
  \BibitemOpen
  \bibfield  {author} {\bibinfo {author} {\bibfnamefont {G.~J.~A.}\
  \bibnamefont {Edge}}, \bibinfo {author} {\bibfnamefont {R.}~\bibnamefont
  {Anderson}}, \bibinfo {author} {\bibfnamefont {D.}~\bibnamefont {Jervis}},
  \bibinfo {author} {\bibfnamefont {D.~C.}\ \bibnamefont {McKay}}, \bibinfo
  {author} {\bibfnamefont {R.}~\bibnamefont {Day}}, \bibinfo {author}
  {\bibfnamefont {S.}~\bibnamefont {Trotzky}}, \ and\ \bibinfo {author}
  {\bibfnamefont {J.~H.}\ \bibnamefont {Thywissen}},\ }\bibfield  {title}
  {\enquote {\bibinfo {title} {Imaging and addressing of individual fermionic
  atoms in an optical lattice},}\ }\href {\doibase 10.1103/PhysRevA.92.063406}
  {\bibfield  {journal} {\bibinfo  {journal} {Phys. Rev. A}\ }\textbf {\bibinfo
  {volume} {92}},\ \bibinfo {pages} {063406} (\bibinfo {year}
  {2015})}\BibitemShut {NoStop}%
\bibitem [{\citenamefont {Gross}\ and\ \citenamefont
  {Bloch}(2017)}]{GrossScience2017}%
  \BibitemOpen
  \bibfield  {author} {\bibinfo {author} {\bibfnamefont {Christian}\
  \bibnamefont {Gross}}\ and\ \bibinfo {author} {\bibfnamefont {Immanuel}\
  \bibnamefont {Bloch}},\ }\bibfield  {title} {\enquote {\bibinfo {title}
  {Quantum simulations with ultracold atoms in optical lattices},}\ }\href
  {\doibase 10.1126/science.aal3837} {\bibfield  {journal} {\bibinfo  {journal}
  {Science}\ }\textbf {\bibinfo {volume} {357}},\ \bibinfo {pages} {995--1001}
  (\bibinfo {year} {2017})}\BibitemShut {NoStop}%
\bibitem [{\citenamefont {Hartke}\ \emph {et~al.}(2020)\citenamefont {Hartke},
  \citenamefont {Oreg}, \citenamefont {Jia},\ and\ \citenamefont
  {Zwierlein}}]{Hartke:2020}%
  \BibitemOpen
  \bibfield  {author} {\bibinfo {author} {\bibfnamefont {Thomas}\ \bibnamefont
  {Hartke}}, \bibinfo {author} {\bibfnamefont {Botond}\ \bibnamefont {Oreg}},
  \bibinfo {author} {\bibfnamefont {Ningyuan}\ \bibnamefont {Jia}}, \ and\
  \bibinfo {author} {\bibfnamefont {Martin}\ \bibnamefont {Zwierlein}},\
  }\href@noop {} {\enquote {\bibinfo {title} {Measuring total density
  correlations in a fermi-hubbard gas via bilayer microscopy},}\ } (\bibinfo
  {year} {2020}),\ \Eprint {http://arxiv.org/abs/2003.11669} {arXiv:2003.11669
  [cond-mat.quant-gas]} \BibitemShut {NoStop}%
\bibitem [{\citenamefont {Sordi}\ \emph {et~al.}(2019)\citenamefont {Sordi},
  \citenamefont {Walsh}, \citenamefont {S\'emon},\ and\ \citenamefont
  {Tremblay}}]{Giovanni:PRBcv}%
  \BibitemOpen
  \bibfield  {author} {\bibinfo {author} {\bibfnamefont {G.}~\bibnamefont
  {Sordi}}, \bibinfo {author} {\bibfnamefont {C.}~\bibnamefont {Walsh}},
  \bibinfo {author} {\bibfnamefont {P.}~\bibnamefont {S\'emon}}, \ and\
  \bibinfo {author} {\bibfnamefont {A.-M.~S.}\ \bibnamefont {Tremblay}},\
  }\bibfield  {title} {\enquote {\bibinfo {title} {Specific heat maximum as a
  signature of mott physics in the two-dimensional hubbard model},}\ }\href
  {\doibase 10.1103/PhysRevB.100.121105} {\bibfield  {journal} {\bibinfo
  {journal} {Phys. Rev. B}\ }\textbf {\bibinfo {volume} {100}},\ \bibinfo
  {pages} {121105} (\bibinfo {year} {2019})}\BibitemShut {NoStop}%
\bibitem [{\citenamefont {Xu}\ \emph {et~al.}(2005)\citenamefont {Xu},
  \citenamefont {Kumar}, \citenamefont {Buldyrev}, \citenamefont {Chen},
  \citenamefont {Poole}, \citenamefont {Sciortino},\ and\ \citenamefont
  {Stanley}}]{water1}%
  \BibitemOpen
  \bibfield  {author} {\bibinfo {author} {\bibfnamefont {Limei}\ \bibnamefont
  {Xu}}, \bibinfo {author} {\bibfnamefont {Pradeep}\ \bibnamefont {Kumar}},
  \bibinfo {author} {\bibfnamefont {S.~V.}\ \bibnamefont {Buldyrev}}, \bibinfo
  {author} {\bibfnamefont {S.-H.}\ \bibnamefont {Chen}}, \bibinfo {author}
  {\bibfnamefont {P.~H.}\ \bibnamefont {Poole}}, \bibinfo {author}
  {\bibfnamefont {F.}~\bibnamefont {Sciortino}}, \ and\ \bibinfo {author}
  {\bibfnamefont {H.~E.}\ \bibnamefont {Stanley}},\ }\bibfield  {title}
  {\enquote {\bibinfo {title} {{Relation between the Widom line and the dynamic
  crossover in systems with a liquid liquid phase transition}},}\ }\href
  {\doibase 10.1073/pnas.0507870102} {\bibfield  {journal} {\bibinfo  {journal}
  {Proc. Natl. Acad. Sci. USA}\ }\textbf {\bibinfo {volume} {102}},\ \bibinfo
  {pages} {16558--16562} (\bibinfo {year} {2005})}\BibitemShut {NoStop}%
\bibitem [{\citenamefont {McMillan}\ and\ \citenamefont
  {Stanley}(2010)}]{supercritical}%
  \BibitemOpen
  \bibfield  {author} {\bibinfo {author} {\bibfnamefont {Paul~F.}\ \bibnamefont
  {McMillan}}\ and\ \bibinfo {author} {\bibfnamefont {H.~Eugene}\ \bibnamefont
  {Stanley}},\ }\bibfield  {title} {\enquote {\bibinfo {title} {Fluid phases:
  Going supercritical},}\ }\href {\doibase doi:10.1038/nphys1711} {\bibfield
  {journal} {\bibinfo  {journal} {Nat Phys}\ }\textbf {\bibinfo {volume} {6}},\
  \bibinfo {pages} {479--480} (\bibinfo {year} {2010})}\BibitemShut {NoStop}%
\bibitem [{\citenamefont {Walsh}\ \emph
  {et~al.}(2019{\natexlab{c}})\citenamefont {Walsh}, \citenamefont {S\'emon},
  \citenamefont {Sordi},\ and\ \citenamefont {Tremblay}}]{CaitlinOpalescence}%
  \BibitemOpen
  \bibfield  {author} {\bibinfo {author} {\bibfnamefont {C.}~\bibnamefont
  {Walsh}}, \bibinfo {author} {\bibfnamefont {P.}~\bibnamefont {S\'emon}},
  \bibinfo {author} {\bibfnamefont {G.}~\bibnamefont {Sordi}}, \ and\ \bibinfo
  {author} {\bibfnamefont {A.-M.~S.}\ \bibnamefont {Tremblay}},\ }\bibfield
  {title} {\enquote {\bibinfo {title} {Critical opalescence across the
  doping-driven mott transition in optical lattices of ultracold atoms},}\
  }\href {\doibase 10.1103/PhysRevB.99.165151} {\bibfield  {journal} {\bibinfo
  {journal} {Phys. Rev. B}\ }\textbf {\bibinfo {volume} {99}},\ \bibinfo
  {pages} {165151} (\bibinfo {year} {2019}{\natexlab{c}})}\BibitemShut
  {NoStop}%
\bibitem [{\citenamefont {Alloul}(2014)}]{Alloul2013}%
  \BibitemOpen
  \bibfield  {author} {\bibinfo {author} {\bibfnamefont {Henri}\ \bibnamefont
  {Alloul}},\ }\bibfield  {title} {\enquote {\bibinfo {title} {What is the
  simplest model that captures the basic experimental facts of the physics of
  underdoped cuprates?}}\ }\href {\doibase 10.1016/j.crhy.2014.02.007}
  {\bibfield  {journal} {\bibinfo  {journal} {Comptes Rendus Physique}\
  }\textbf {\bibinfo {volume} {15}},\ \bibinfo {pages} {519 -- 524} (\bibinfo
  {year} {2014})}\BibitemShut {NoStop}%
\bibitem [{\citenamefont {Haule}\ and\ \citenamefont
  {Kotliar}(2007)}]{hauleDOPING}%
  \BibitemOpen
  \bibfield  {author} {\bibinfo {author} {\bibfnamefont {Kristjan}\
  \bibnamefont {Haule}}\ and\ \bibinfo {author} {\bibfnamefont {Gabriel}\
  \bibnamefont {Kotliar}},\ }\bibfield  {title} {\enquote {\bibinfo {title}
  {Strongly correlated superconductivity: A plaquette dynamical mean-field
  theory study},}\ }\href {\doibase 10.1103/PhysRevB.76.104509} {\bibfield
  {journal} {\bibinfo  {journal} {Phys. Rev. B}\ }\textbf {\bibinfo {volume}
  {76}},\ \bibinfo {eid} {104509} (\bibinfo {year} {2007})}\BibitemShut
  {NoStop}%
\bibitem [{\citenamefont {Mikelsons}\ \emph {et~al.}(2009)\citenamefont
  {Mikelsons}, \citenamefont {Khatami}, \citenamefont {Galanakis},
  \citenamefont {Macridin}, \citenamefont {Moreno},\ and\ \citenamefont
  {Jarrell}}]{MikelsonThermodynamics:2009}%
  \BibitemOpen
  \bibfield  {author} {\bibinfo {author} {\bibfnamefont {K.}~\bibnamefont
  {Mikelsons}}, \bibinfo {author} {\bibfnamefont {E.}~\bibnamefont {Khatami}},
  \bibinfo {author} {\bibfnamefont {D.}~\bibnamefont {Galanakis}}, \bibinfo
  {author} {\bibfnamefont {A.}~\bibnamefont {Macridin}}, \bibinfo {author}
  {\bibfnamefont {J.}~\bibnamefont {Moreno}}, \ and\ \bibinfo {author}
  {\bibfnamefont {M.}~\bibnamefont {Jarrell}},\ }\bibfield  {title} {\enquote
  {\bibinfo {title} {Thermodynamics of the quantum critical point at finite
  doping in the two-dimensional hubbard model studied via the dynamical cluster
  approximation},}\ }\href {\doibase 10.1103/PhysRevB.80.140505} {\bibfield
  {journal} {\bibinfo  {journal} {Phys. Rev. B}\ }\textbf {\bibinfo {volume}
  {80}},\ \bibinfo {pages} {140505} (\bibinfo {year} {2009})}\BibitemShut
  {NoStop}%
\bibitem [{\citenamefont {Knaute}\ and\ \citenamefont
  {K\"ampfer}(2017)}]{eeQCD2017}%
  \BibitemOpen
  \bibfield  {author} {\bibinfo {author} {\bibfnamefont {J.}~\bibnamefont
  {Knaute}}\ and\ \bibinfo {author} {\bibfnamefont {B.}~\bibnamefont
  {K\"ampfer}},\ }\bibfield  {title} {\enquote {\bibinfo {title} {Holographic
  entanglement entropy in the qcd phase diagram with a critical point},}\
  }\href {\doibase 10.1103/PhysRevD.96.106003} {\bibfield  {journal} {\bibinfo
  {journal} {Phys. Rev. D}\ }\textbf {\bibinfo {volume} {96}},\ \bibinfo
  {pages} {106003} (\bibinfo {year} {2017})}\BibitemShut {NoStop}%
\bibitem [{\citenamefont {Larsson}\ and\ \citenamefont
  {Johannesson}(2005{\natexlab{b}})}]{LarssonScalingHubbard:2005}%
  \BibitemOpen
  \bibfield  {author} {\bibinfo {author} {\bibfnamefont {Daniel}\ \bibnamefont
  {Larsson}}\ and\ \bibinfo {author} {\bibfnamefont {Henrik}\ \bibnamefont
  {Johannesson}},\ }\bibfield  {title} {\enquote {\bibinfo {title}
  {{Entanglement Scaling in the One-Dimensional Hubbard Model at
  Criticality}},}\ }\href {\doibase 10.1103/PhysRevLett.95.196406} {\bibfield
  {journal} {\bibinfo  {journal} {Phys. Rev. Lett.}\ }\textbf {\bibinfo
  {volume} {95}},\ \bibinfo {pages} {196406} (\bibinfo {year}
  {2005}{\natexlab{b}})}\BibitemShut {NoStop}%
\bibitem [{\citenamefont {Lu}\ and\ \citenamefont
  {Grover}(2019{\natexlab{a}})}]{Grover:PRB2019}%
  \BibitemOpen
  \bibfield  {author} {\bibinfo {author} {\bibfnamefont {Tsung-Cheng}\
  \bibnamefont {Lu}}\ and\ \bibinfo {author} {\bibfnamefont {Tarun}\
  \bibnamefont {Grover}},\ }\bibfield  {title} {\enquote {\bibinfo {title}
  {Singularity in entanglement negativity across finite-temperature phase
  transitions},}\ }\href {\doibase 10.1103/PhysRevB.99.075157} {\bibfield
  {journal} {\bibinfo  {journal} {Phys. Rev. B}\ }\textbf {\bibinfo {volume}
  {99}},\ \bibinfo {pages} {075157} (\bibinfo {year}
  {2019}{\natexlab{a}})}\BibitemShut {NoStop}%
\bibitem [{\citenamefont {Lu}\ and\ \citenamefont
  {Grover}(2019{\natexlab{b}})}]{Grover:arXiv2019}%
  \BibitemOpen
  \bibfield  {author} {\bibinfo {author} {\bibfnamefont {Tsung-Cheng}\
  \bibnamefont {Lu}}\ and\ \bibinfo {author} {\bibfnamefont {Tarun}\
  \bibnamefont {Grover}},\ }\href@noop {} {\enquote {\bibinfo {title}
  {Structure of quantum entanglement at a finite temperature critical point},}\
  } (\bibinfo {year} {2019}{\natexlab{b}}),\ \Eprint
  {http://arxiv.org/abs/1907.01569} {arXiv:1907.01569 [cond-mat.str-el]}
  \BibitemShut {NoStop}%
\bibitem [{\citenamefont {Yeo}\ and\ \citenamefont
  {Phillips}(2019)}]{Phillips:PRD2019}%
  \BibitemOpen
  \bibfield  {author} {\bibinfo {author} {\bibfnamefont {Luke}\ \bibnamefont
  {Yeo}}\ and\ \bibinfo {author} {\bibfnamefont {Philip~W.}\ \bibnamefont
  {Phillips}},\ }\bibfield  {title} {\enquote {\bibinfo {title} {Local
  entropies across the mott transition in an exactly solvable model},}\ }\href
  {\doibase 10.1103/PhysRevD.99.094030} {\bibfield  {journal} {\bibinfo
  {journal} {Phys. Rev. D}\ }\textbf {\bibinfo {volume} {99}},\ \bibinfo
  {pages} {094030} (\bibinfo {year} {2019})}\BibitemShut {NoStop}%
\bibitem [{\citenamefont {Keimer}\ \emph {et~al.}(2015)\citenamefont {Keimer},
  \citenamefont {Kivelson}, \citenamefont {Norman}, \citenamefont {Uchida},\
  and\ \citenamefont {Zaanen}}]{keimerRev}%
  \BibitemOpen
  \bibfield  {author} {\bibinfo {author} {\bibfnamefont {B.}~\bibnamefont
  {Keimer}}, \bibinfo {author} {\bibfnamefont {S.~A.}\ \bibnamefont
  {Kivelson}}, \bibinfo {author} {\bibfnamefont {M.~R.}\ \bibnamefont
  {Norman}}, \bibinfo {author} {\bibfnamefont {S.}~\bibnamefont {Uchida}}, \
  and\ \bibinfo {author} {\bibfnamefont {J.}~\bibnamefont {Zaanen}},\
  }\bibfield  {title} {\enquote {\bibinfo {title} {From quantum matter to
  high-temperature superconductivity in copper oxides},}\ }\href {\doibase
  10.1038/nature14165} {\bibfield  {journal} {\bibinfo  {journal} {Nature}\
  }\textbf {\bibinfo {volume} {518}},\ \bibinfo {pages} {179--186} (\bibinfo
  {year} {2015})}\BibitemShut {NoStop}%
\bibitem [{\citenamefont {Kyung}\ \emph {et~al.}(2006)\citenamefont {Kyung},
  \citenamefont {Kancharla}, \citenamefont {S\'{e}n\'{e}chal}, \citenamefont
  {Tremblay}, \citenamefont {Civelli},\ and\ \citenamefont {Kotliar}}]{kyung}%
  \BibitemOpen
  \bibfield  {author} {\bibinfo {author} {\bibfnamefont {B.}~\bibnamefont
  {Kyung}}, \bibinfo {author} {\bibfnamefont {S.~S.}\ \bibnamefont
  {Kancharla}}, \bibinfo {author} {\bibfnamefont {D.}~\bibnamefont
  {S\'{e}n\'{e}chal}}, \bibinfo {author} {\bibfnamefont {A.-M.~S.}\
  \bibnamefont {Tremblay}}, \bibinfo {author} {\bibfnamefont {M.}~\bibnamefont
  {Civelli}}, \ and\ \bibinfo {author} {\bibfnamefont {G.}~\bibnamefont
  {Kotliar}},\ }\bibfield  {title} {\enquote {\bibinfo {title} {{Pseudogap
  induced by short-range spin correlations in a doped Mott insulator}},}\
  }\href {\doibase 10.1103/PhysRevB.73.165114} {\bibfield  {journal} {\bibinfo
  {journal} {Phys. Rev. B}\ }\textbf {\bibinfo {volume} {73}},\ \bibinfo {eid}
  {165114} (\bibinfo {year} {2006})}\BibitemShut {NoStop}%
\bibitem [{\citenamefont {Zhao}\ \emph {et~al.}(2017)\citenamefont {Zhao},
  \citenamefont {Belvin}, \citenamefont {Liang}, \citenamefont {Bonn},
  \citenamefont {Hardy}, \citenamefont {Armitage},\ and\ \citenamefont
  {Hsieh}}]{Zhao:2017}%
  \BibitemOpen
  \bibfield  {author} {\bibinfo {author} {\bibfnamefont {L.}~\bibnamefont
  {Zhao}}, \bibinfo {author} {\bibfnamefont {C.~A.}\ \bibnamefont {Belvin}},
  \bibinfo {author} {\bibfnamefont {R.}~\bibnamefont {Liang}}, \bibinfo
  {author} {\bibfnamefont {D.~A.}\ \bibnamefont {Bonn}}, \bibinfo {author}
  {\bibfnamefont {W.~N.}\ \bibnamefont {Hardy}}, \bibinfo {author}
  {\bibfnamefont {N.~P.}\ \bibnamefont {Armitage}}, \ and\ \bibinfo {author}
  {\bibfnamefont {D.}~\bibnamefont {Hsieh}},\ }\bibfield  {title} {\enquote
  {\bibinfo {title} {{A global inversion-symmetry-broken phase inside the
  pseudogap region of YBa2Cu3Oy}},}\ }\href
  {http://dx.doi.org/10.1038/nphys3962} {\bibfield  {journal} {\bibinfo
  {journal} {Nature Physics}\ }\textbf {\bibinfo {volume} {13}},\ \bibinfo
  {pages} {250--254} (\bibinfo {year} {2017})}\BibitemShut {NoStop}%
\bibitem [{\citenamefont {Scheurer}\ \emph {et~al.}(2018)\citenamefont
  {Scheurer}, \citenamefont {Chatterjee}, \citenamefont {Wu}, \citenamefont
  {Ferrero}, \citenamefont {Georges},\ and\ \citenamefont
  {Sachdev}}]{Scheurer:PNAS2019}%
  \BibitemOpen
  \bibfield  {author} {\bibinfo {author} {\bibfnamefont {Mathias~S.}\
  \bibnamefont {Scheurer}}, \bibinfo {author} {\bibfnamefont {Shubhayu}\
  \bibnamefont {Chatterjee}}, \bibinfo {author} {\bibfnamefont {Wei}\
  \bibnamefont {Wu}}, \bibinfo {author} {\bibfnamefont {Michel}\ \bibnamefont
  {Ferrero}}, \bibinfo {author} {\bibfnamefont {Antoine}\ \bibnamefont
  {Georges}}, \ and\ \bibinfo {author} {\bibfnamefont {Subir}\ \bibnamefont
  {Sachdev}},\ }\bibfield  {title} {\enquote {\bibinfo {title} {Topological
  order in the pseudogap metal},}\ }\href {\doibase 10.1073/pnas.1720580115}
  {\bibfield  {journal} {\bibinfo  {journal} {Proceedings of the National
  Academy of Sciences}\ }\textbf {\bibinfo {volume} {115}},\ \bibinfo {pages}
  {E3665--E3672} (\bibinfo {year} {2018})}\BibitemShut {NoStop}%
\bibitem [{\citenamefont {Zaanen}(2019)}]{Zaanen:SciPost2019}%
  \BibitemOpen
  \bibfield  {author} {\bibinfo {author} {\bibfnamefont {Jan}\ \bibnamefont
  {Zaanen}},\ }\bibfield  {title} {\enquote {\bibinfo {title} {{Planckian
  dissipation, minimal viscosity and the transport in cuprate strange
  metals}},}\ }\href {\doibase 10.21468/SciPostPhys.6.5.061} {\bibfield
  {journal} {\bibinfo  {journal} {SciPost Phys.}\ }\textbf {\bibinfo {volume}
  {6}},\ \bibinfo {pages} {61} (\bibinfo {year} {2019})}\BibitemShut {NoStop}%
\bibitem [{\citenamefont {Bagrov}\ \emph {et~al.}(2019)\citenamefont {Bagrov},
  \citenamefont {Danilov}, \citenamefont {Brener}, \citenamefont {Harland},
  \citenamefont {Lichtenstein},\ and\ \citenamefont
  {Katsnelson}}]{Bagrov:arXiv2019}%
  \BibitemOpen
  \bibfield  {author} {\bibinfo {author} {\bibfnamefont {Andrey~A.}\
  \bibnamefont {Bagrov}}, \bibinfo {author} {\bibfnamefont {Mikhail}\
  \bibnamefont {Danilov}}, \bibinfo {author} {\bibfnamefont {Sergey}\
  \bibnamefont {Brener}}, \bibinfo {author} {\bibfnamefont {Malte}\
  \bibnamefont {Harland}}, \bibinfo {author} {\bibfnamefont {Alexander~I.}\
  \bibnamefont {Lichtenstein}}, \ and\ \bibinfo {author} {\bibfnamefont
  {Mikhail~I.}\ \bibnamefont {Katsnelson}},\ }\href@noop {} {\enquote {\bibinfo
  {title} {Detecting quantum critical points in the $t-t'$ fermi-hubbard model
  via complex network theory},}\ } (\bibinfo {year} {2019}),\ \Eprint
  {http://arxiv.org/abs/1904.11463} {arXiv:1904.11463 [cond-mat.str-el]}
  \BibitemShut {NoStop}%
\end{thebibliography}
%

\end{document}